\documentclass[11pt,tightenlines,eqsecnum,floats,aps,amsmath,superscriptaddress,amssymb,nofootinbib,longbibliography]{revtex4-1}

\usepackage{amsmath}
\usepackage{amssymb}
\usepackage{graphicx}
\usepackage{float}
\usepackage{tensor}
\usepackage{physics}
\usepackage{babel}
\usepackage{csquotes}
\usepackage{mathtools}
\usepackage{placeins}
\usepackage{xcolor}
\usepackage[colorlinks=true,linkcolor=blue,citecolor=blue,urlcolor=cyan,plainpages=false,pdfpagelabels]{hyperref}
\usepackage{bm}
\usepackage[bottom]{footmisc}
\usepackage{pifont}
\usepackage[normalem]{ulem}
\usepackage{xspace}
\usepackage{cleveref}
\usepackage{multirow}

\newcommand{\nocontentsline}[3]{}
\let\origcontentsline\addcontentsline
\newcommand\stoptoc{\let\addcontentsline\nocontentsline}
\newcommand\resumetoc{\let\addcontentsline\origcontentsline}

\newcommand{\uS}{{\mathrm{S}}}
\newcommand{\uE}{{\mathrm{E}}}
\newcommand{\uc}{{\mathrm{c}}}
\newcommand{\uin}{{\mathrm{in}}}
\newcommand{\ie}{i.e.~}
\newcommand{\etc}{etc.\xspace}
\newcommand{\tin}{t_{\uin}}
\newcommand{\tO}{t_0}
\newcommand{\xic}{\xi_\uc}
\newcommand{\xio}{\xi_0}
\newcommand{\gp}{g_{\mathrm{p}}}
\newcommand{\bea}{\begin{equation}\begin{aligned}}
\newcommand{\eea}{\end{aligned}\end{equation}}

\newcommand{\T}{\mathrm{T}}

\newcommand{\om}{\omega}
\newcommand{\Om}{\Omega}

\newcommand{\mH}{\mathcal{H}}
\newcommand{\mI}{\mathcal{I}}
\newcommand{\mO}{\mathcal{O}}
\newcommand{\s}{\sigma}

\newcommand{\tdiag}{\mathrm{diag}}
\newcommand{\yes}{\ding{51}}
\newcommand{\no}{\ding{55}}

\newcommand{\pr}{\prime}

\newcommand{\tad}{\mathrm{ad}}
\newcommand{\tnad}{\mathrm{nad}}

\newcommand{\tT}{\mathrm{T}}

\begin{document}

\date{\today}

\title{Recoherence, adiabaticity, and Markovianity in Gaussian maps}

\author{Dimitrios Kranas}
\email{dimitrios.kranas@ens.fr}
\affiliation{Laboratoire de Physique de l’Ecole Normale Sup\'erieure, ENS, CNRS, Universit\'e PSL, Sorbonne Universit\'e, Universit\'e Paris Cit\'e, 75005 Paris, France}

\author{Julien Grain}
\email{
julien.grain@universite-paris-saclay.fr}
\affiliation{Universit\'e Paris-Saclay, CNRS, Institut d'Astrophysique Spatiale, 91405, Orsay, France
}

\author{Vincent Vennin}
\email{vincent.vennin@phys.ens.fr}
\affiliation{Laboratoire de Physique de l’Ecole Normale Sup\'erieure, ENS, CNRS, Universit\'e PSL, Sorbonne Universit\'e, Universit\'e Paris Cit\'e, 75005 Paris, France}

\begin{abstract}
Motivated by the recent discovery of situations where cosmological fluctuations recohere during inflation, we investigate the relationship between quantum recoherence (late-time purification after a transient phase of decoherence), adiabaticity, and Markovianity. To that end, we study a simple setup of two linearly-coupled harmonic oscillators, and compute the purity of one oscillator when the interaction is switched off. We find that there exists a critical value for the coupling strength below which the purity oscillates and above which it decays exponentially. This decay cannot be captured by perturbation theory; hence, decoherence is always a non-perturbative phenomenon. When the interaction is turned off, the purity either freezes to its value prior to the turn-off, or it smoothly goes back to a value very close to one (recoherence). This depends on the rate at which the turn-off occurs. We thus develop a new adiabatic-expansion scheme and find complete recoherence at any finite order in the inverse turn-off time. Therefore, decoherence is always a non-adiabatic effect. The critical value of the turn-off time above which recoherence takes place is then expressed in terms of the other time scales of the problem. Finally, we show that the dynamics of the system is never Markovian, even when decoherence takes place. We introduce a new measure of Markovianity dubbed the Bures velocity and use it to optimise Markovian approximations.
\end{abstract}

\maketitle

\tableofcontents

\section{Introduction}
\label{sec:intro}

The quantum origin of inhomogeneities at cosmic scales has garnered increasing interest in recent years. In the standard model of cosmology, all cosmic structures observed today (from galaxies to galaxy clusters and cosmic filaments) are thought to grow from primordial seeds extracted out of the quantum vacuum in an early phase of accelerated expansion dubbed inflation. The production of such primeval density fluctuations proceeds through a typical mechanism of quantum fields in a strong external field \cite{birrell1982}, which, in this situation, is given by the classical gravitational field driving the quasi-exponential cosmic expansion. In practice, cosmological fluctuations are described by a set of time-dependent harmonic oscillators. As inflation proceeds, their physical wavelength takes over the curvature radius of the universe, and the oscillators become unstable. As a result, primordial inhomogeneities transit from the vacuum state to a $2n$-mode squeezed state~\cite{Grishchuk:1990bj,Albrecht:1992kf,Martin:2007bw,Grain:2019vnq,Colas_2022}. Yet, current astronomical observations \cite{Planck:2018jri,Planck:2019kim,eBOSS:2021owp,DES:2022qpf} could be equally well explained if these inhomogeneities arose from classical statistical fluctuations, which raises two related questions. First, how do such quantum fluctuations acquire classical properties \cite{Polarski_1996,KIEFER_1998,Kiefer:2008ku,Markovian_Starobinsky,SUDARSKY_2011,Burgess:2014eoa,Martin_2016,Martin_2022,Chandran_2024} and second, is there genuine signatures of their quantum-mechanical origin \cite{Campo_2006,Maldacena_2015,Martin_2016b,Choudhury_2017,Choudhury_2017b,Martin_2017,Ando_2020,Green:2020whw,Espinosa_Portal_s_2022,tejerinaperez2024entangleduniverse,Sou_2024}. 

In this context, decoherence plays a key role in the set of obstructions for probing quantum properties of primordial inhomogeneities~\cite{Martin_2017}. Decoherence is the process by which an observed quantum system transits to a statistical mixture when coupled to an unobserved environment \cite{Zurek:1981xq,Zurek:1982ii,Joos:1984uk}. Characterising the channels by which primordial fluctuations experience decoherence is essential not only to assess whether this obstruction can be circumvented, but also because it is commonly thought to come with modifications of inflationary observables and of the emergence of cosmic structures \cite{Martin:2018zbe,Martin:2018lin,DaddiHammou:2022itk,Lopez:2025arw}. This is why it has become important to evaluate the amount of entanglement present in quantum fluctuations and how decoherence proceeds in cosmological spacetimes \cite{Brandenberger:1990bx,Barvinsky:1998cq,Lombardo:2004fr,Lombardo:2005iz,Martineau:2006ki,Prokopec:2006fc,Burgess:2006jn,Sharman:2007gi,Campo:2008ju,Nelson:2016kjm,Burgess:2022nwu,Ning:2023ybc,Colas:2022kfu,Colas:2022hlq,Colas:2024xjy,Burgess:2024eng,Colas:2024ysu}. 

Quantum environments are ubiquitous in the primordial universe. For instance, they arise because of the non-linear self-interaction of gravity, in which case the environment is made of the unobserved modes at short wavelengths (see, for instance \cite{Burgess:2022nwu}). Another situation is inflation with multiple fields where density fluctuations interact with entropic perturbations \cite{Achucarro:2010da,Cespedes:2012hu,Achucarro:2012sm,Assassi:2013gxa,Tong:2017iat,Shiu:2011qw,Pinol:2020kvw,Pinol:2021aun,Jazayeri:2022kjy,Turok:1987pg,Damour:1995pd,Kachru:2003sx,Krause:2007jk,Chen:2009zp,Pi:2012gf,Chen:2012ge,Tolley:2009fg,Baumann:2014nda}. Here, entropic fluctuations are massive enough to be inaccessible, yet sufficiently coupled to density fluctuations for the latter not be in isolation; hence qualifying entropic perturbations as a quantum environment to be traced over \cite{Colas:2022kfu,Burgess:2024heo}. However, cosmological environments strongly depart from the very idea of a thermal bath. Indeed, both cosmic inhomogeneities and cosmic environments are out of equilibrium since they evolve in a dynamical background spacetime that constantly pumps new pairs of quanta out of the vacuum fluctuations. Moreover, such environments might not be large but instead composed of a few eigenfrequencies, meaning that information is not necessarily redistributed on short time scales. From a theoretical standpoint, this leads to an open quantum field theory \cite{breuerTheoryOpenQuantum2002,Calzetta:2008iqa,Boyanovsky:2015xoa,Burgess:2022rdo,Burgess:2020tbq,Colas:2023wxa,Salcedo:2024smn,Salcedo:2024nex} with many applications in cosmology \cite{Fukuma:2013uxa,Boyanovsky:2015jen,Boyanovsky:2015tba,Burgess:2015ajz,Hollowood:2017bil,Shandera:2017qkg,Choudhury:2017qyl,Choudhury:2018ppd,Boyanovsky:2018soy,Brahma:2020zpk,Rai:2020edx,Martin:2021xml,Martin:2021qkg,Banerjee:2021lqu,Brahma:2022yxu,Brahma:2023hki,Brahma:2021mng,Alicki:2023tfz,Alicki:2023rfv,Creminelli:2023aly,Bhattacharyya:2024duw} and in the broader context of open quantum field theory in relativistic gravitational backgrounds, including accelerating spacetimes and black holes (see for example \cite{Anastopoulos_2013,Burrage:2018pyg,Cheung:2018cwt,Kaplanek:2019dqu,Akhtar:2019qdn,Kaplanek:2019vzj,Kaplanek:2020iay,Kaplanek:2021fnl,Burgess:2021luo,Kaplanek:2022xrr,Cao:2022kjn,chaykovLoopCorrectionsMinkowski2022a,Prudhoe:2022pte,Kading:2022hhc,Kading:2023mdk,Bowen:2024emo,Belfiglio:2024qsa,cho2025nonmarkovianquantummasterfokkerplanck}).  From the perspective of decoherence of cosmological fluctuations, embedding quantum fields in a cosmological background leads to a variety of situations. It ranges from full decoherence to almost full recoherence depending on the type of coupling, the mass of the environmental degree of freedom, and the peculiar dynamics of the background spacetime \cite{Colas:2022kfu,Colas:2022hlq,Colas:2024xjy,Brahma:2024ycc,Bhattacharyya:2025cxv}. Interestingly, this variety of behaviours arises already in the simple situation of one environmental degree of freedom and linear evolution, which was shown to be robust against perturbative non-linearities \cite{Colas:2024ysu}. This strongly departs from the flat-space case, where alternating phases of decoherence and recoherence are expected. 

The fact that cosmological fluctuations recohere in some situations shows that their effective dynamics is highly non-Markovian. In contrast, situations where full decoherence proceeds suggest that Markovianity can emerge even for small environments.\footnote{Note that, {\it per se}, full decoherence does not imply Markovianity. Yet, it remains a strong indication of emergent Markovianity; see \cite{Burgess:2022nwu,Brahma:2022yxu} in the cosmological context.} The reason for such diversity is the time dependence of the frequencies of both the system and the environment, and of their coupling strength.  This means that the quantumness of cosmological inhomogeneities at the end of inflation is not solely fixed by the parameters of the model but strongly depends on their entire temporal evolution, for instance, whether they vary slowly or rapidly. In that respect, WKB approaches exploit the slow variation of such parameters to approximately solve for the dynamics. These have been extensively used in quantum field theory in curved spaces and in primordial cosmology \cite{Winitzki:2005rw,Grain:2006dg,Grain:2007yv}. In the latter case, slow variations of the parameters of the dynamics are to be understood against time, and this is why the WKB approximation is sometimes referred to as an adiabatic approximation. At a technical level, it allows us to get analytical insights and to predict cosmological observables in situations where the dynamics cannot be exactly solved for because of the intricate time-dependence of the parameters \cite{Martin:2002vn,Casadio:2004ru,Casadio:2005em}. At a more fundamental level, this approximation is essential to define the so-called adiabatic vacua of quantum fields embedded in cosmological spacetimes and to regularise higher-order cosmological correlators thanks to adiabatic renormalisation \cite{PhysRevD.9.341,FULLING1974176,TSBunch_1980,PhysRevD.36.2963,Animali:2022lig}.\\

Mostly motivated by such a cosmological context, we investigate the link between decoherence, Markovianity, and adiabaticity. To this end, we consider a situation where the system is a harmonic oscillator coupled to an environment made of one single harmonic oscillator too, and both are placed in their vacuum state initially. However, we here consider an explicitly time-dependent coupling strength. Though we are not in the realm of quantum fields, this simple setup is in fact a convenient proxy for cosmological inhomogeneities. This is because scales decouple at leading order in cosmological perturbation theory, and fluctuations during inflation are thus given by a collection of such time-dependent harmonic oscillators that can be treated separately. Moreover, the system plus the environment following linear evolution, their quantum state falls in the class of Gaussian states \cite{AlessioSerafini_2004,Colas_2022,Martin_2023}. These are entirely described by their covariance matrix\footnote{Gaussian states are fully characterised by the vector of first moments, containing the expectation values of the operators associated with the system, and the covariance matrix, containing the two-point functions. For non-displaced quantum states, as those emerging in cosmology and also considered here, the vector of first moments vanishes and thus, the covariance matrix fully characterises the quantum state.}, which is easily related to cosmological observables. This also means that tracers of entanglement are all related to the symplectic eigenvalues of the covariance matrix \cite{AlessioSerafini_2004}. Here, we will use the purity parameter as a measure of decoherence since it is readily related to other measures of entanglement such as the Renyi-2 entropy \cite{renyi1961measures,Kudler-Flam:2022zgm,Kudler-Flam:2023kph}. Finally, this model gives us full control of the parameter space. This allows us to cover the various situations encountered in the cosmological context within a single framework, including the cases of unstable directions and of rapid-versus-slow variation of the parameters. This latter case is an ideal playground to further develop WKB-like approaches in the context of decoherence and paves the way to connect techniques employed in cosmology to quantum information theory. 

The paper is organised as follows. In \cref{sec:Model}, we introduce our model and show that the diversity of situations arising for cosmological perturbations is recovered within that framework. In \cref{sec:Criticality:Perturbativity}, we explore in greater detail the parameter space of the model. We identify several phases depending on the coupling strength and the hierarchy between the frequencies of the system and the environment, and showcase their impact on decoherence, considering an instantaneous switch-on and switch-off of the coupling. In particular, we highlight that perturbativity is not solely controlled by the coupling strength, meaning that reliable predictions can be obtained from perturbation theory in the supercritical regime where the coupling strength is large enough for an unstable direction to emerge. In \cref{sec:Adiabaticity}, we investigate the role of adiabatic variations of the parameters of the model on decoherence. To this end, we introduce an adiabatic-expansion scheme that consists of a perturbative expansion in the time derivative of the coupling rather than its strength. This allows us to determine whether late-time decoherence can be achieved or not under adiabatic evolution and considering the subcritical regime in which the two coupled harmonic oscillators have only stable directions. In \cref{sec:Markov}, we explore under which circumstances a Markovian dynamics emerges. Purity is a measure of decoherence but not of Markovianity since decoherence can be achieved even in non-Markovian evolution. Hence, we introduce the Bures velocity as a measure of non-Markovianity. It is obtained by separating the Markovian direction from the non-Markovian one, and we establish its relation to the purity parameter. Finally, we conclude in \cref{sec:Conclusions} by summarising our results and opening on several perspectives. The paper ends with a couple of appendices, to which the most technical details of our investigations are deferred. 

\section{The model}
\label{sec:Model}

\subsection{General framework}
We consider two quantum harmonic oscillators $(\hat{x}_\uS, \hat{p}_\uS)$ and $(\hat{x}_\uE, \hat{p}_\uE)$ of unit mass and eigenfrequencies $\om_\uS$ and $\om_\uE$, respectively. We are interested in the situation where the two oscillators decouple in the asymptotic past and future, and linearly interact during a finite time interval. The interaction we consider is thus of the form $\xi(t)x_\uS x_\uE $, where $\xi(t)$ is a coupling function such that
$\underset{t \rightarrow \pm \infty}{\lim} \xi(t) = 0$. The Hamiltonian of the system is
\begin{equation}
\hat{H}=\frac{1}{2}\hat{p}_\uS^2+\frac{1}{2}\om_\uS^2\hat{x}_\uS^2+\frac{1}{2}\hat{p}_\uE ^2+\frac{1}{2}\om_\uE ^2\hat{x}_\uE ^2+\xi(t)\hat{x}_\uS\hat{x}_\uE\, , \label{Ham.2SHOs}
\end{equation}
resulting in the equations of motion
\begin{align}
&\ddot{\hat{x}}_\uS +\om_\uS ^2\hat{x}_\uS =-\xi(t)\hat{x}_\uE \, ,\label{eom.xS} \\
&\ddot{\hat{x}}_\uE +\om_\uE ^2\hat{x}_\uE =-\xi(t)\hat{x}_\uS \, , \label{eom.xE}
\end{align}
where a dot denotes derivation with respect to time.
Note that, although the present analysis is motivated by questions arising in cosmology and discussed in \cref{sec:intro}, this model is not meant to mimic known early-universe setups. In particular, in cosmology the eigenfrequencies also depend on time, which we do not assume here. Our goal is rather to identify which features of the coupling function are relevant to determine the fate of the purity of the system, on generic grounds.

Note that any other linear interaction can be written in the above form by time-independent local canonical transformations (hereafter, ``local'' refers to operations acting solely on the $\uS$ or $\uE$ degrees of freedom). For example, the momentum-position interaction $\xi(t)p_\uS x_\uE $ can be brought into the form~(\ref{Ham.2SHOs}) by performing the canonical transformation $x_\uS =-\tilde{p}_\uS /\om_\uS $, $p_\uS =\om_\uS \tilde{x}_\uS $, where the $\uE$ sector is left invariant. This transformation can be viewed as a rotation by $90^\circ$ in the local phase space of $\uS$, followed by a rescaling (squeezing) by a factor $\om_\uS $. The new effective coupling function is $\tilde{\xi}(t)=\om_\uS \psi(t)$. 
Without loss of generality, hereafter we thus focus on the position-position interaction (note that our results do not directly apply if the frequencies are time dependent or if several interaction terms are present). Reference \cite{Bhattacharyya:2025cxv} explores the different effects of position-position and position-momentum interactions in the cosmological context.

\subsection{Transport equation}
It is convenient to define the vector of operators $\hat{\bm{R}}=(\hat{x}_\uS ,\hat{p}_\uS ,\allowbreak \hat{x}_\uE ,\hat{p}_\uE )^\T$. Then, the Hamiltonian operator can be expressed in the compact form
\begin{equation}
\hat{H}=\frac{1}{2}\hat{\bm{R}}^\T\bm{\mH}\hat{\bm{R}} \, , \label{Ham.oper}
\end{equation}
where the Hamiltonian matrix reads
\begin{equation}
\bm{\mH}=
\begin{pmatrix}
\om_\uS ^2 & 0 & \xi(t) & 0\\
0 & 1 & 0 & 0\\
\xi(t) & 0 & \om_\uE ^2 & 0\\
0 & 0 & 0 & 1\\
\end{pmatrix}.
\end{equation}
To study the quantum decoherence of the ``system'' oscillator $\uS$ induced by its interaction with the ``environment'' oscillator $\uE$, we evaluate the evolution of the purity $\gamma_\uS $ associated with the quantum state $\hat{\rho}_\uS $ of the system. In general, the purity of a quantum state $\hat{\rho}$ is defined as $\gamma=\Tr(\hat{\rho}^2)$. It equals unity for pure states and takes values between $0$ and $1$ for mixed states, with $\gamma\to 0$ corresponding to strongly mixed (\ie fully decohered) states. For Gaussian states, the purity reduces to~\cite{serafini}
\begin{equation}
\gamma_=\frac{1}{\sqrt{\det(\bm{\sigma})}}, \label{purity.def}
\end{equation}
where $\bm{\sigma}$ is the covariance matrix of the state under consideration, defined by
\begin{equation}
\bm{\sigma} = \langle\{\hat{\bm{R}}-\bm{\mu}, \hat{\bm{R}}^\T-\bm{\mu}^\T\}\rangle .\label{cov.mat.def}
\end{equation}
Here, $\bm{\mu}=\langle \hat{\bm{R}} \rangle$ is the vector of first moments and we employ the notation $\langle\hat{O}\rangle\equiv \mathrm{Tr}(\hat{\rho} \hat{O})$. The curly brackets $\{\cdot,\cdot\}$ represent the anticommutator.

Both oscillators are prepared in their vacua, which are Gaussian states. Furthermore, quadratic Hamiltonians, such as the one under consideration in this model, preserve the Gaussian nature of the state. Note also that $\bm{\mu}=\bm{0}$ for the vacuum state and throughout the evolution generated by \cref{Ham.oper}. The covariance matrix of the joint system-and-environment setup can be written as
\begin{equation}
\bm{\sigma}=
\begin{pmatrix}
\bm{\sigma}_\uS &\bm{\sigma}_{\uS \uE}\\
\bm{\sigma}_{\uS \uE}^\T&\bm{\sigma}_\uE 
\end{pmatrix}, \label{sigma_tot}
\end{equation}
where
\begin{equation}
\bm{\sigma}_{\mathrm{I}}=
\begin{pmatrix}
2\left<\hat{x}_{\mathrm{I}}^2\right> & \left<\hat{x}_{\mathrm{I}}\hat{p}_{\mathrm{I}}+\hat{p}_{\mathrm{I}}\hat{x}_{\mathrm{I}}\right>\\
\left<\hat{x}_{\mathrm{I}}\hat{p}_{\mathrm{I}}+\hat{p}_{\mathrm{I}}\hat{x}_{\mathrm{I}}\right> & 2\left<\hat{p}_{\mathrm{I}}^2\right>
\end{pmatrix} \label{sigma_I}
\end{equation}
is the covariance matrix describing the quantum state of the ${\mathrm{I}}$-oscillator $({\mathrm{I}}=\uS,\uE)$
and 
\begin{equation}
\bm{\sigma}_{\uS \uE}=
2\begin{pmatrix}
\left<\hat{x}_\uS\hat{x}_\uE\right> & \left<\hat{x}_\uS \hat{p}_\uE \right>\\
\left<\hat{p}_\uS \hat{x}_\uE \right> & \left<\hat{p}_\uS \hat{p}_\uE \right>
\end{pmatrix} \label{sigmaSE}
\end{equation}
is the matrix containing the cross-correlations between the system and the environment. 

The covariance matrix satisfies the transport equation (see, for instance, \cite{Benchmarking}) 
\begin{equation}
\frac{\dd \bm{\s}}{\dd t}=\bm{\Om}\bm{\mH}\bm{\s}-\bm{\s}\bm{\mH}\bm{\Om}, \label{transport.eq}
\end{equation}
where
\begin{equation}
\bm{\Om}=
\begin{pmatrix}
0&1&0&0\\
-1&0&0&0\\
0&0&0&1\\
0&0&-1&0
\end{pmatrix}.
\end{equation}
From the solution of this equation, one can extract the upper-left block $\bm{\sigma}_\uS (t)$, which, inserted into \cref{purity.def}, gives $\gamma_\uS(t)$.\footnote{Note that, since the whole setup is in a pure state, the system and environment have the same purity, $\gamma_\uS(t)=\gamma_\uE(t)$. \label{footnote:gammaS=gammaE}} Our goal is to solve \cref{transport.eq} for different model parameters and study the evolution of the system's purity in different regimes.

\subsection{The adiabatic basis}\label{sec:rotated:basis}

When the coupling function $\xi$ is a constant, the above setup is equivalent to two uncoupled oscillators under a global rotation. In the basis where the oscillators are uncoupled, the dynamics is trivial, which makes it convenient to use in intermediate steps of the purity calculation. If $\xi(t)$ depends on time, the two oscillators couple even in that basis, but they do so via time derivatives of the coupling function, which is why it will be referred to as the ``adiabatic'' basis. The goal of this section is to introduce it explicitly.

Starting from $\hat{\bm{R}}$, let us consider the rotated phase-space vector $\hat{\bm{Y}}$ defined as
\bea 
\hat{\bm{R}} = \bm{S}_\theta \hat{\bm{Y}}
\quad\text{where}\quad
\bm{S}_\theta=
\begin{pmatrix}
\cos{\theta} & 0 & \sin{\theta} & 0\\
0 & \cos{\theta} & 0 & \sin{\theta}\\
-\sin{\theta} & 0 & \cos{\theta} & 0\\
0 & -\sin{\theta} & 0 & \cos{\theta}
\end{pmatrix} . \label{S.theta}
\eea 
In the rotated basis, the Hamiltonian operator~\eqref{Ham.oper} is given by
\begin{equation}
\hat{H}=\frac{1}{2}\bm{Y}^{\mathrm{T}}\bm{\mH}_{\bm{Y}}\bm{Y}
\quad\text{where}\quad
\bm{\mH}_{\bm{Y}}=\bm{S}_{\theta}^{\mathrm{T}}\bm{\mH}\bm{S}_{\theta}+\bm{S}_{\theta}^{\T}\bm{\Om}\dot{\bm{S}}_\theta
\, .
\end{equation}
When $\xi$ is constant, the off-diagonal blocks of $\bm{\mH}_{\bm{Y}}$ vanish if one sets
\begin{equation}
\theta=\frac{1}{2}\arctan{\left(\frac{2\xi}{\om_\uE ^2-\om_\uS ^2}\right)}. \label{theta}
\end{equation}
This choice for the rotation angle corresponds to the adiabatic basis and leads to
\bea
\label{eq:Hy}
\bm{\mH}_{\bm{Y}} = 
\begin{pmatrix}
\om_1^2(t) & 0 & 0 & \dot{\theta}(t)\\
0 & 1 & -\dot{\theta}(t) & 0\\
0 & -\dot{\theta}(t) & \om_2^2(t) & 0\\
\dot{\theta}(t) & 0 & 0 & 1\\
\end{pmatrix}\, ,
\eea 
where the two normal frequencies $\om_1$ and $\om_2$ are given by 
\bea
\label{eq:om1:om2:def}
\om_1=&\frac{1}{\sqrt{2}}\sqrt{\om_\uS ^2+\om_\uE ^2-\sqrt{4\xi^2+(\om_\uE ^2-\om_\uS ^2)^2}}\, , \\
\om_2=&\frac{1}{\sqrt{2}}\sqrt{\om_\uS ^2+\om_\uE ^2+\sqrt{4\xi^2+(\om_\uE ^2-\om_\uS ^2)^2}}\, . 
\eea 

In the adiabatic basis, the two oscillators couple through $\dot{\theta}$, hence $\dot{\xi}$, as announced above. More explicitly, decomposing $\hat{\bm{Y}}=\left(\hat{x}_1,\hat{p}_1,\hat{x}_2,\hat{p}_2\right)^\mathrm{T}$, the interacting part of the Hamiltonian in the adiabatic basis reads $\dot{\theta}(t)(\hat{x}_1\hat{p}_2-\hat{p}_1\hat{x}_2)$, hence the interaction is of the position-momentum type. When $\xi$ is constant, the equations of motion read $\ddot{\hat{x}}_i+\om_i^2\hat{x}_i=0$, hence 
\bea
\left.\hat{x}_i(t)\right\vert_{\dot{\xi}=0}=&\hat{b}_i\frac{1}{\sqrt{2|\om_i|}}e^{-i\om_1 (t-\tin)}+\hat{b}^\dagger_i\frac{1}{\sqrt{2|\om_i|}}e^{+i\om_i (t-\tin)}, 
\label{eq:xi:free:sol}\\
\eea 
where $i=1,2$ and $\tin$ is a reference time. From \cref{eq:om1:om2:def}, one can check that $\om_2$ is always real, while $\om_1$ is real only when $\xi\leq \om_\uS \om_\uE $ and is purely imaginary otherwise. In that latter case, $e^{\pm i\om_1t}=e^{\mp|\om_1|t}$ become exponentially decaying and growing with time. We therefore anticipate that
\bea
\label{eq:xic:def}
\xic = \om_\uS \om_\uE
\eea 
plays the role of a critical value for the coupling strength $\xi$, and we will see various manifestations of this property. The regime $\xi<\xic$ will be referred to as ``subcritical'', while $\xi>\xic$ will be described as ``supercritical''.

The Hamiltonian~\eqref{eq:Hy} can be further simplified by rescaling the position and momentum variables according to $\hat{x}_i\rightarrow \hat{X}_i/\sqrt{\om_i}$ and $\hat{p}_i\rightarrow \sqrt{\om_i} \hat{P}_i$, \ie introduce the rescaled vector $\hat{\bm{Z}}=(\hat{X}_1,\hat{P}_1,\hat{X}_2,\hat{P}_2)^{\mathrm{T}}$ defined as
\begin{equation}
\hat{\bm{Y}}=\bm{S}_{\om_1,\om_2} \hat{\bm{Z}} \, ,\label{inverse.rescaling}
\end{equation}
with
\begin{equation}
\bm{S}_{\om_1,\om_2}=\begin{pmatrix}
\bm{s}({\om_1}) & \bm{0}\\
\bm{0} & \bm{s}({\om_2})
\end{pmatrix}
\quad\text{where}\quad
\bm{s}({\om})=\begin{pmatrix}
1/\sqrt{\om} & 0\\
0 & \sqrt{\om}
\end{pmatrix}
\end{equation}
is a local squeezing matrix. In the $\bm{Z}$-basis, the Hamiltonian is $\hat{H}=\frac{1}{2}\hat{\bm{Z}}^{\mathrm{T}}\bm{\mH}_{\bm{Z}}\hat{\bm{Z}}$
where $\bm{\mH}_{\bm{Z}}=\bm{S}_{\om_1,\om_2}^{\mathrm{T}}\bm{\mH_Y}\bm{S}_{\om_1,\om_2}+\bm{S}_{\om_1,\om_2}^{\T}\bm{\Om}\dot{\bm{S}}_{\om_1,\om_2}$, which leads to
\begin{equation}
\label{eq:HZ}
\bm{\mH}_{\bm{Z}}=
\begin{pmatrix}
\om_1(t) & \beta_1(t) & 0 & \dot{\theta}(t)\sqrt{\frac{\om_2(t)}{\om_1(t)}}\\
\beta_1(t) & \om_1(t) & -\dot{\theta}(t)\sqrt{\frac{\om_1(t)}{\om_2(t)}} & 0\\
0 & -\dot{\theta}(t)\sqrt{\frac{\om_1(t)}{\om_2(t)}} & \om_2(t) & \beta_2(t)\\
\dot{\theta}(t)\sqrt{\frac{\om_2(t)}{\om_1(t)}} & 0 & \beta_2(t) & \om_2(t)\\
\end{pmatrix}
\quad\text{where}\quad\beta_i=\frac{\dot{\om}_i(t)}{2\om_i(t)}\, .
\end{equation}
The rescaled basis has the advantage of treating position and momentum on the same footing, which will be useful below.

\subsection{Criticality and recoherence}
\label{sec:Criticality:and:Recoherence}

\begin{figure}[t]
\centering 
\includegraphics[width=0.48\textwidth]{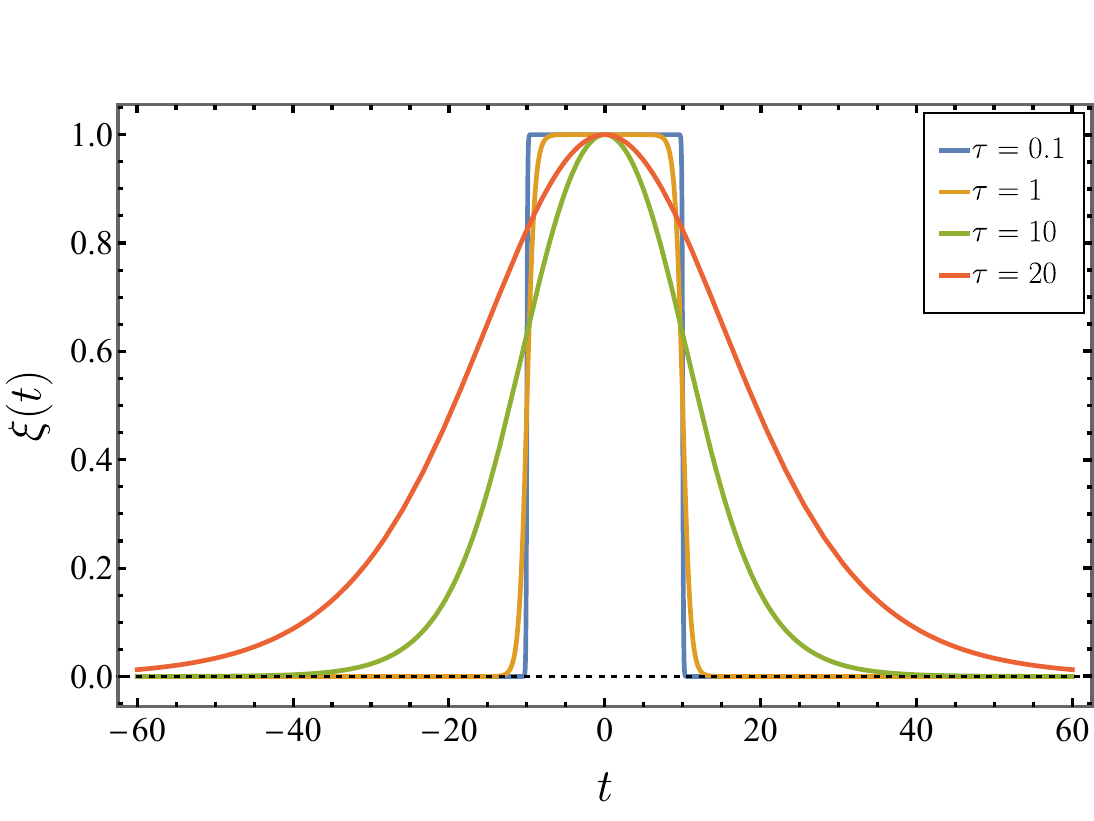} 
\includegraphics[width=0.48\textwidth]{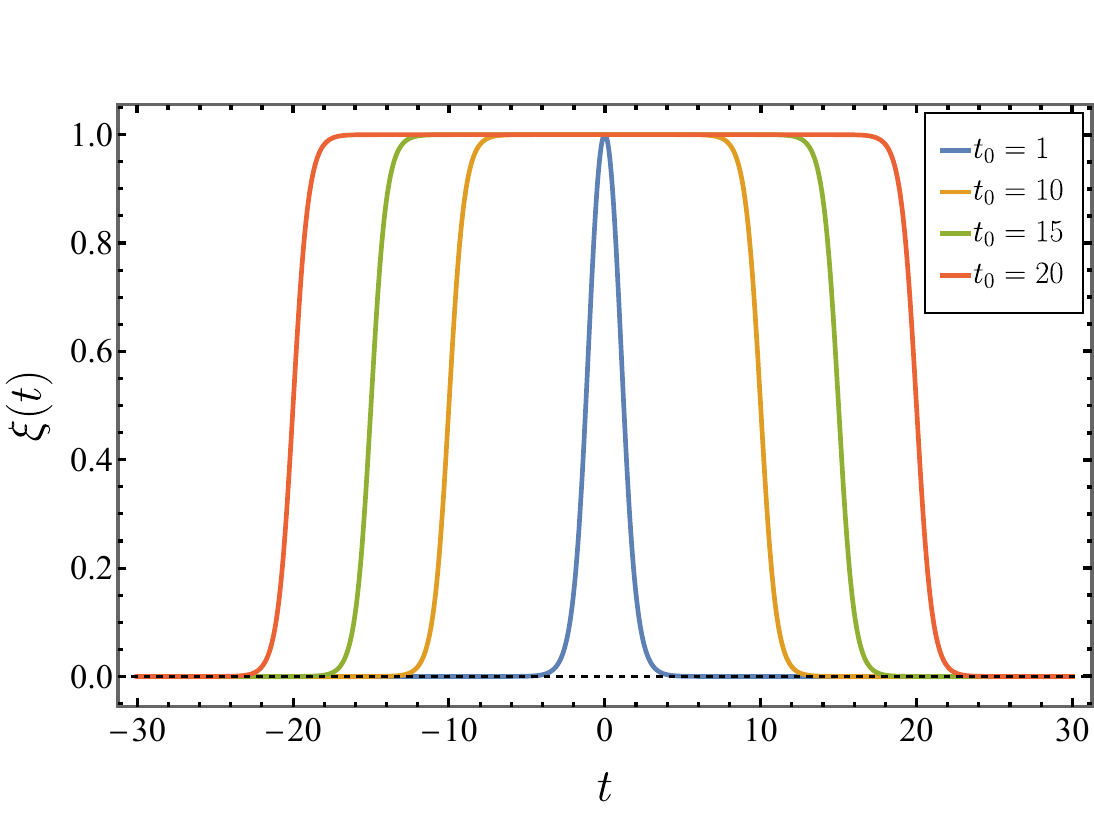}
\caption{Coupling function for a few values of $\tau$ with $\xio=1$ and $\tO =10$ (left panel), and for a few values of $\tO$ with $\xio=1$ and $\tau=1$ (right panel).
}\label{fig.xi} 
\end{figure}

The covariance matrix being a $4\times 4$ symmetric matrix, it contains $10$ independent elements, and the transport equation~\eqref{transport.eq} is thus a set of 10 coupled ordinary differential equations. They are solved with initial conditions corresponding to the vacuum state $\ket{0}=\ket{0}_\uS \otimes \ket{0}_\uE $, namely $\bm{\sigma}(\tin)=\tdiag\left(\om_\uS ^{-1},\om_\uS ,\om_\uE ^{-1},\om_\uE \right)$, where $\tin$ is some initial time at which the coupling effectively vanishes.  
In practice, the coupling function is taken to be of the form
\begin{equation}
\xi(t)=\xio\frac{1}{1+\tanh^2(\frac{\tO }{\tau})}\left[1+\tanh(\frac{\tO +t}{\tau})\tanh(\frac{\tO -t}{\tau})\right]. \label{xi}
\end{equation}
This function vanishes in the asymptotic past and future, it takes positive values of order $\xio$ over a time interval of duration $2\tO$ centred around $t=0$, and the rate at which the coupling is turned on and off is set by the timescale $\tau$. It is displayed in \cref{fig.xi} for a few values of $\tau$ and $\tO$. The interplay between $\tau$ and $\tO $ dictates the shape of the interaction: if $\tau\ll \tO $, it is of the top-hat type, while it becomes sharply peaked if $\tau\gg \tO $. In order to make sure that numerical integration starts when $\xi$ is still negligible, in practice we set $\tin=-\tO-20\tau$. Finally, note that the setup is invariant under a global rescaling of time, which allows us to set $\om_\uS=1$ without loss of generality and have only four free parameters: $\om_\uE $, $\xio$, $\tau$, and $\tO $.\\

\begin{figure}[t]
\centering
\includegraphics[width=0.7\textwidth]{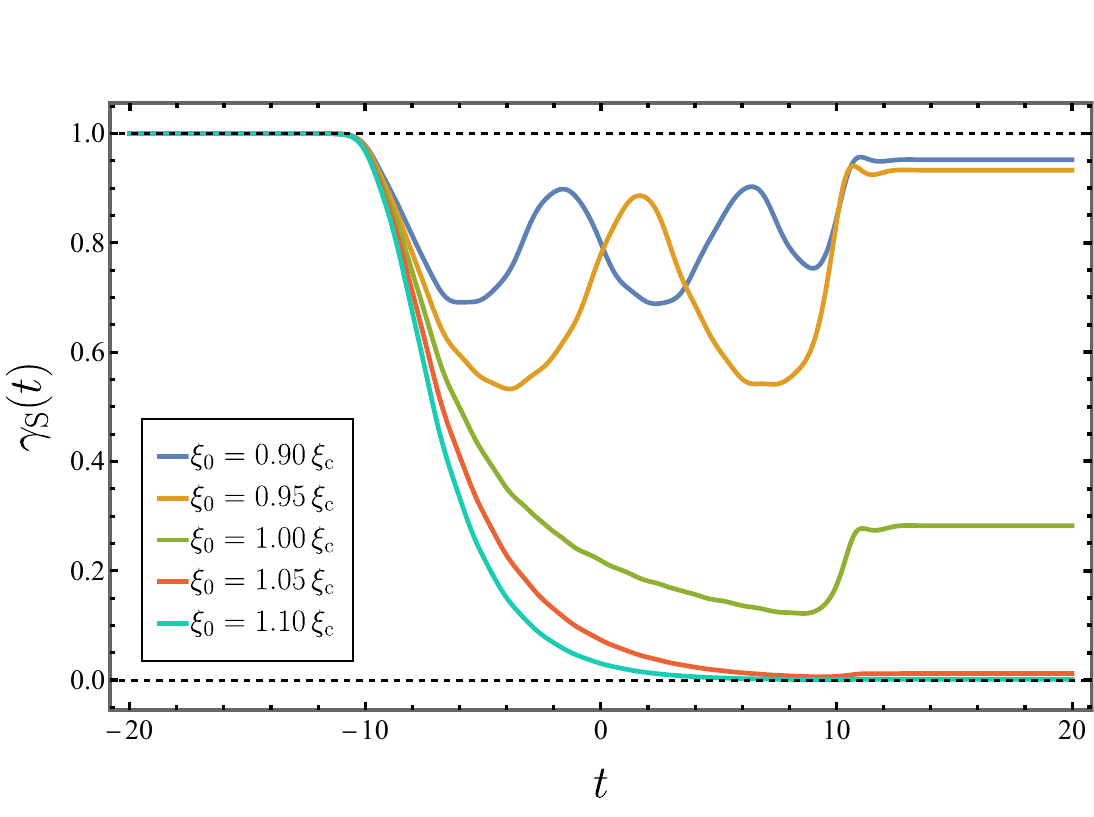}
\caption{Purity as a function of time for a few values of the coupling strength $\xio$, with $\om_\uS =1$, $\om_\uE =2$, and with $\tau=1$ and $\tO =10$ (this corresponds to the orange line in the right panel of \cref{fig.xi}).
}\label{fig.gs.cases} 
\end{figure}

Let us start by building some intuition about how the system's purity behaves in such a setup, before gaining analytical insight into its different regimes. In \cref{fig.gs.cases}, we show the time behaviour of $\gamma_\uS$ for different values of $\xio/\xic$. This confirms that $\xic$, defined in \cref{eq:xic:def}, plays the role of a critical value for the coupling strength. In the supercritical regime, $\xio>\xic$, the purity decreases monotonically during the interaction. The two oscillators become entangled, resulting in \textit{decoherence} of their individual quantum states. In the subcritical regime, the purity oscillates during the interaction, signalling that information moves back and forth between the two quantum systems and their mutual correlations. This allows for \textit{recoherence}, \ie backflow of information to the individual oscillators, which is associated with an increase of their purity. 

\begin{figure}[t]
\centering 
\includegraphics[width=0.48\textwidth]{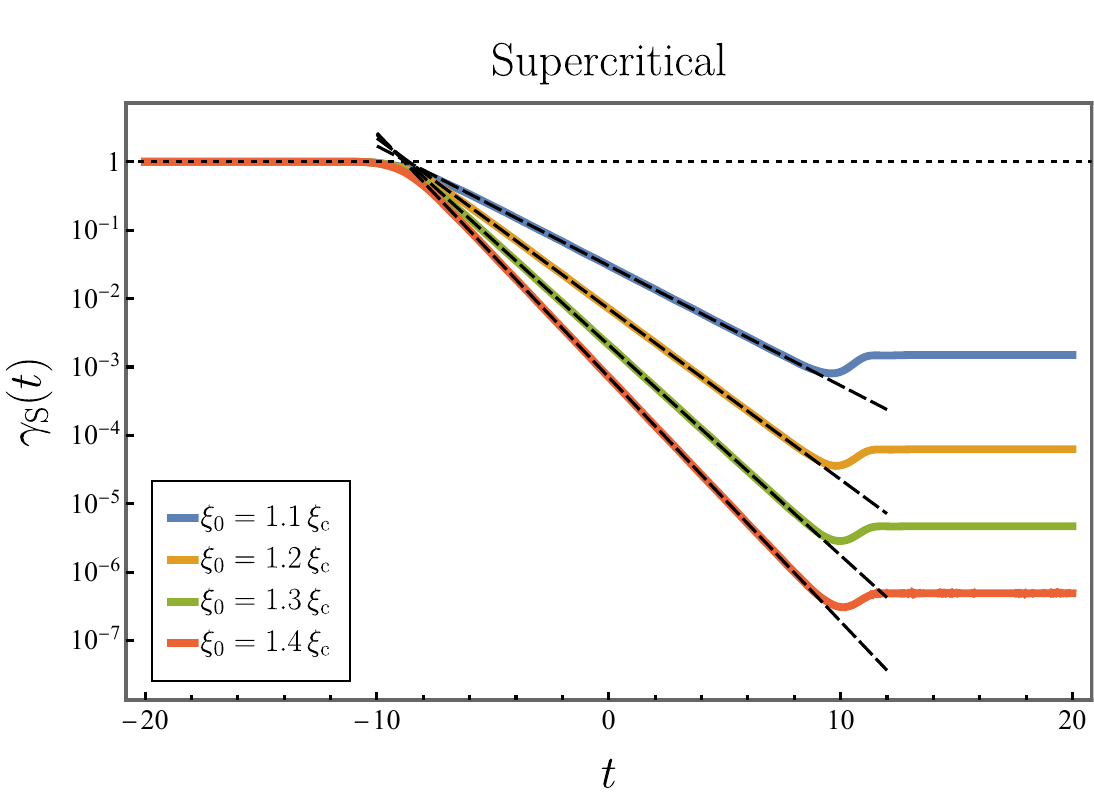} 
\includegraphics[width=0.48\textwidth]{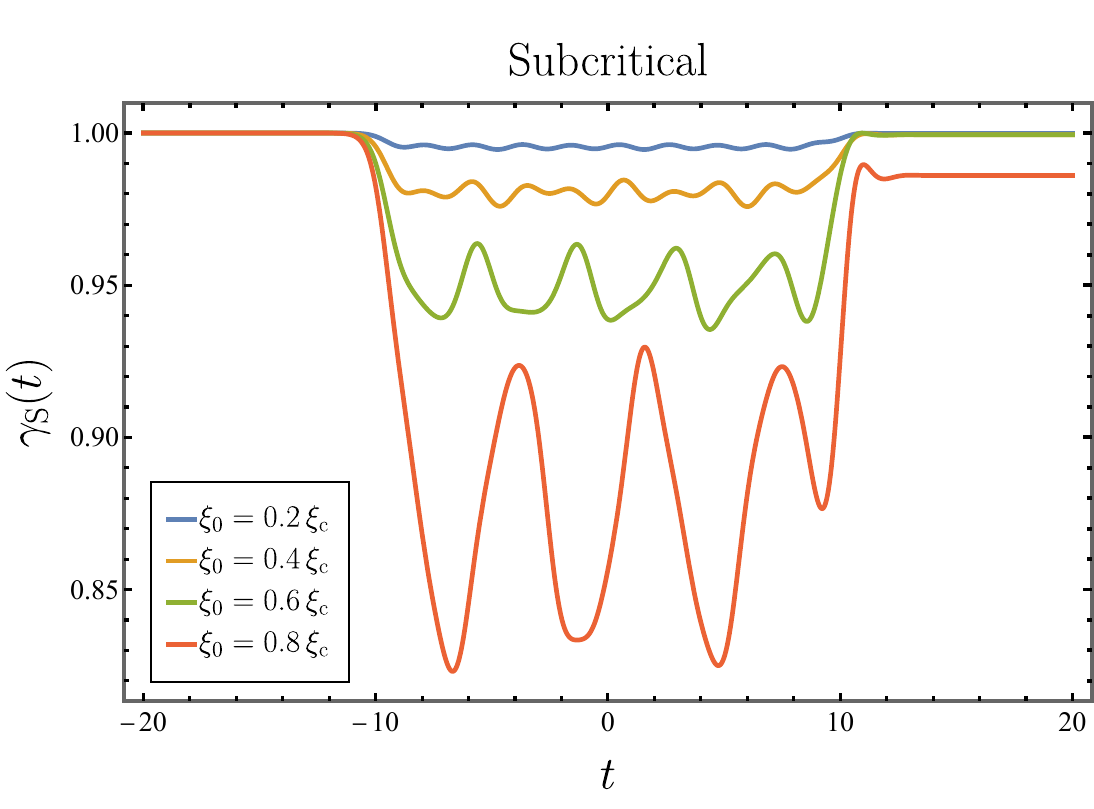}
\caption{Same as in \cref{fig.gs.cases} where the supercritical cases are displayed in the left panel with a logarithmic scale, and the subcritical cases in the right panel. In the supercritical regime, the purity decays as $e^{-\vert\om_1\vert t}$.
}\label{fig.gS.var.xio}
\end{figure}

The super- and subcritical cases are displayed separately in \cref{fig.gS.var.xio}, where a logarithmic scale is used for the supercritical cases to highlight that purity decays exponentially with time in that case, with a rate $\vert\om_1\vert$. This will be shown explicitly in \cref{sec:ISOSO:discussion}. In the subcritical regime, the amplitude of the oscillations increases with the coupling strength, in a way that will also be characterised below. 

\begin{figure}[t]
\centering 
\includegraphics[width=0.48\textwidth]{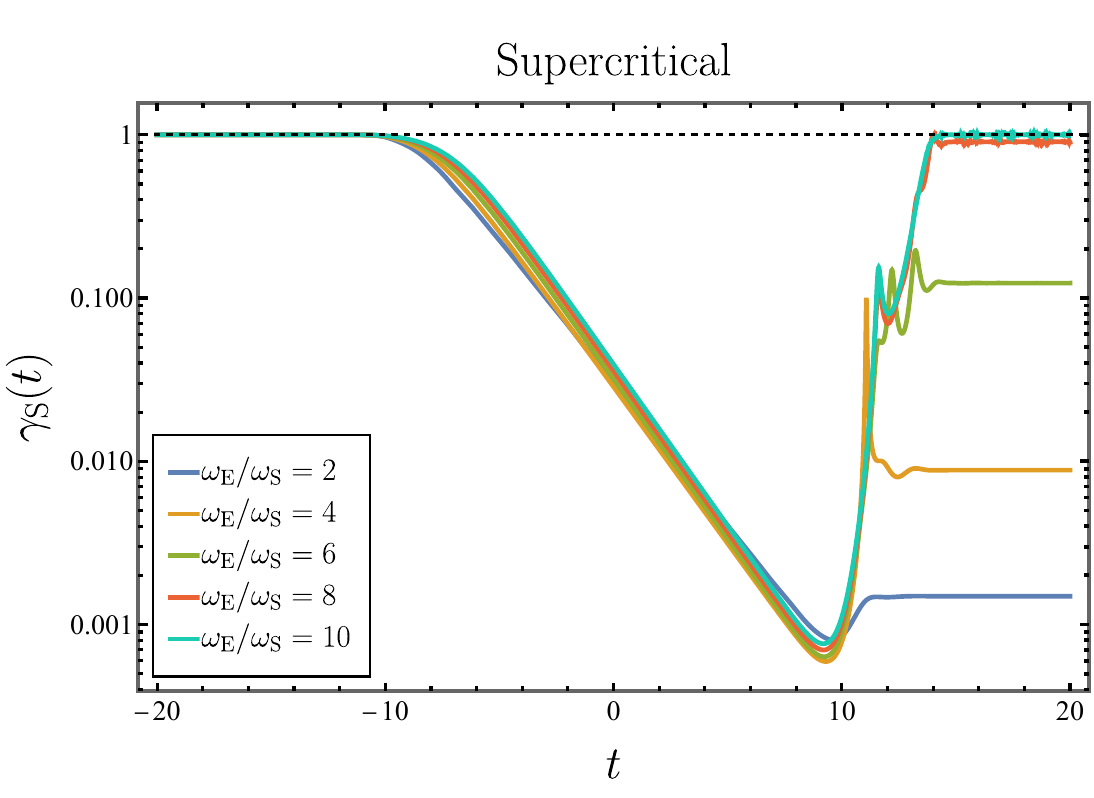}
\includegraphics[width=0.48\textwidth]{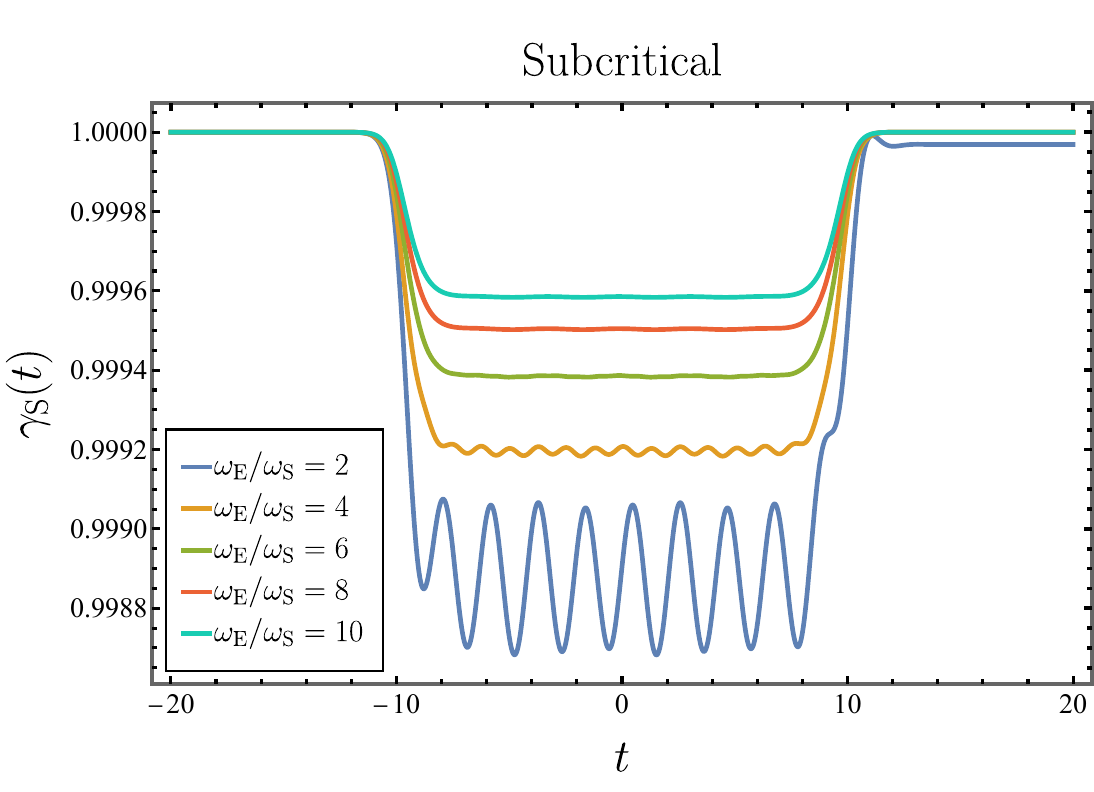}
\caption{Purity as a function of time for a few values of $\om_\uE/\om_\uS$, with $\tau=1$ and $\tO =10$, for $\xio=1.1\xic$ (supercritical regime, left panel) and $\xio=0.1\xic$ (subcritical regime, right panel)
}\label{fig.gS.var.omE} 
\end{figure}

Another parameter of interest for the behaviour of the purity is the ratio $\om_\uE/\om_\uS$ between the frequencies of the environment and of the system. The purity is displayed for a few values of this ratio in \cref{fig.gS.var.omE}. In the supercritical regime, $\om_\uE/\om_\uS$ does not alter the decohering phase much, but it strongly affects the amount by which the system recoheres as the interaction is turned off, hence the asymptotic value of the purity at late times. In the subcritical regime, a smaller frequency separation increases the amplitude of the purity oscillations. This can be interpreted as a consequence of the fact that the closer the two frequencies, the more similar the two oscillators are and, thus, the more efficiently they ``communicate". These features will be analytically described and further discussed in what follows.

\section{Criticality and perturbativity}
\label{sec:Criticality:Perturbativity}

\subsection{Parameter space}

As mentioned above, the parameter space of the model is four-dimensional, and before exploring it in a systematic way, let us try to organise it by means of simple criteria. We already saw that the value of $\xio$ with respect to $\xic$ determines whether the dynamics proceeds in the super or subcritical regime and is therefore an important separatrix. For this reason, we introduce
\bea
\psi\equiv \frac{\xi_0}{\xi_\uc}\, .
\eea 
We also noticed that the ratio between the system and environment frequencies plays an important role, and this is why we introduce
\bea 
w\equiv \frac{\om_\uS}{\om_\uE}\, .
\eea 

Another relevant scale for $\xio$ is the one that determines whether the influence of the environment on the system is perturbative or not. If the Hamiltonian matrix is split between a free (\ie~block diagonal) part and an interacting (\ie~block off-diagonal) part, $\bm{\mH_Q}=\bm{\mH}_{\bm{Q},\mathrm{free}}+\bm{V}$, the relative importance of the interaction can be assessed by taking the ratio between the Frobenius norm\footnote{The Frobenius norm of an $N\times M$ matrix $\bm{A}$, with matrix elements $a_{ij}$, is defined as 
$|\bm{A}|_{\mathrm{F}}:=\sqrt{\sum_{i=1}^N\sum_{j=1}^M|a_{ij}|^2}$.} of these two parts,
\begin{equation}
\gp=\frac{|\bm{V}|_{\mathrm{F}}}{|\bm{\mH}_{\bm{Q},\mathrm{free}}|_{\mathrm{F}}}=\frac{\xio}{\sqrt{2\om_\uS \om_\uE (\om_\uS ^2+\om_\uE ^2)}}\, .\label{gp}
\end{equation}
This ratio has been computed in the rescaled (but not rotated) basis $\hat{\bm{Q}}=\bm{S}_{\om_\uS,\om_\uE}^{-1} \hat{\bm{R}}$, such that all elements of $\bm{\mH_Q}=\bm{S}_{\om_\uS,\om_\uE}^{\mathrm{T}} \bm{\mH} \bm{S}_{\om_\uS,\om_\uE}$ have the same physical dimension and the Frobenius norm can be evaluated. 

\begin{figure}[t]
\centering
\includegraphics[width=0.7\textwidth]{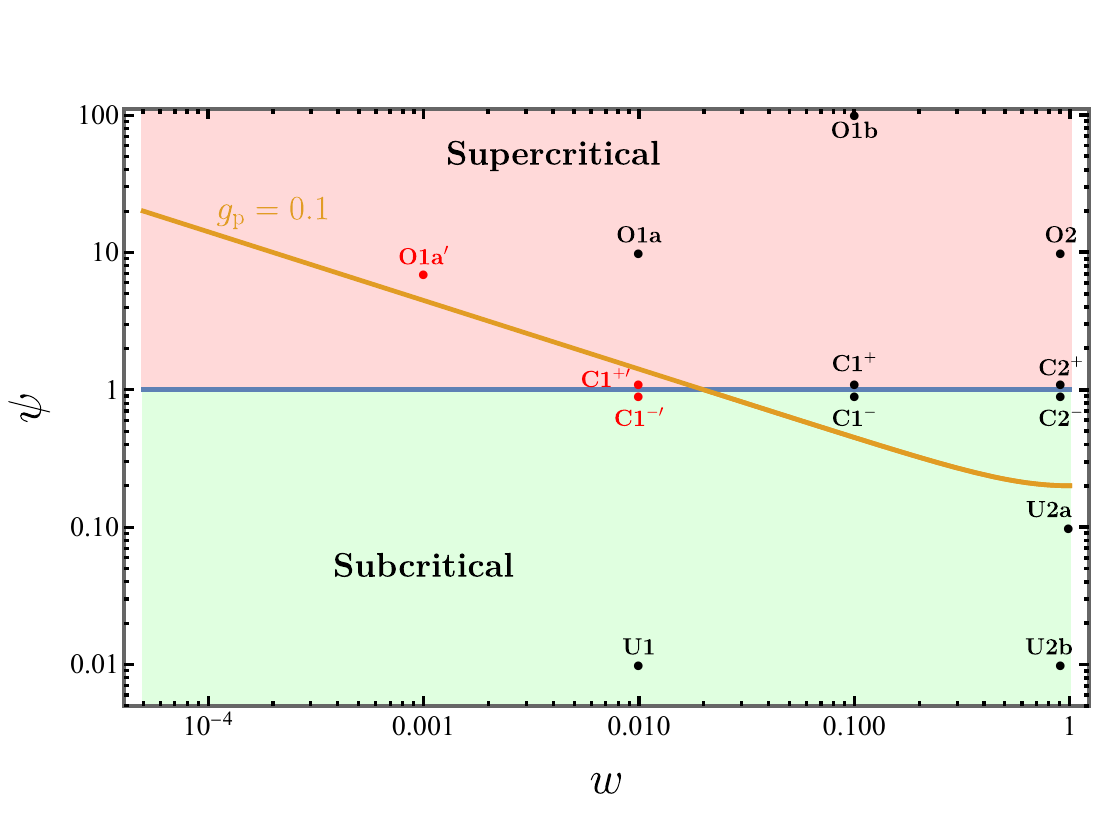}
\caption{Criticality ($\psi=1$, blue line) and perturbativity ($\gp=0.1$, orange line) in the parameter space $(w,\psi)$. Several points are labelled, which correspond to the specific cases discussed in \cref{sec:purity:expansion:instantaneous} and \cref{sec:ISOSO:discussion}.
}\label{fig.phase.diagram}
\end{figure}

The contour $\gp=0.1$ is displayed in the plane $(w,\psi)$ in \cref{fig.phase.diagram}. Note that, in terms of $w$ and $\psi$, 
\bea 
\label{eq:gp:w:psi}
\gp=\psi\sqrt{\frac{w}{2(1+w^2)}}\, .
\eea
Hereafter, we assume that $\om_\uS\leq \om_\uE$, since the case where $\om_\uS>\om_\uE$ can be treated by swapping $\uS$ and $\uE$ given that the purity of the system and the environment are the same, see \cref{footnote:gammaS=gammaE}. One can see that there exist four regimes:
\begin{itemize}
\item a subcritical, perturbative regime ($\psi\ll 1$, $\gp\ll 1$),
\item a supercritical, non-perturbative regime ($\psi\gg 1$, $\gp\gg 1$),
\item a mildly subcritical, mildly non-perturbative regime ($\psi\lesssim 1$, $\gp\gtrsim 1$) that requires $w\lesssim 1$,
\item a supercritical, perturbative regime ($\psi\gg 1$, $\gp\ll 1$) that requires $w\ll 1$.
\end{itemize}
The last two regimes are interesting since they suggest that subcriticality does not guarantee that perturbation theory will perform well, and conversely, perturbation theory should be able to probe some of the supercritical regime.

\subsection{Perturbation theory}
\label{sec:Perturbation:theory}

To make the above discussion more explicit, let us unfold the perturbative calculation of the purity. In the rescaled basis $\hat{\bm{Q}}$, the Hamiltonian generates the dynamics
\begin{equation}
\frac{\dd \hat{\bm{Q}}}{\dd t}=\bm{K}\hat{\bm{Q}}\, ,
\quad\text{where}\quad
\bm{K}= \bm{\Omega} \bm{\mH}_{\bm{Q}}\, . \label{K}
\end{equation}
Using the decomposition $\bm{\mH_Q}=\bm{\mH}_{\bm{Q},\mathrm{free}}+\bm{V}$, one can write $\bm{K}=\bm{K}_0+\bm{K}_V$ where
\bea
\bm{K}_0=  \bm{\Omega} \bm{\mH}_{\bm{Q},\mathrm{free}}=
\begin{pmatrix}
0 & \om_\uS  & 0 & 0\\
-\om_\uS  & 0 & 0 & 0\\
0 & 0 & 0 & \om_\uE \\
0 & 0 & -\om_\uE  & 0\\
\end{pmatrix}
\quad\text{and}\quad
\bm{K}_V= \bm{\Omega} \bm{V}=
\begin{pmatrix}
0 & 0 & 0 & 0\\
0 & 0 & -\lambda(t) & 0\\
0 & 0 & 0 & 0\\
-\lambda(t) & 0 & 0 & 0\\
\end{pmatrix}\, .
\eea
For later convenience, we have introduced the frequency $\lambda(t):=\xi(t)/\sqrt{\om_\uS \om_\uE }$.
In the absence of interactions, $\dd \hat{\bm{Q}}_0/\dd t =\bm{K}_0\hat{\bm{Q}}_0$ admits the solution
\begin{equation}
\hat{\bm{Q}}_0(t)=\bm{G}_0(t,\tin)\hat{\bm{Q}}(\tin),\label{Q.sol.o}
\end{equation}
where 
\begin{equation}
\label{eq:G0}
\bm{G}_0(t,\tin)=\exp\left[\int_{\tin}^t \dd t^\pr\,K_0(t^\pr)\right]=\begin{pmatrix}
\cos(\om_\uS \Delta t) & \sin(\om_\uS \Delta t) & 0 & 0\\
-\sin(\om_\uS \Delta t) & \cos(\om_\uS \Delta t) & 0 & 0\\
0 & 0 & \cos(\om_\uE \Delta t) & \sin(\om_\uE \Delta t)\\
0 & 0 & -\sin(\om_\uE \Delta t) & \cos(\om_\uE \Delta t)
\end{pmatrix}
\end{equation}
is a symplectic (and, in this instance, orthogonal) matrix, \ie~$\bm{G}_0^{\mathrm{T}}\bm{\Om}\bm{G}_0=\bm{\Om}$ and $\bm{G}_0^{\mathrm{T}}\bm{G}_0=\bm{1}$. Here, to ease the notation, we have introduced $\Delta t=t-{\tin}$.

The free dynamics can be factored out by going to the interaction picture
\begin{equation}
\hat{\widetilde{\bm{Q}}}(t)=\bm{G}_0^{-1}(t,\tin)\hat{\bm{Q}}(t), \label{Qtilde}
\end{equation}
leading to the equation of motion
\begin{equation}
\frac{\dd \hat{\widetilde{\bm{Q}}}}{\dd t}=\widetilde{\bm{K}}_V(t)\hat{\widetilde{\bm{Q}}}, \label{Qtilde.eom}
\end{equation}
where 
\bea
\widetilde{\bm{K}}_V(t)=&\bm{G}_0^{-1}(t,\tin)\bm{K}_V(t)\bm{G}_0(t,{\tin})\\
=&
\begin{small}
\lambda(t)
\begin{pmatrix}
0 & 0 & \sin(\om_\uS \Delta t)\cos(\om_\uE \Delta t) & \sin(\om_\uS \Delta t)\sin(\om_\uE \Delta t)\\
0 & 0 & -\cos(\om_\uS \Delta t)\cos(\om_\uE \Delta t) & -\cos(\om_\uS \Delta t)\sin(\om_\uE \Delta t)\\
\cos(\om_\uS \Delta t)\sin(\om_\uE \Delta t) & \sin(\om_\uS \Delta t)\sin(\om_\uE \Delta t) & 0 & 0\\
-\cos(\om_\uS \Delta t)\cos(\om_\uE \Delta t) & -\sin(\om_\uS \Delta t)\cos(\om_\uE \Delta t) & 0 & 0
\end{pmatrix}
\end{small}.
\eea
The solution to \cref{Qtilde.eom} is given by
\begin{equation}
\hat{\widetilde{\bm{Q}}}(t)=\widetilde{\bm{G}}_V(t,{\tin})\hat{\widetilde{\bm{Q}}}(\tin) \, ,\label{Qtilde.solution}
\end{equation}
where 
\begin{equation}
\widetilde{\bm{G}}_V(t,\tin)=\mathrm{T}\exp\left[\int_{\tin}^t \dd t^\pr\,\widetilde{\bm{K}}_V(t^\pr)\right] .\label{G.tilde}
\end{equation}
Here, $\mathrm{T}$ denotes the time-ordering operator, which is required since $[\widetilde{\bm{K}}_V(t),\widetilde{\bm{K}}_V(t^\pr)]\neq 0$ in general. Taylor expanding the exponential function in the above corresponds to performing a Dyson series, where each $\widetilde{\bm{K}}_V$ contributes a power $\lambda$. At order $\lambda$ we find no correction to the purity, so we need to expand up to order $\lambda^2$, where
\begin{equation}
\widetilde{\bm{G}}_V^{(2)}(t,{\tin})=\bm{I}+\bm{\delta\widetilde{G}}_V^{(1)}(t,{\tin})+\bm{\delta\widetilde{G}}^{(2)}_V(t,{\tin}),
\end{equation}
with
\begin{align}
\bm{\delta\widetilde{G}}^{(1)}_V(t,{\tin})=&\int_{\tin}^t \dd t^\pr\,\widetilde{\bm{K}}_V(t^\pr),\\
\bm{\delta\widetilde{G}}^{(2)}_V(t,{\tin})=&\int_{\tin}^t \dd t^\pr\int_{\tin}^{t^\pr}\dd t^{\pr\pr}\,\widetilde{\bm{K}}_V(t^\pr)\widetilde{\bm{K}}_V(t^{\pr\pr}).
\end{align}

Going back to the original basis $\hat{\bm{R}}(t)=\bm{S}_{\om_\uS,\om_\uE}\hat{\bm{Q}}(t)= \bm{S}_{\om_\uS,\om_\uE} \bm{G}_0(t,\tin)\hat{\widetilde{\bm{Q}}}(t)$, \cref{Qtilde.solution} leads to $\hat{\bm{R}}(t) = \bm{S}(t,\tin) \hat{\bm{R}}(\tin)$ where
\bea
\bm{S}(t,\tin) =
\bm{S}_{\om_\uS,\om_\uE}
\bm{G}_0(t,\tin)
\widetilde{\bm{G}}_V(t,\tin)
\bm{S}^{-1}_{\om_\uS,\om_\uE}\, .
\eea 
This implies that the covariance matrix evolves according to $\bm{\sigma}(t)=\bm{S}(t,\tin)\bm{\sigma}({\tin})\bm{S}^{\mathrm{T}}(t,\tin)$, where we recall that $\bm{\sigma}({\tin})=\text{diag}\left(\om_\uS ^{-1},\om_\uS ,\om_\uE ^{-1},\om_\uE \right)=\bm{S}_{\om_\uS,\om_\uE}^2$. Up to second order, this leads to 
\begin{align}
\bm{\sigma}^{(2)}(t)=\bm{\sigma}^{(0)}+\bm{\delta\sigma}^{(1)}(t)+\bm{\delta\sigma}^{(2)}(t)
\end{align}
where
\begin{align}
\bm{\sigma}^{(0)}(t)=&\bm{\sigma}(\tin)\\
\bm{\delta\sigma}^{(1)}(t)=&\bm{S}_{\om_\uS,\om_\uE}\bm{G}_0(t,\tin)\left[\bm{\delta\widetilde{G}}_V^{(1)}(t,\tin)+\bm{\delta\widetilde{G}}_V^{(1)\tT}(t,\tin)\right]\bm{G}_0^\tT(t,\tin)\bm{S}_{\om_\uS,\om_\uE},\\
\bm{\delta\sigma}^{(2)}(t)=&\bm{S}_{\om_\uS,\om_\uE}\bm{G}_0(t,\tin)\left[\bm{\delta\widetilde{G}}_V^{(1)}(t,\tin)\bm{\delta\widetilde{G}}_V^{(1)\tT}(t,\tin)
\right. \nonumber \\ & \left.
+\bm{\delta\widetilde{G}}_V^{(2)}(t,\tin)+\bm{\delta\widetilde{G}}_V^{(2)\tT}(t,\tin)\right]\bm{G}_0^\tT(t,\tin)\bm{S}_{\om_\uS,\om_\uE}\, .
\end{align}
For the $\uS$ oscillator, only the $2\times 2$ upper-left block needs to be considered. At zeroth order, this leads to $\bm{\sigma}^{(0)}_\uS =\text{diag}(\om_\uS ^{-1},\om_\uS )$, while $\bm{\delta\sigma}_\uS^{(1)}(t)=0$, at first order, and 
\begin{align}
\bm{\delta\sigma}_\uS^{(2)}(t)=\int_{\tin}^t \dd t^\pr\int_{\tin}^t \dd t^{\pr\pr}\,\lambda(t^\pr)\lambda(t^{\pr\pr})\begin{pmatrix}
{\om_\uS^{-1} }A(t,t^\pr,t^{\pr\pr}) & B(t,t^\pr,t^{\pr\pr})\\
C(t,t^\pr,t^{\pr\pr}) & \om_\uS D(t,t^\pr,t^{\pr\pr})
\end{pmatrix}
\end{align}
at second order. Here we have introduced the functions
\bea
A(t,t^\pr,t^{\pr\pr}):=\frac{1}{2}\sin\left[\om_\uS (t-t^\pr)\right]\bigg\{&\left[1-2\Theta(t^\pr-t^{\pr\pr})\right]\sin\left[\om_\uS t-\om_\uE t^\pr+(\om_\uE -\om_\uS )t^{\pr\pr}\right]\\
&+\left[1+2\Theta(t^\pr-t^{\pr\pr})\right]\sin\left[\om_\uS t+\om_\uE t^\pr-(\om_\uS +\om_\uE )t^{\pr\pr}\right]\bigg\}\, ,
\eea
\bea
B(t,t^\pr,t^{\pr\pr}):=\frac{1}{4}\bigg\{&\sin\left[(\om_\uE -\om_\uS )(t^\pr-t^{\pr\pr})\right]-\sin\left[(\om_\uS +\om_\uE )(t^\pr-t^{\pr\pr})\right]\\
&+\left[1-2\Theta(t^\pr-t^{\pr\pr})\right]\sin\left[2\om_\uS t-(\om_\uS +\om_\uE )t^\pr+(\om_\uE -\om_\uS )t^{\pr\pr}\right]\\
&+\left[1+2\Theta(t^\pr-t^{\pr\pr})\right]\sin\left[2\om_\uS t+(\om_\uE -\om_\uS )t^\pr-(\om_\uS +\om_\uE )t^{\pr\pr}\right]\bigg\}\, ,
\eea
\bea
C(t,t^\pr,t^{\pr\pr}):=\frac{1}{4}\bigg\{&\sin\left[(\om_\uS +\om_\uE )(t^\pr-t^{\pr\pr})\right]-\sin\left[(\om_\uE -\om_\uS )(t^\pr-t^{\pr\pr})\right]\\
&+\left[1-2\Theta(t^\pr-t^{\pr\pr})\right]\sin\left[2\om_\uS t-(\om_\uS +\om_\uE )t^\pr+(\om_\uE -\om_\uS )t^{\pr\pr}\right]\\
&+\left[1+2\Theta(t^\pr-t^{\pr\pr})\right]\sin\left[2\om_\uS t+(\om_\uE -\om_\uS )t^\pr-(\om_\uS +\om_\uE )t^{\pr\pr}\right]\bigg\}\, ,
\eea
\bea
D(t,t^\pr,t^{\pr\pr}):=\frac{1}{2}\cos\left[\om_\uS (t-t^\pr)\right]\bigg\{&\left[1-2\Theta(t^\pr-t^{\pr\pr})\right]\cos\left[\om_\uS t-\om_\uE t^\pr+(\om_\uE -\om_\uS )t^{\pr\pr}\right]\\
&+\left[1+2\Theta(t^\pr-t^{\pr\pr})\right]\cos\left[\om_\uS t+\om_\uE t^\pr-(\om_\uS +\om_\uE )t^{\pr\pr}\right]\bigg\}\, .
\eea
The covariance matrix in the $\uS$ sector thus reads
\begin{align}
\bm{\sigma}_\uS ^{(2)}(t)=
\begin{pmatrix}
{\om_\uS }^{-1}\left[1+I_A(t)\right] & I_B(t)\\
I_C(t) & \om_\uS I_D(t)
\end{pmatrix},
\end{align}
where $I_A(t)=\int_{{\tin}}^t\dd t^\pr\int_{{\tin}}^t\dd t^{\pr\pr}\,\lambda(t^\pr)\lambda(t^{\pr\pr})A(t,t^\pr,t^{\pr\pr})$ and similar expressions for $I_B(t)$, $I_C(t)$, $I_D(t)$. These double integrals are of order $\lambda^2$, so at second order the purity reads
\begin{align}
\nonumber
\gamma_\uS ^{(2)}(t)=1-\frac{1}{4}\int_{{\tin}}^tdt^\pr\int_{{\tin}}^tdt^{\pr\pr}\,\lambda(t^\pr)\lambda(t^{\pr\pr})\bigg\{&\left[1-2\Theta(t^\pr-t^{\pr\pr})\right]\cos\left[(\om_\uE -\om_\uS )(t^\pr-t^{\pr\pr})\right]\\
&\left[1+2\Theta(t^\pr-t^{\pr\pr})\right]\cos\left[(\om_\uS +\om_\uE )(t^\pr-t^{\pr\pr})\right]\bigg\} \, ,\label{gS.perturbative}
\end{align}
see \cref{purity.def}. As we will see below, this expression implies that perturbative corrections to the purity are of order $\lambda^2/(\om_\uS^2+\om_\uE^2)=\gp^2$, in agreement with the above Frobenius estimate.

\subsection{The instantaneous switch-on/switch-off limit}
\label{sec:ISOSO}

Our next goal is to explore the different regions identified in the parameter space pictured in \cref{fig.phase.diagram}. We will do so in the regime where the interaction is instantaneously switched on and off (ISOSO), $\tau\ll t_0$. The reason is twofold. First, \cref{sec:Markov} is dedicated to a detailed analysis of the effects associated with a slow turn on and off of the interaction, and it is convenient to discuss these two limits separately. Second, in the ISOSO limit, the coupling function is piecewise constant, and the model becomes analytically tractable. 

Indeed, let us consider the case where $\xi(t)=\xio$ if $-\tO<t<\tO$, and vanishes otherwise. When $t< -\tO$, the two oscillators decouple, and the equation of motion $\ddot{\hat{x}}_{\mathrm{I}} +\om_{\mathrm{I}} ^2\hat{x}_{\mathrm{I}} =0$ has solution
\bea
\label{eq:ISOSO:x:A}
\left. \hat{x}_{\mathrm{I}} (t)\right\vert_{t< -\tO}=&\frac{e^{-i\om_{\mathrm{I}}  (t-\tin)}}{\sqrt{2\om_{\mathrm{I}} }}\hat{a}_{\mathrm{I}}  +\frac{e^{i\om_{\mathrm{I}}  (t-\tin)}}{\sqrt{2\om_{\mathrm{I}} }}\hat{a}_{\mathrm{I}} ^\dagger 
\eea
where we recall that $\mathrm{I}=\uS,\uE$. When $-\tO<t<\tO$, the two oscillators couple, but they decouple in the adiabatic basis introduced in \cref{sec:rotated:basis}, where
\bea
\label{eq:xi:ISOSO}
\left. \hat{x}_{i} (t)\right\vert_{-\tO<t<\tO}=&\frac{e^{-i\om_{i}  (t+\tO)}}{\sqrt{2\om_{i} }}\hat{b}_{i}  +\frac{e^{i\om_{i}  (t+\tO)}}{\sqrt{2\om_{i} }}\hat{b}_{i}^\dagger \, .
\eea
Here, $\hat{b}_i$ and $\hat{b}^\dagger_i$ are annihilation and creation operators (since rotations are canonical transformations), hence $[\hat{b}_i,\hat{b}_j^\dagger]=\delta_{i,j}$, and we recall that $i=1,2$. Undoing the rotation, \cref{eq:xi:ISOSO} leads to
\bea
\label{eq:ISOSO:x:B}
\left.\hat{x}_\uS (t)\right\vert_{-\tO<t<\tO}=&\cos{\theta}\hat{x}_1(t)+\sin{\theta}\hat{x}_2(t)\\
=&\hat{b}_1\cos\theta \frac{e^{-i\om_1(t+\tO )}}{\sqrt{2|\om_1|}}+\hat{b}_2\sin\theta \frac{e^{-i\om_2(t+\tO )}}{\sqrt{2\om_2}}+\hat{b}^\dagger_1\cos\theta \frac{e^{i\om_1(t+\tO )}}{\sqrt{2|\om_1|}}+\hat{b}^\dagger_2\sin\theta \frac{e^{i\om_2(t+\tO )}}{\sqrt{2\om_2}},\\ 
\left.\hat{x}_\uE (t)\right\vert_{-\tO<t<\tO}=&-\sin{\theta}\hat{x}_1(t)+\cos{\theta}\hat{x}_2(t)\\
=&-\hat{b}_1\sin\theta \frac{e^{-i\om_1(t+\tO )}}{\sqrt{2|\om_1|}}+\hat{b}_2\cos\theta \frac{e^{-i\om_2(t+\tO )}}{\sqrt{2\om_2}}-\hat{b}^\dagger_1\sin\theta \frac{e^{i\om_1(t+\tO )}}{\sqrt{2|\om_1|}}+\hat{b}^\dagger_2\cos\theta \frac{e^{i\om_2(t+\tO )}}{\sqrt{2\om_2}}.
\eea

\begin{figure}[t]
\centering 
\includegraphics[width=0.48\textwidth]{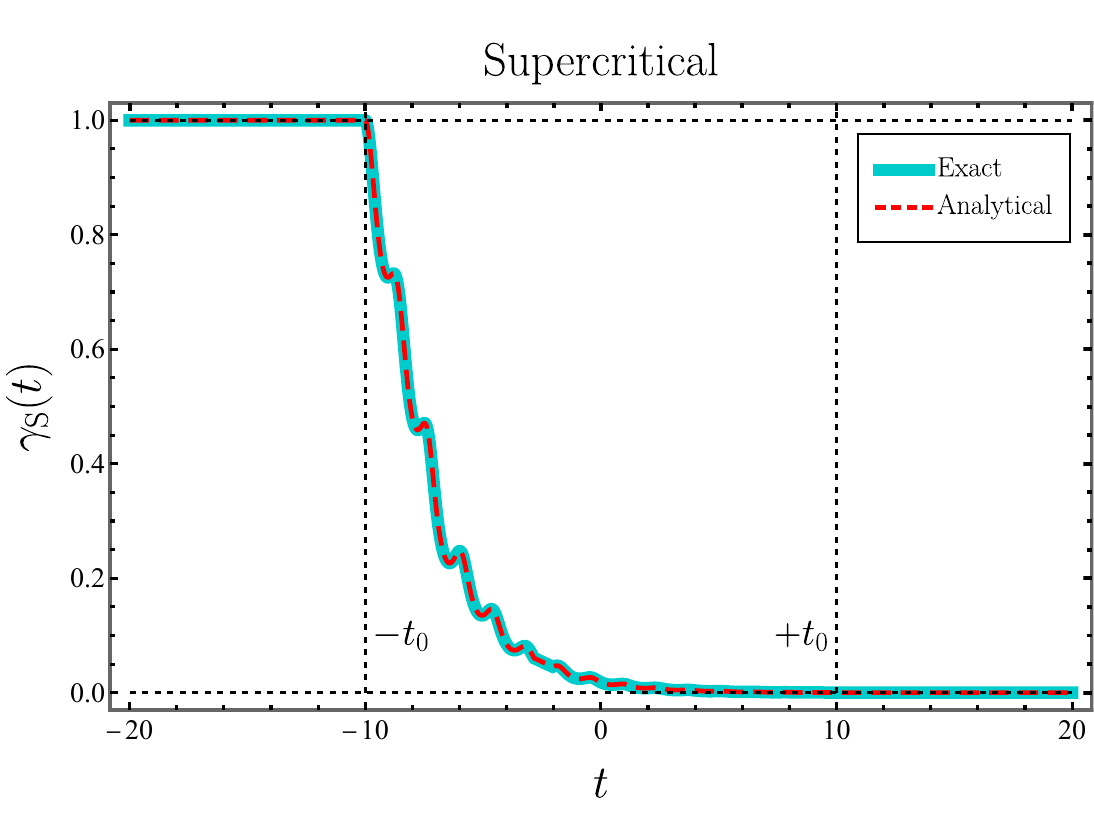}
\includegraphics[width=0.48\textwidth]{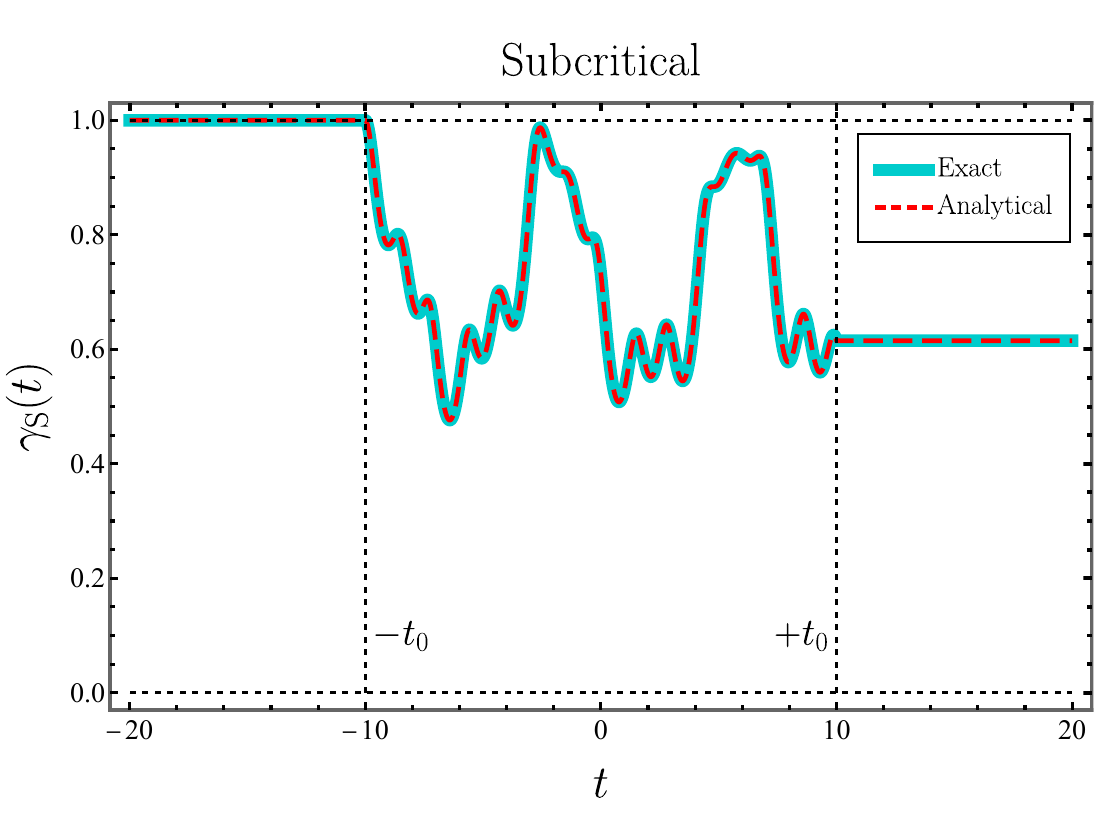} 
\caption{Comparison between the purity computed analytically in the ISOSO limit (see \cref{sec:ISOSO}) and its numerical counterpart (see \cref{sec:Criticality:and:Recoherence}, where we took $\tau/\tO=10^{-4}$ in \cref{xi} to be close to the ISOSO limit). We have used $\om_\uS=1$, $\om_\uE=2$, $\tO=10$, $\xio=1.1\xic$ in the left panel (supercritical regime) and $\xio=0.9\xic$ in the right panel (subcritical regime).
}
\label{fig.gS.ANCOMP} 
\end{figure}

Since the coupling function remains finite at all times, the position and momentum operators are continuous at the transition,
\bea
\label{eq:ISOSO:matching}
\left.\hat{x}_{\mathrm{I}}(-\tO)\right\vert_{t<\tO}=\left.\hat{x}_{\mathrm{I}}(-\tO)\right\vert_{t>\tO}
\quad\text{and}\quad
\left.\hat{p}_{\mathrm{I}}(-\tO)\right\vert_{t<\tO}=\left.\hat{p}_{\mathrm{I}}(-\tO)\right\vert_{t>\tO}
\eea 
where $\hat{p}_{\mathrm{I}}=\dot{\hat{x}}_{\mathrm{I}}$. This matching condition allows one to express the $\hat{b}_i$ operators in terms of the $\hat{a}_i$ operators, and the corresponding expression can be found in \cref{app:ISOSO}. This, in turn, leads to the two-point function of the $\hat{b}_i$ operators when evaluated on the vacuum, hence to the two-point function of the system's position and momentum during the phase $-\tO<t<\tO$ when the interaction is on. These expressions are given in \cref{app:ISOSO} and lead to an analytical formula for the purity $\gamma_\uS$, using \cref{purity.def}. Note that, in the phases $t<-\tO$ and $t>\tO$, $\gamma_\uS$ is constant since the interaction is off, hence $\gamma_\uS=1$ at $t<-\tO$ and $\gamma_\uS = \gamma_\uS(t_0)$ at $t>\tO$.
To check the validity of the obtained formula for $\gamma_\uS$, it is compared with a numerical integration of the transport equation in \cref{fig.gS.ANCOMP}, where we set $\tau/\tO = 10^{-4}$ to be close to the ISOSO limit. One can check that the agreement is indeed excellent. 

Note that, in the ISOSO limit, the perturbative formula~\eqref{gS.perturbative} also becomes explicit, and reduces to
\bea
\gamma_\uS ^{(2)}(t)=1-4\gp^2
\frac{1+w^2}{(1+w)^2}
\sin^2\left[\frac{1}{2}(\om_\uS +\om_\uE )(t-{\tin})\right]. \label{gS.gP}
\eea
As anticipated above, the correction to the purity is thus controlled by $\gp^2$ (note that $(1+w^2)/(1+w)^2$ is always between $1/2$ and $1$).

\subsection{Exploration of the parameter space}
\label{sec:purity:expansion:instantaneous}

We are now in a position to explore the parameter map displayed in \cref{fig.phase.diagram} in a systematic way. Models belonging to all possible regimes are displayed with black dots, using the following terminology: ``U'' (under criticality) corresponds to $\psi\ll 1$, ``O'' (over criticality) corresponds to $\psi\gg 1$ while ``C'' (near criticality) corresponds to $\psi\simeq 1$. Moreover, we distinguish models with $w\ll 1$, labelled with ``1'', from models with $w\lesssim 1$, labelled with $2$. Thus, in our nomenclature, $\mathrm{U}1$ refers to $\psi \ll 1$ and $w\ll 1$, $\mathrm{C}2$ refers to $\psi\simeq 1$ and $w\lesssim 1$, \etc. We now review these cases one by one.

\stoptoc
\subsubsection{Case U1: $\psi\ll 1$ and $w\ll 1$}
\resumetoc

When expanding the ISOSO purity in this limit, at leading order, one finds
\begin{equation}
\gamma_\uS \simeq 1-2w\psi^2\sin^2\left[\frac{1}{2}\left(\om_1+\om_2\right)\Delta t\right]
\label{gS.proxy.U1} 
\end{equation}
during the phase $-\tO<t<\tO$, where $\om_1\simeq \om_\uS$ and $\om_2\simeq \om_\uE$. Here, $\Delta t = t+\tO$. The above expression coincides with the prediction from perturbation theory~\eqref{gS.gP}, since in the U1 limit \cref{eq:gp:w:psi} reduces to $\gp^2\simeq w\psi^2/2 $. This is expected since U1 lies deep in the perturbative regime. Note also that, when expanding \cref{eq:om1:om2:def} at next-to-leading order, $\om_1= \om_\uS(1-\psi^2/2)$ and $\om_2=\om_\uE(1+w^2\psi^2/2)$, hence the expansion of the frequencies $\om_1$ and $\om_2$ holds until $\Delta t$ is of order $\Delta t_\mathrm{sec} = 1/(\om_\uS \psi^2)$. Past that point, secular effects induce a slow dephasing, which perturbation theory does not account for at leading order. 

\begin{figure}[t]
\centering 
\includegraphics[width=0.48\textwidth]{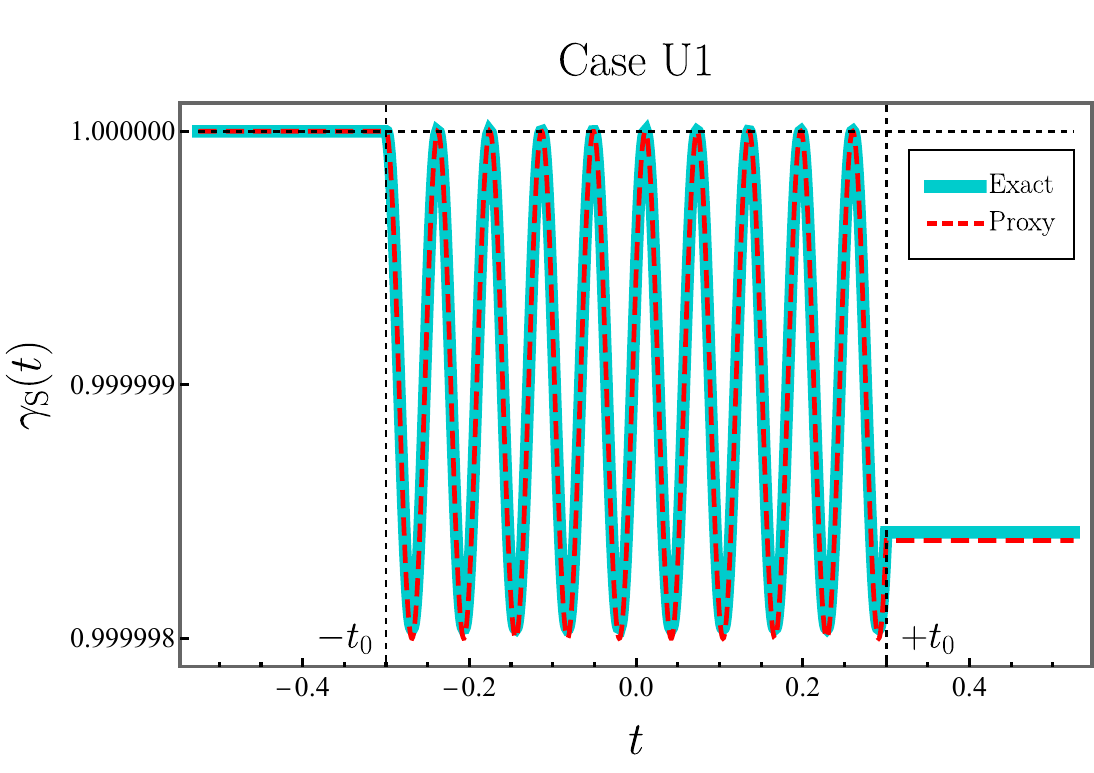} 
\includegraphics[width=0.48\textwidth]{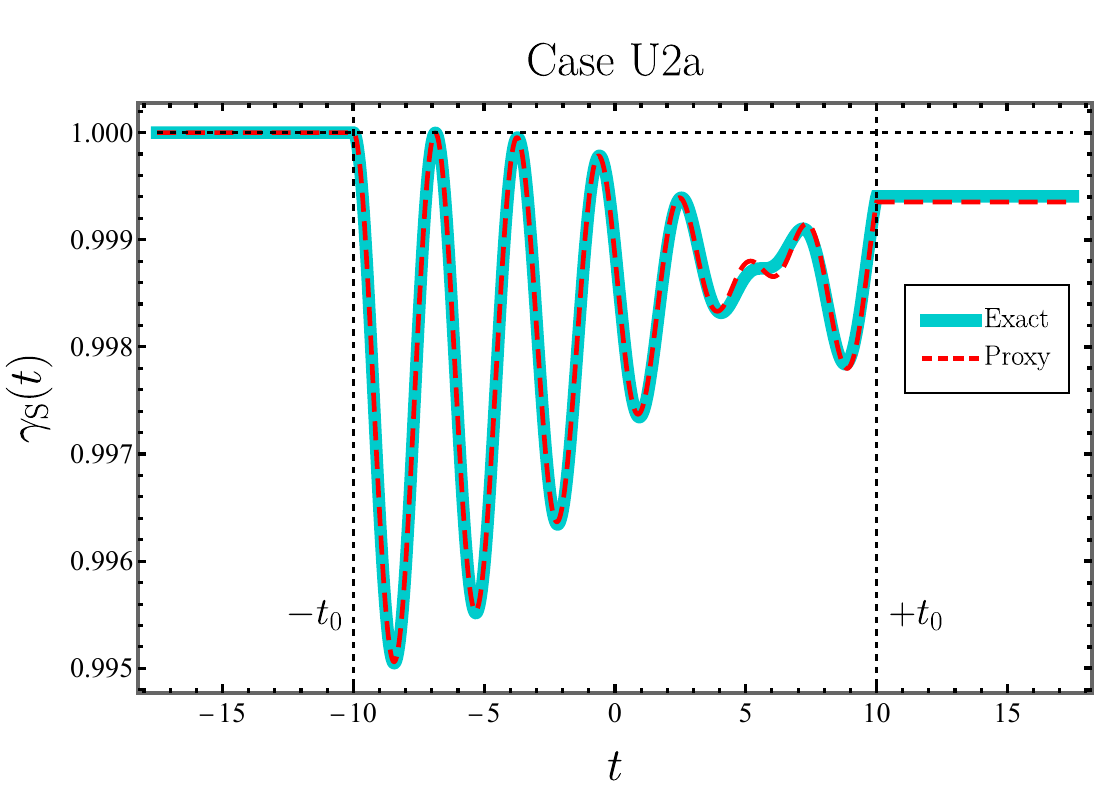} 
\includegraphics[width=0.48\textwidth]{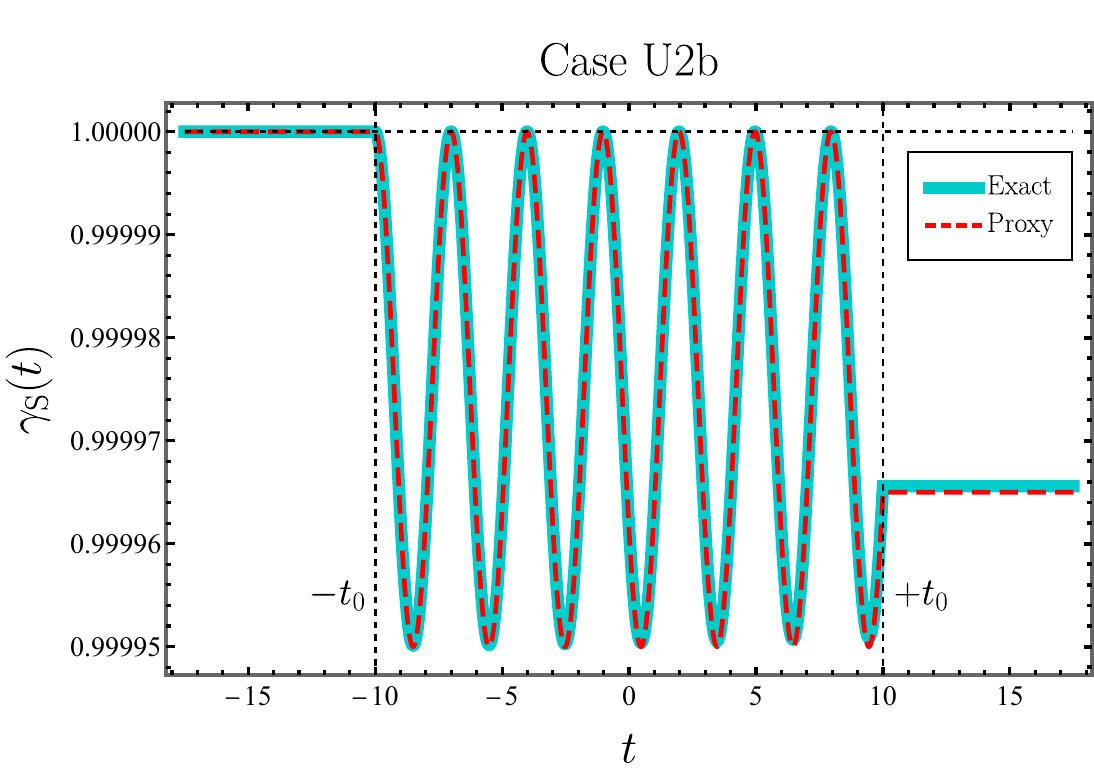} 
\caption{Comparison between the exact and expanded purities~\eqref{gS.proxy.U1}, \eqref{gS.proxy.U2.wpsi} and \eqref{gS.proxy.U2.psiw}, in the U1, U2a and U2b cases respectively. The value $\om_\uS=1$ is used in all panels, and the exact result is obtained from numerically solving the transport equations with $\tau=10^{-4}\tO$. The values for $(\tO,w,\psi)$ are as follows: $\mathrm{U}1:(0.3,10^{-2},10^{-2})$, $\mathrm{U}2a:(10,1/1.01,0.1)$ and $\mathrm{U}2b:(10,1/1.1,0.01)$. They correspond to the points labeled in \cref{fig.phase.diagram}.
}\label{fig:U123} 
\end{figure}

The formula~\eqref{gS.proxy.U1} is compared with a full numerical integration of the transport equations in the first panel of \cref{fig:U123}, where the agreement is excellent. In the U1 regime, the purity oscillates at frequency $\om_\uE$ and never substantially departs from one. In all figures of this section, the purity approximations are displayed with $\om_1$ and $\om_2$ given by \cref{eq:om1:om2:def}, \ie~these frequencies are not re-expanded. With the parameter values used in \cref{fig:U123}, $\om_\uS \Delta t_\mathrm{sec} =10^4 \gg \om_\uS \tO$, hence the secular dephasing mentioned above remains negligible anyway. 

\stoptoc
\subsubsection{Case U2: $\psi\ll 1$ and $w\simeq 1$}
\resumetoc

In this case, contrary to U1, the ordering of the expansion in the two small parameters $\psi$ and $\delta w = 1/w-1$ matters; hence, the two subcases need to be distinguished.

\paragraph{Sub-case U2a: $\delta w\ll \psi$}

When first expanding the ISOSO purity in $\delta w$ and then in $\psi$, one obtains
\bea
\gamma_\uS \simeq 1
&- \frac{1}{16} \psi^2 \left\lbrace[3 - 2 \cos(2 \omega_1 \Delta t)- 
2 \cos(2 \omega_2 \Delta t) + 
\cos\left[2(\omega_2 - \omega_1) \Delta t\right]\right\rbrace ,
\label{gS.proxy.U2.wpsi}
\eea
where $\om_1\simeq \om_\uS (1+\delta w/2-\psi/2)$ and $\om_2\simeq \om_\uS (1+\delta w/2+\psi/2)$. This implies that $\om_2-\om_1$ is of order $\om_\uS \psi$. Hence, at leading order, the above reduces to the perturbative result~\eqref{gS.gP}, as expected. At late times, since the leading correction to $\om_1$ and $\om_2$ is of order $\om_\uS \psi$, secular dephasing occurs past $\Delta t_\mathrm{sec}=1/(\om_\uS \psi)$.
The formula~\eqref{gS.proxy.U2.wpsi} is displayed in the second panel of \cref{fig:U123}, where it shows good agreement with the numerical integration. 

\paragraph{Sub-case U2b: $\psi \ll \delta w$}
When first expanding in $\psi$ and then in $\delta w$, one finds
\begin{equation}
\gamma_\uS \simeq  1-\frac{1}{2}\psi^2\sin^2\left(\frac{\om_1+\om_2}{2}\Delta t\right), \label{gS.proxy.U2.psiw}
\end{equation}
where $\om_1\simeq \om_\uS[1-\psi^2/(4\delta w)]$ and $\om_2\simeq \om_\uE[1+\psi^2/(4\delta w)]$. This coincides with perturbation theory, see \cref{gS.gP}, with secular dephasing occurring at $\om_\uS \Delta t_{\mathrm{sec}}=\delta w/\psi^2$. The formula~\eqref{gS.proxy.U2.psiw} is compared with numerical integration in the third panel of \cref{fig:U123}, where the agreement is again very good. 

The two subcases U2a and U2b thus coincide at early times, when they reduce to the perturbative formula, and only differ at late times, when they are subject to different secular dephasings.

\stoptoc
\subsubsection{Case C1: $w\ll 1$ and $\psi \simeq 1$}
\resumetoc

\begin{figure}[t]
\centering 
\includegraphics[width=0.48\textwidth]{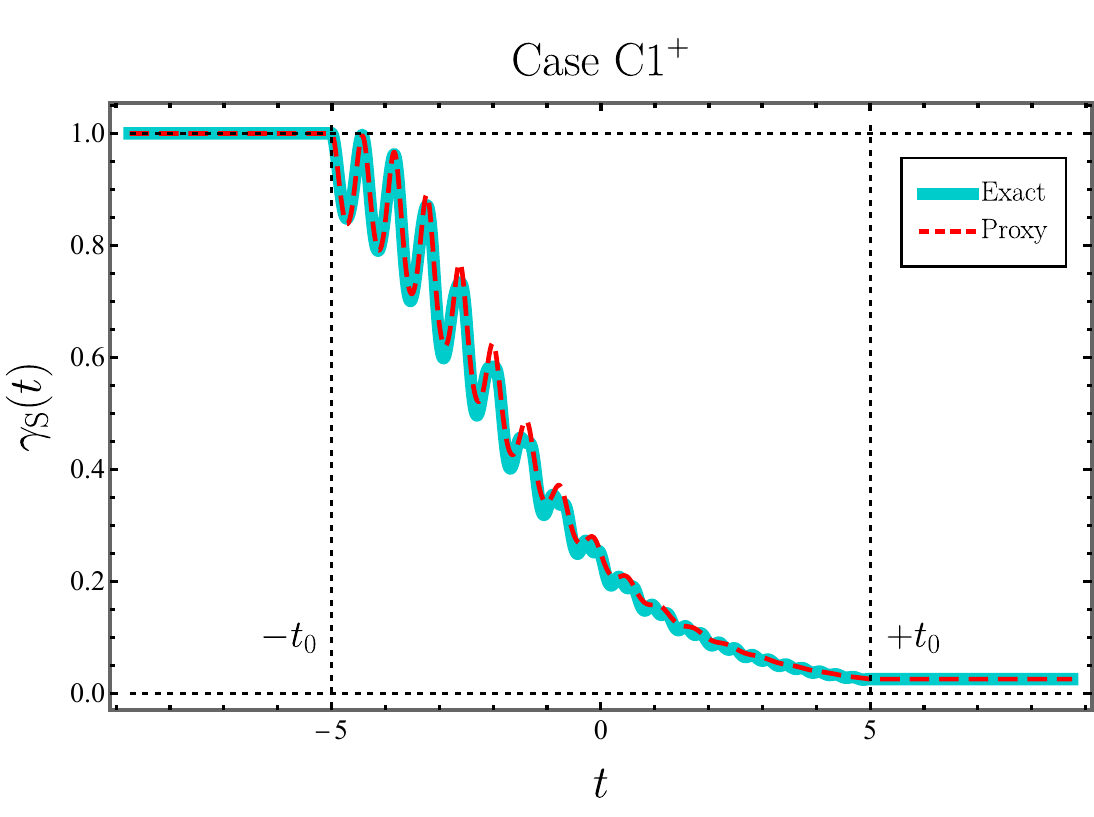} 
\includegraphics[width=0.48\textwidth]{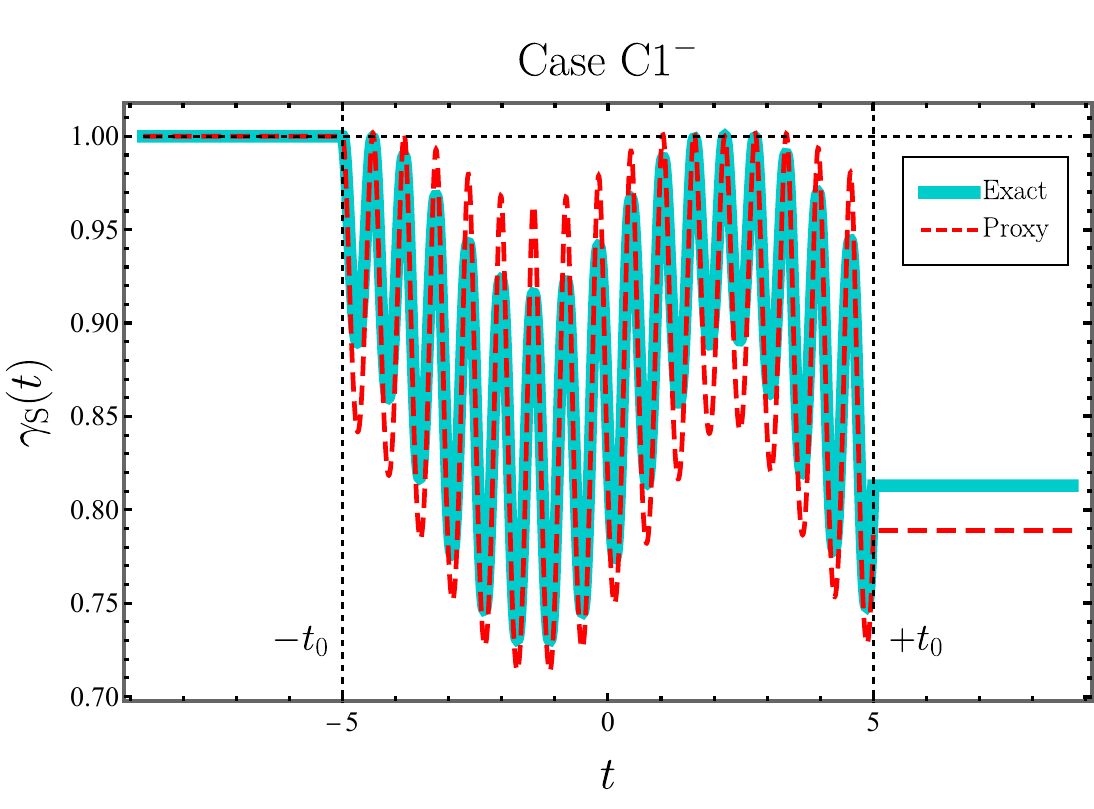} 
\includegraphics[width=0.48\textwidth]{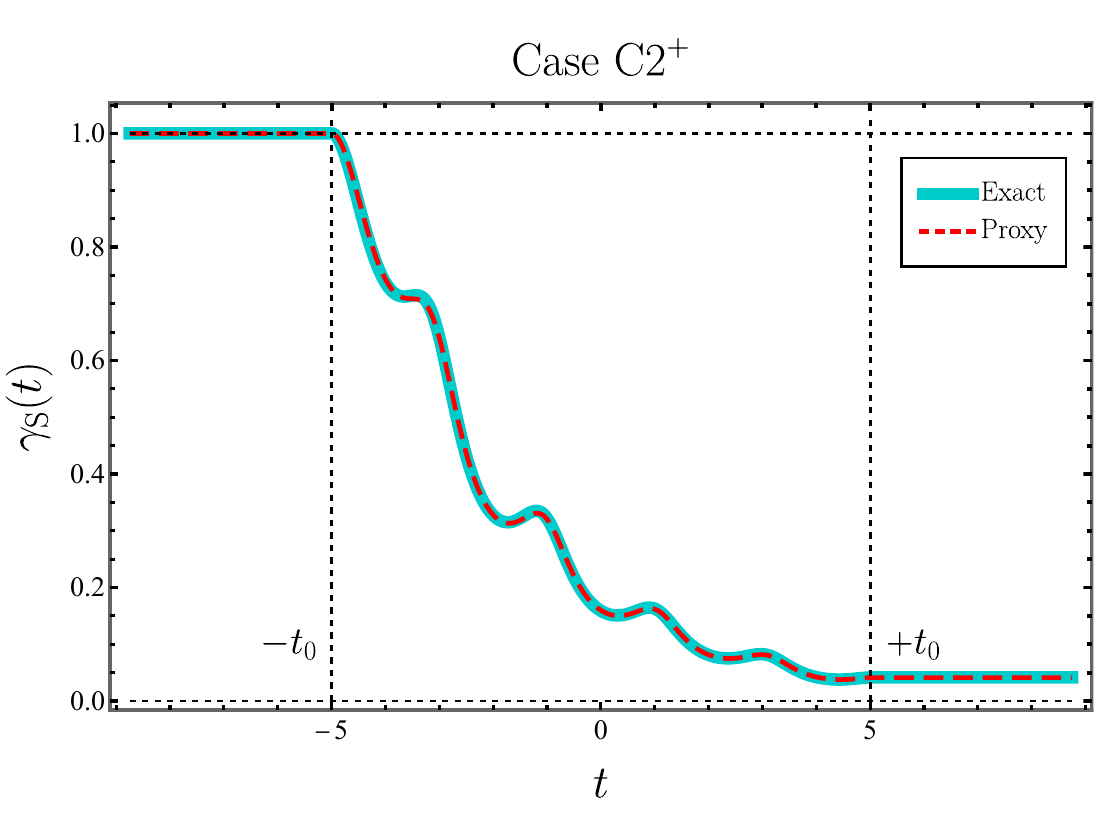} 
\includegraphics[width=0.48\textwidth]{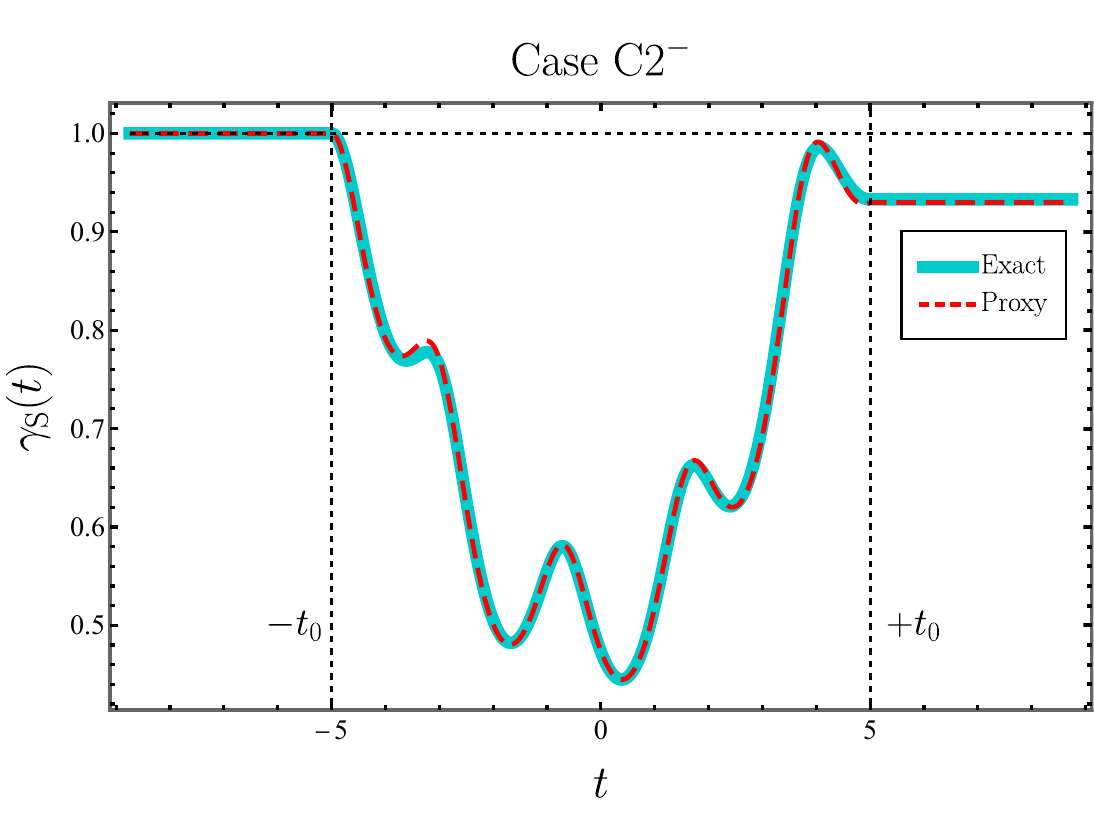} 
\caption{Comparison between the exact and expanded purities~\eqref{gS.proxy.C1} and \eqref{gS.proxy.C2}, in the C1 and C2 cases respectively. The value $\om_\uS=1$ is used in all panels, and the exact result is obtained from numerically solving the transport equations with $\tau=t_010^{-4}$. The values for $(\tO,w,\psi)$ are as follows: $\mathrm{C}1^+:(5,10^{-1},1.1)$, $\mathrm{C}1^-:(5,10^{-1},0.9)$, $\mathrm{C}2^+:(5,1/1.1,1.1)$ and $\mathrm{C}2^-:(5,1/1.1,0.9)$. They correspond to the points labelled in \cref{fig.phase.diagram}.
}\label{fig:C1:C2} 
\end{figure}

This case corresponds to a large frequency separation close to the critical point. Expanding the ISOSO purity in $w$ and $\delta\psi=1-\psi$, at leading order one finds
\bea
\gamma_\uS \simeq \biggl\{1 &+\frac{w}{2\delta\psi}\sin^2(\om_1\Delta t)+\sqrt{2} \frac{w}{\sqrt{\abs{\delta\psi}}} \sin(\omega_1 \Delta t) \sin(\omega_2 \Delta t)\\
&+\frac{w}{8}\left[9+7\cos\big(2\om_1\Delta t\big)-16\cos(\om_1\Delta t)\cos(\om_2\Delta t)\right]\biggr\}^{-1/2},
\label{gS.proxy.C1} 
\eea
where $\omega_1\simeq \sqrt{2 \delta\psi }\omega_\uS$ and $\om_2\simeq \om_\uE (1+w^2/2)$.  
Note that the above expression is valid regardless of the sign of $\delta\psi$, \ie it applies both in the sub- and supercritical regimes. In the supercritical regime, since $\delta\psi<0$, $\om_1=i\vert\om_1\vert$, the circular functions become hyperbolic functions and the purity decays exponentially. This explains the behaviour seen in the two upper panels of \cref{fig:C1:C2}, in the super- and subcritical regimes respectively, where the agreement with the full numerical integration is also confirmed. 

From the expanded expressions of $\om_1$ and $\om_2$, one can see that secular effects arise at a time $\om_\uS\Delta t_{\mathrm{sec}}=\min(1/w,1/\sqrt{\vert\delta\psi\vert})$, but they are of a different natures in the sub- and supercritical case: in the former, it consists in secular dephasing, in the latter, it is a secular growth. This is the reason why the inverse square root appearing in \cref{purity.def} has not been re-expanded above, since the correction to the purity can be large at late times.

At early times, \ie when $\Delta t\ll \Delta t_{\mathrm{sec}}$, \cref{gS.proxy.C1} reduces to $\gamma_\uS \simeq [1+4w\sin^2(\om_2\Delta t/2)]^{-1/2}$, which can be re-expanded in $w$ and one recovers the perturbative result~\eqref{gS.gP}, noticing that $\gp^2\simeq w/2$ in the C1 regime. This explains why perturbation theory is formally correct even in the supercritical regime, provided $w$ is small enough: it correctly describes the oscillations that take place with large frequency $\om_2\simeq \om_\uE$ and that dominate the dynamics at early times, but at late times, it fails to correctly account for the secular decay of the purity, \ie~for decoherence (this is further discussed below, in \cref{sec:ISOSO:discussion}).  

Note finally that even in the subcritical regime, the correction to the purity can be large at late times, since it is controlled by $w/\delta\psi$ there. Therefore, if $\delta\psi\ll w\ll 1$, the purity oscillations have a large amplitude and are not properly described by perturbation theory. We conclude that perturbation theory may always fail at late times, not only for the phase of the purity but also for its overall amplitude. 

\stoptoc
\subsubsection{Case C2: $w\simeq 1$ and $\psi \simeq 1$}
\resumetoc

We now consider the regime where the two frequencies are close, and $\xio$ is near the critical value. Expanding the ISOSO purity in $\delta w$ and $\delta\psi$, one finds
\bea
\gamma_\uS(t) \simeq \biggl\{& \frac{1}{8\delta\psi}\sin^2(\om_1 \Delta t)\left[3-\cos(2\om_2\Delta t)\right]+\frac{1}{8 \sqrt{2\abs{\delta\psi}}}\sin(2\om_1\Delta t)\sin(2\om_2\Delta t)\\
&+\frac{23+\cos(2\om_1\Delta t)\left[11-3\cos(2\om_2\Delta t)\right]+\cos(2\om_2\Delta t)}{32}\biggl\}^{-1/2}, \label{gS.proxy.C2}
\eea
where $\om_1\simeq \sqrt{\delta\psi}\om_\uS$ and $\omega_2\simeq \sqrt{2}\omega_\uS(1+\delta w/2-\delta\psi/4)$. Similarly to the case C1, in the supercritical regime, $\delta\psi<0$ and $\om_1$ becomes imaginary, hence the purity decays exponentially and decoherence takes place. At early times, the second line in \cref{gS.proxy.C2} dominates (this is why the formula is expanded up to that order), but contrary to case C1, at leading order it does not reduce to the perturbative formula~\eqref{gS.gP}. This is because, in the C2 regime, $\gp$ given in \cref{eq:gp:w:psi} is of order one, hence C2 never falls in the perturbative regime. 

The above formula is compared with the full numerical integration of the transport equations in the bottom panels of \cref{fig:C1:C2}, where one can check that the agreement is indeed excellent.

\stoptoc
\subsubsection{Case O1: $w\ll 1$ and $\psi \gg 1$}
\resumetoc

Let us now turn our attention to the case of large frequency separation and large coupling. The behaviour of the purity depends on the ordering according to which the expansion in the two small parameters $w$ and $1/\psi$ is performed, hence the two subcases need to be distinguished. 

\begin{figure}[t]
\centering 
\includegraphics[width=0.48\textwidth]{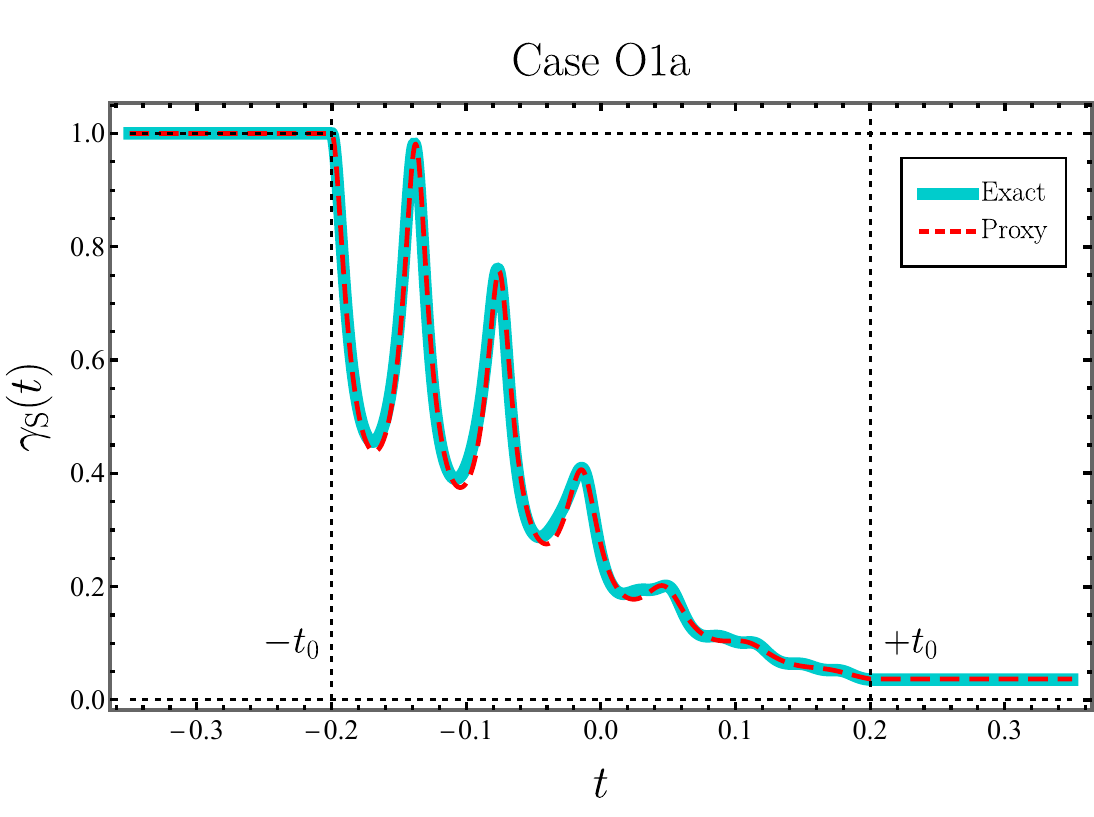} 
\includegraphics[width=0.493\textwidth]{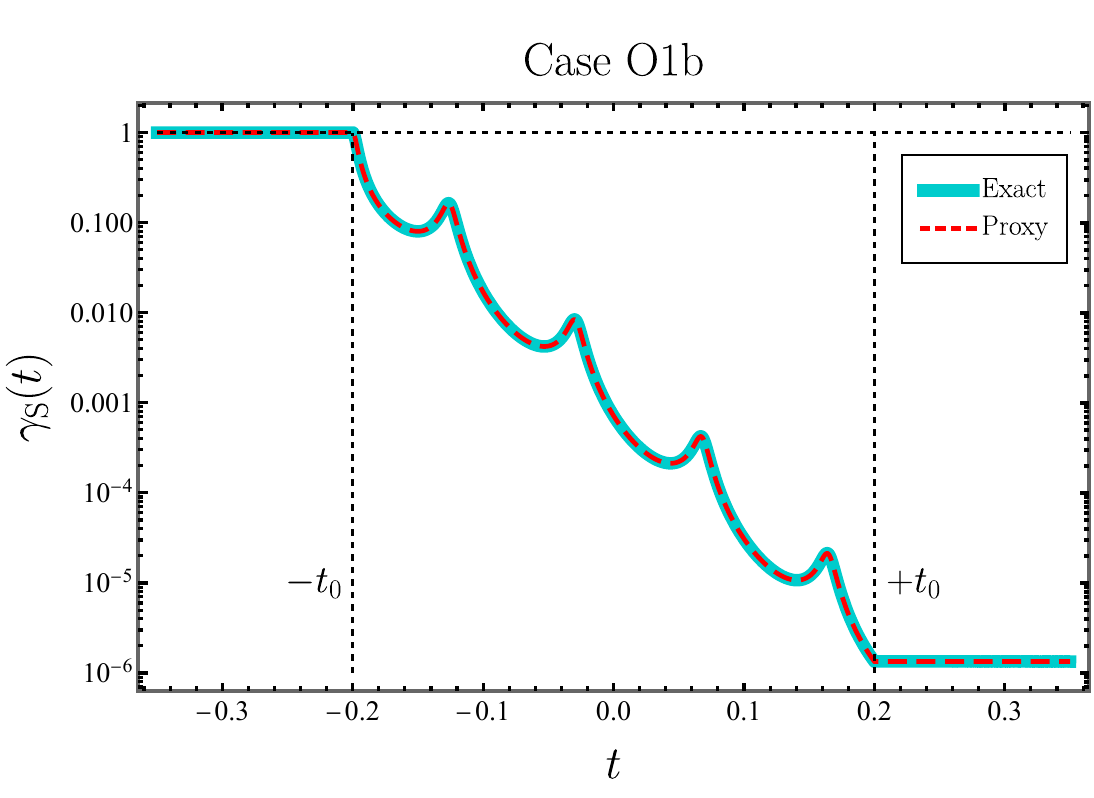} 
\includegraphics[width=0.48\textwidth]{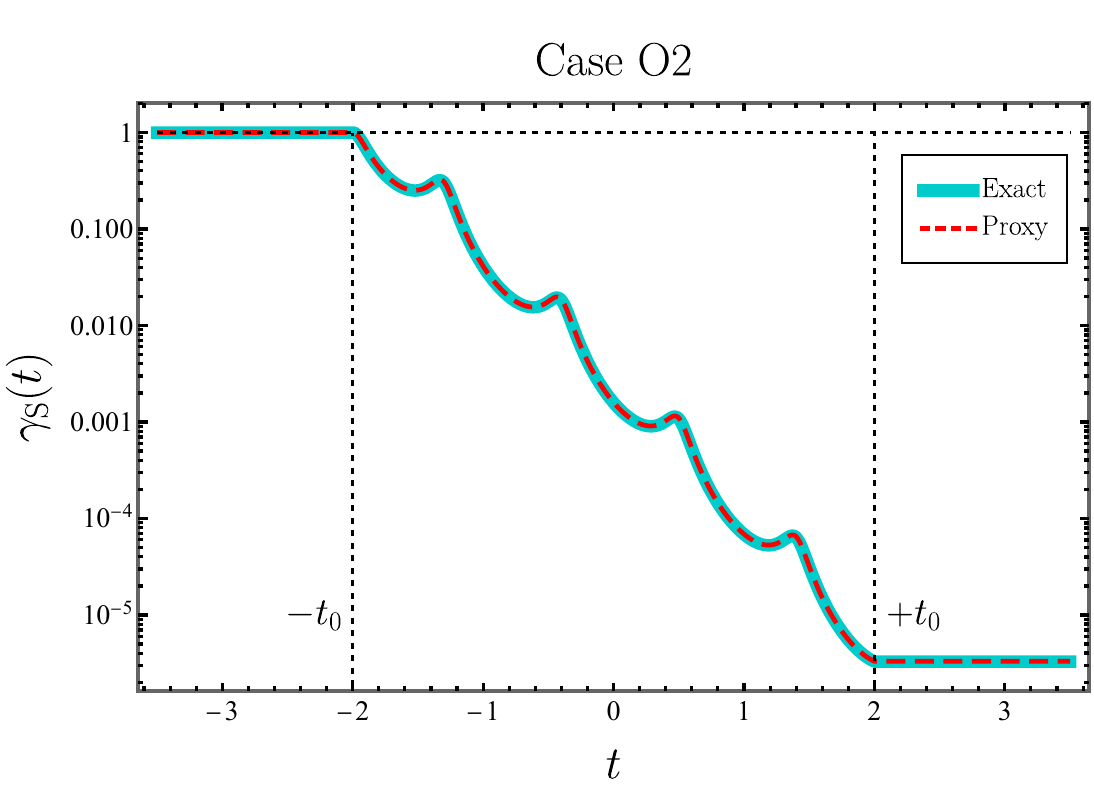} 
\caption{Comparison between the exact and expanded purities~\eqref{gS.proxy.O1a}, \eqref{gS.proxy.O1b} and \eqref{gS.proxy.O2}, in the O1a, O1b and O2 cases respectively. The value $\om_\uS=1$ is used in all panels, and the exact result is obtained from numerically solving the transport equations with $\tau=10^{-4} \tO$. The values for $(t_0,w,\psi)$ are as follows: $\mathrm{O}1a:(0.2,10^{-2},10)$, $\mathrm{O}1b:(0.2,10^{-1},10^2)$ and $\mathrm{O}2:(2,1/1.1,10)$. They correspond to the points labelled in \cref{fig.phase.diagram}.
}
\label{fig:O1a:O1b:O2} 
\end{figure}

\paragraph{Sub-case O1a: $w \ll 1/\psi\ll 1$}
When expanding the ISOSO purity first in $w$ and then in $1/\psi$, one finds
\bea
\gamma_\uS(t) \simeq\left\lbrace1+ w\psi^2\left[\frac{1}{2}\cosh(2|\om_1|\Delta t)-2\cosh(|\om_1|\Delta t)\cos(\om_2\Delta t)+\frac{3}{2}\right]\right\rbrace^{-1/2}, \label{gS.proxy.O1a}
\eea
where $\om_1\simeq i\om_\uS \psi[1-1/(2\psi^2)]$ and $\om_2\simeq \om_\uE (1+w^2\psi^2/2)$. Note that, since $w \psi\ll 1$, one has $\vert\om_1\vert \ll \om_2$, hence decoherence (driven by $\om_1$) proceeds more slowly than the oscillations (driven by $\om_2$). Case O1a thus corresponds to slow decoherence that modulates rapid oscillations, as can be checked in \cref{fig:O1a:O1b:O2}.

\paragraph{Sub-case O1b: $1/\psi \ll w \ll 1$}
When expanding the ISOSO purity first in $1/\psi$ and then in $w$, one finds
\bea
\gamma_\uS(t)\simeq \biggl\{ &\frac{\psi}{8}\left[\cosh(2|\om_1|\Delta t)+\sinh(2|\om_1|\Delta t)\sin(2\om_2\Delta t)-\cos(2\om_2\Delta t)\right]\\
&+\frac{1}{16w}\left[\cosh(2|\om_1|\Delta t)-2\cosh(2|\om_1|\Delta t)\cos(2\om_2\Delta t)+\cos(2\om_2\Delta t)\right]\\
&+ \frac{1}{8} \left[5+\cos(2\om_2\Delta t)+2\cosh(2|\om_1|\Delta t)\cos^2(\om_2\Delta t)\right]\\
&{ +\frac{1}{128\psi w^2}\left[3\cos(2\om_2\Delta t) -3\cosh(2|\om_1|\Delta t)+32\sinh(|\om_1|\Delta t)\sin(\om_2\Delta t)\right.}\\
&{\left.-13\sinh(2|\om_1|\Delta t)\sin(2\om_2\Delta t)\right]}
\biggl\}^{-1/2}
, 
\label{gS.proxy.O1b}
\eea
where $\om_1\simeq i \om_\uS \sqrt{\psi/w}[1-1/(4w\psi)]$ and $\om_2\simeq \om_\uS \sqrt{\psi/w}[1+1/(4w\psi)]$. The above formula is expanded at next-to-leading orders since those are relevant at early times, and are needed to recover $\gamma_\uS\to 1$ when $\Delta t\to 0$. At late times, however, the first term dominates, and decoherence proceeds at an exponential rate, being further enhanced by $\psi$. It is thus very efficient, as can be checked in \cref{fig:O1a:O1b:O2}.

\stoptoc
\subsubsection{Case O2: $w\simeq 1$ and $\psi \gg 1$}
\resumetoc

Finally, let us consider the case of small frequency separation but large coupling. Expanding the ISOSO purity in $\delta w$ and $1/\psi$, one finds
\bea
\gamma_\uS \simeq \biggl(&\frac{\psi}{8}\left[\cosh(2|\om_1|\Delta t)+\sinh(2|\om_1|\Delta t)\sin(2\om_2\Delta t)-\cos(2\om_2\Delta t)\right]\\
&+\frac{1}{8}\left\lbrace 5+2\cos(2\om_2\Delta t)+\cosh(2|\om_1|\Delta t)\left[2-\cos(2\om_2 t)\right]\right\rbrace \biggr)^{-1/2}, \label{gS.proxy.O2}
\eea
where $\om_1\simeq i \sqrt{\psi}\om_\uS[1+\delta w/2 - 1/(2\psi)]$ and $\om_2\simeq \sqrt{\psi}\om_\uS[1+\delta w/2 + 1/(2\psi)]$. In the above expression, the second line is of higher order, but it dominates at early times, where it ensures that $\gamma_\uS\to 1$ when $\Delta t\to 0$. Since the first line is the same as in \cref{gS.proxy.O1b}, the late-time behaviour of the purity is the same as in case O1b, given that the expressions for $\om_1$ and $\om_2$ are also similar. This explains the similarities between the two last panels in \cref{fig:O1a:O1b:O2}.

\subsection{Discussion}
\label{sec:ISOSO:discussion}

\begin{table}[t]
\centering
  \begin{tabular}{|c|c|c|c|c|c|}
    \hline
    \multirow{2}{*}{Case} & \multirow{2}{*}{Conditions} & \multirow{2}{*}{$\gamma_\uS^{-2}$} & \multirow{2}{*}{Perturbativity} & Oscillation & Decoherence \\  
          &   &   &   & frequency & rate \\ 
    \hline     
    U1 &
    $w \ll 1$ \& $\psi \ll 1$ & \multirow{3}{*}{$1+\mO\left(w\psi^2\right)$} & \multirow{4}{*}{\yes} & \multirow{6}{*}{$\om_\uE$} & \multirow{3}{*}{\textbf{---}} \\ 
    \cline{1-2}  
    U2a &
    $\delta w\ll \psi \ll 1$  &  &   &   & \\ 
    \cline{1-2} 
    U2b &
    $\psi\ll \delta w \ll 1$   &   &   &   & \\ 
    \cline{1-3} \cline{6-6}
    C1 &
    $w \ll 1$ \& $\delta \psi \ll 1$  & \multirow{2}{*}{$1+\mO(w/\delta\psi)$} &   &   & $\sqrt{\vert \delta\psi\vert }\om_\uS$\\ 
    \cline{1-2}  \cline{4-4}  
    C2 &
    $\delta w \ll 1$ \& $\delta \psi \ll 1$  &   & \no &  & if $\delta\psi<0$ \\ 
    \cline{1-4} \cline{6-6}
    O1a &
    $w \ll 1/\psi \ll 1$ & $1+\mO(w\psi^2)$ &  only if $w\psi^2\ll 1$ &   & $\xi_0/\om_\uE$\\
     \cline{1-4} \cline{5-6}
     O1b &
    $1/\psi \ll w\ll 1 $ & \multirow{2}{*}{$\mO(\psi)$} & \multirow{2}{*}{\no}  & \multirow{2}{*}{$\sqrt{\xio}$} & \multirow{2}{*}{$\sqrt{\xio}$}\\
    \cline{1-2}  
    O2 & $\delta w \ll 1$ \& $1/\psi \ll 1$ &   &  &   &  \\ 
    \hline
  \end{tabular}
\caption{Behaviour of the purity in the different regimes identified in parameter space. The third column reports the order at which $\det(\bm{\sigma}_\uS)=\gamma_\uS^{-2}$ starts. The fourth column indicates whether or not perturbation theory holds. The fifth and sixth columns give the frequency of the purity oscillations and the exponential rate $\vert\om_1\vert$ at which decoherence takes place (when it does take place), respectively.}
\label{tab.cases}
\end{table}

The above results are summarised in \cref{tab.cases}.  The subcritical cases all take place in the perturbative regime, where the purity oscillates at frequency $\om_\uE$ with an amplitude controlled by $w\psi^2$. In the near-critical cases, perturbativity holds only for large frequency separation at early times, and decoherence takes place at supercritical coupling with a rate $\sqrt{\vert\delta\psi\vert}\om_\uS$. The supercritical cases are non-perturbative in general and undergo decoherence at a rate that depends on the hierarchy between $w$ and $1/\psi$.

Let us further comment on the validity of the perturbative approach in the supercritical regime. As mentioned above, this is the case for the early stage of the near-critical case C1. In fact, even overcritical cases can behave perturbatively at early times, provided $w$ is small enough: in case O1a, since $\gp^2\simeq w\psi^2/2$, perturbativity is recovered if $w\ll 1/\psi^2$. This can be checked explicitly in \cref{gS.proxy.O1a}: if $w \psi^2\ll 1$, at early times one recovers \cref{gS.gP}.

\begin{figure}[t]
\centering 
\includegraphics[width=0.48\textwidth]{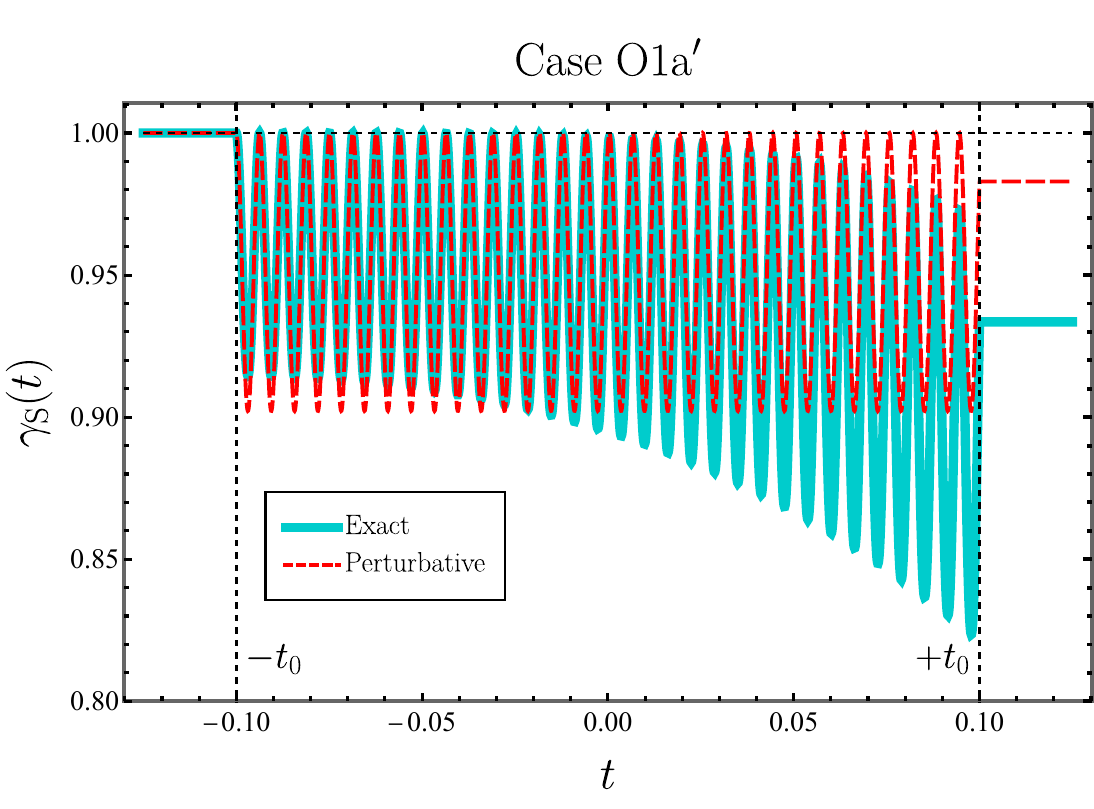} 
\includegraphics[width=0.48\textwidth]{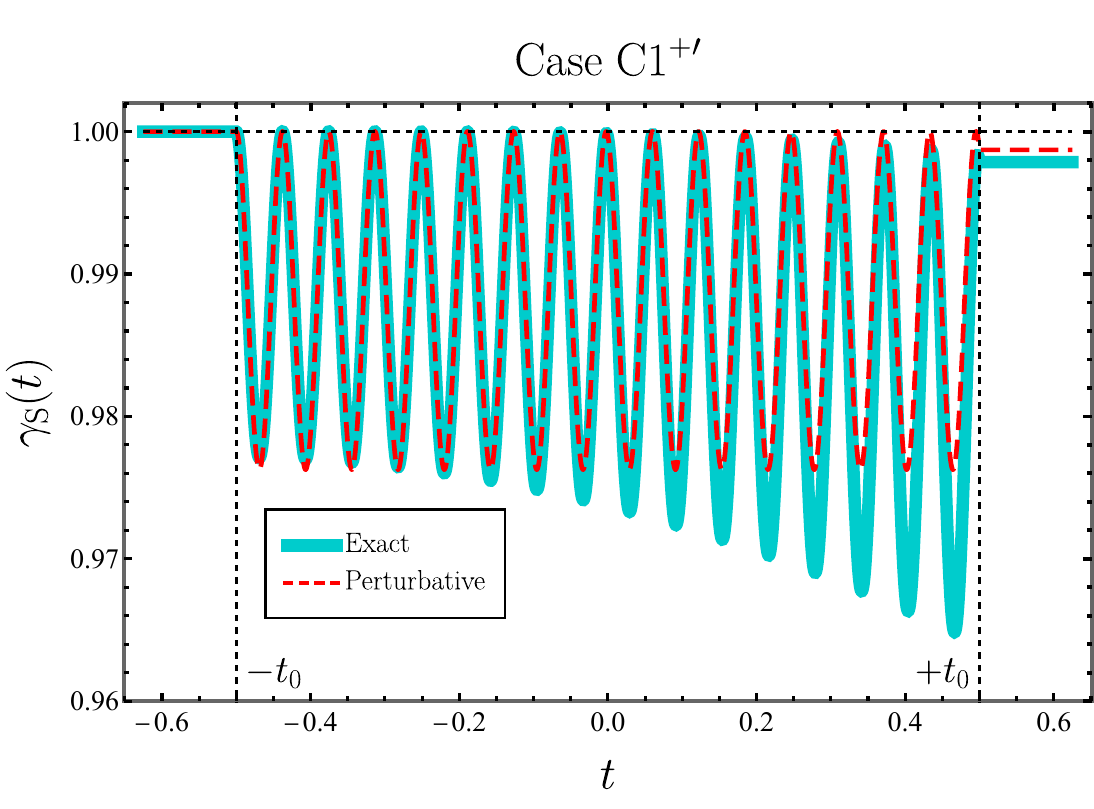} 
\caption{Comparison between the exact and perturbative purity~\eqref{gS.gP} in the $\mathrm{O}1\mathrm{a}'$ ($\om_\uS=1$, $\om_\uE=1000$, $\xio=7\xic$, $\tO=0.1$, leading to $\gp=0.16$) and $\mathrm{C}1^{+\prime}$ ($\om_\uS=1$, $\om_\uE=100, \tO=0.5$, $\xio=1.1\xic$, leading to $\gp\simeq 0.078$) cases. Perturbation theory captures the rapid oscillations but not the secular decay of the purity (a similar behaviour is found for the $\mathrm{C}1^{-\prime}$ case labelled on \cref{fig.phase.diagram} but not displayed here).
}
\label{fig:C1bar:O1abar} 
\end{figure}

For this reason, we show the cases C1 and O1a in \cref{fig:C1bar:O1abar} again, with smaller values of $w$ such that $\gp\simeq 0.1$ and they fall in the perturbative regime. We are referring to the perturbative version of these cases using a prime, \ie $\mathrm{O}1\mathrm{a}^\prime$, $\mathrm{C}1^{+\prime}$, and $\mathrm{C}1^{-\prime}$. One can see that perturbation theory is indeed valid at early times, when the dynamics of the purity is dominated by rapid oscillations that \cref{gS.gP} tracks correctly. At late times, however, the slow decay of the purity is not properly captured, so perturbation theory breaks down due to secular effects. Therefore, in the model studied in this work, decoherence is always a non-perturbative phenomenon.\footnote{This does not preclude perturbative effects to be resummed in non-perturbative master equations~\cite{breuerTheoryOpenQuantum2002,Calzetta:2008iqa} that successfully account for decoherence.}

Let us also stress that decoherence takes place in and only in the supercritical regime, and proceeds exponentially as $\gamma_\uS\propto e^{-\vert\om_1\vert t}$. This exponential profile has been added to the left panel of \cref{fig.gS.var.xio}, where one can check that it provides a good fit to the purity decay indeed. The rate $\vert\om_1\vert$ is reported in the last column of \cref{tab.cases}, where one can see that it increases with the coupling strength as expected, and also that it decays with $\om_\uE$ in the case O1a. In the subcritical cases, the amplitude of the purity oscillations also increases with $\xio$ and decreases with $\om_\uE$. This can be interpreted as a consequence of the fact that, at large frequency separation, the system and the environment ``communicate'' less efficiently.

\section{Adiabaticity and late-time purity}
\label{sec:Adiabaticity}

In the instantaneous switch-on and switch-off limit studied in \cref{sec:Criticality:Perturbativity}, when the interaction turns off, the purity freezes to the value it acquired at the end of the interacting phase, without any effect from the transition itself. This is because, as explained above, the covariance matrix (hence the purity) needs to be continuous at the transition if the coupling remains finite. In subcritical regimes where the purity merely oscillates, this implies that the late-time value of the purity is seemingly random, since it depends on the phase of the oscillations at the turn-off time, see \cref{fig:U123}. 

This somehow contrasts with the preliminary exploration conducted in \cref{sec:Criticality:and:Recoherence}, see \cref{fig.gS.var.xio,fig.gS.var.omE}, where the purity increases substantially as the coupling is smoothly turned off. This phenomenon of recoherence, which is at the heart of the present work, thus appears to be related to the rate at which the interaction is switched off. In this section, our goal is to analyse this relationship more closely, by designing an expansion scheme opposite to the one laid out in \cref{sec:ISOSO}, namely work in the limit where the coupling function $\xi(t)$ varies slowly at the transitions, $\tau\gg \tO$. For explicitness, we will assume that the coupling remains subcritical at all times, $\xi(t)<\xic$ (this assumption will be further discussed in \cref{sec:Conclusions}).

In the adiabatic basis introduced in \cref{sec:rotated:basis}, the interaction is mediated by $\dot{\xi}$, which is small if $\xi(t)$ is a slow function. This basis is thus well suited to perform an adiabatic expansion (hence its name). It leads to equations of motion of the form
\bea
\label{eq:Z:eom}
\dot{\hat{\bm{Z}}} = \bm{K} \hat{\bm{Z}}
\quad\text{where}\quad
\bm{K}=\bm{\Omega}\bm{\mathcal{H}_Z}
\eea
and $\bm{\mathcal{H}_Z}$ is given in \cref{eq:HZ}. The matrix $\bm{K}$ can be decomposed as $\bm{K}=\bm{K}_\tad + \bm{K}_\tnad$, where $\bm{K}_\tad$ arises from the diagonal portion of $\bm{\mathcal{H}_Z}$ and $\bm{K}_\tnad$ stems from the off-diagonal elements. This leads to
\bea
\label{Kad:Knad}
\bm{K}_\tad = \begin{pmatrix}
0 & \om_1 & 0 & 0\\
-\om_1 & 0 & 0 & 0\\
0 & 0 & 0 & \om_2\\
0 & 0 & -\om_2 & 0\\
\end{pmatrix}
\quad\text{and}\quad
\bm{K}_\tnad=
\begin{pmatrix}
\beta_1 & 0 & -\dot{\theta}\sqrt{\frac{\om_1}{\om_2}} & 0\\
0 & -\beta_1 & 0 & -\dot{\theta}\sqrt{\frac{\om_2}{\om_1}}\\
\dot{\theta}\sqrt{\frac{\om_2}{\om_1}} & 0 & \beta_2 & 0\\
0 & \dot{\theta}\sqrt{\frac{\om_1}{\om_2}} & 0 & -\beta_2
\end{pmatrix}\, .
\eea 
Since $\bm{K}_\tnad$ contains time derivatives of the coupling in all elements, our strategy is to perform a perturbative expansion in $\bm{K}_\tnad$, similar to the one developed in \cref{sec:Perturbation:theory}. At leading order, this reduces to the usual WKB approximation, but it allows us to organise non-adiabatic corrections in a systematic and efficient way that will prove useful for the problem at hand. 

\subsection{Adiabatic solution}

Keeping only $\bm{K}_\tad$ in the equation of motion~\eqref{eq:Z:eom}, the solution can be written as
\begin{equation}
\hat{\bm{Z}}_{\tad}(t)=\bm{G}_\tad(t,\tin)\hat{\bm{Z}}(\tin)
\quad\text{where}\quad
\bm{G}_\tad(t,\tin)=\exp\left[\int_{\tin}^t \dd t^\pr\,\bm{K}_{\tad}(t^\pr)\right] .
\label{Z.sol.ad}
\end{equation}
For the same reason as in \cref{sec:Perturbation:theory}, $\bm{G}_\tad$ is a symplectic and orthogonal matrix, \ie $\bm{G}_\tad^{\mathrm{T}}\bm{\Omega}\bm{G}_\tad=\bm{\Omega}$ and $\bm{G}_\tad^{\mathrm{T}}\bm{G}_\tad=\bm{1}$. Inserting \cref{Kad:Knad} in the above, $\bm{G}_\tad$ is found to be made of rotation blocks, \ie
\bea
\bm{G}_\tad = \begin{pmatrix}
\bm{r}({W_1}) & 0\\
0 & \bm{r}({W_2})
\end{pmatrix}
\quad\text{where}\quad
\bm{r}(\varphi):=\begin{pmatrix}
\cos\varphi & \sin\varphi\\
-\sin\varphi & \cos\varphi
\end{pmatrix} 
\eea 
and we have defined
\begin{equation}
W_i(t):=\int_{\tin}^t\dd t^\pr\,\om_i(t^\pr) \quad \text{for} \quad i=1,2. \label{W} 
\end{equation}
In the original basis $\hat{\bm{R}} = \bm{S}_\theta \hat{\bm{Y}} = \bm{S}_\theta \bm{S}_{\om_1,\om_2}\hat{\bm{Z}}$, the evolution~\eqref{Z.sol.ad} reads
\bea
\hat{\bm{R}}(t)=\bm{S}_{\theta}(t)\bm{S}_{\om_1,\om_2}(t)
\bm{G}_\tad(t,\tin)
\bm{S}_{\om_\uS,\om_\uE}^{-1}\hat{\bm{R}}({\tin}). \label{R.evolution}
\eea
Here, we have used that $\bm{S}_\theta(\tin)=\bm{1}$ if $\tin$ is chosen in the asymptotic past where the system and the environment decouple, and likewise, $\bm{S}_{\om_1,\om_2}(\tin)=\bm{S}_{\om_\uS,\om_\uE}$. For the covariance matrix~\eqref{cov.mat.def}, this leads to 
\bea
\label{eq:sigma:adiab:1}
\bm{\sigma}(t) = &
\bm{S}_{\theta}(t)\bm{S}_{\om_1,\om_2}(t)\bm{G}_\tad(t,\tin)\bm{S}_{\om_\uS,\om_\uE}^{-1}
\bm{\sigma}(\tin)
\bm{S}_{\om_\uS,\om_\uE}^{-1} 
\bm{G}_\tad^{\mathrm{T}}(t,\tin)
\bm{S}_{\om_1,\om_2}
\bm{S}_{\theta}^{\mathrm{T}}(t)
\\
= & \bm{S}_{\theta}(t)\bm{S}_{\om_1,\om_2}^2(t)\bm{S}_{\theta}^{\mathrm{T}}(t) 
\\
=&\begin{pmatrix}
\cos^2[\theta(t)]\bm{s}^2[\om_1(t)]+\sin^2[\theta(t)]\bm{s}^2[\om_2(t)] & \frac{\sin[2\theta(t)]}{2}\left\lbrace \bm{s}^2[\om_2(t)]-\bm{s}^2[\om_1(t)]\right\rbrace\\
\frac{\sin[2\theta(t)]}{2}\left\lbrace\bm{s}^2[\om_2(t)]-\bm{s}^2[\om_1(t)]\right\rbrace & \cos^2[\theta(t)]\bm{s}^2[\om_2(t)]+\sin^2[\theta(t)]\bm{s}^2[\om_1(t)]
\end{pmatrix}
\eea
where we have used that $\bm{S}_{\om_1,\om_2}$ is symmetric,  that $\bm{\sigma}(\tin)=\tdiag(\om_\uS^{-1},\om_\uS,\om_\uE^{-1},\om_\uE)=\bm{S}_{\om_\uS,\om_\uE}^2$ and that $\bm{G}_\tad$ is orthonormal. The covariance matrix of the $\uS$-oscillator is given by the upper-left $2\times 2$ block of \cref{eq:sigma:adiab:1}, the determinant of which leads to
\begin{equation}
\gamma_{\uS}^{(0)}(t)=\left\lbrace 1-\frac{\sin^2\left[2\theta(t)\right]}{4}\left[2-\frac{\om_1(t)}{\om_2(t)}-\frac{\om_2(t)}{\om_1(t)}\right]\right\rbrace^{-1/2} ,
\label{gS.adiabatic}
\end{equation}
see \cref{purity.def}. Here, the $(0)$ superscript means $0^{\mathrm{th}}$ order in the adiabatic expansion.

\begin{figure}[t]
\centering
\includegraphics[width=0.48\textwidth]{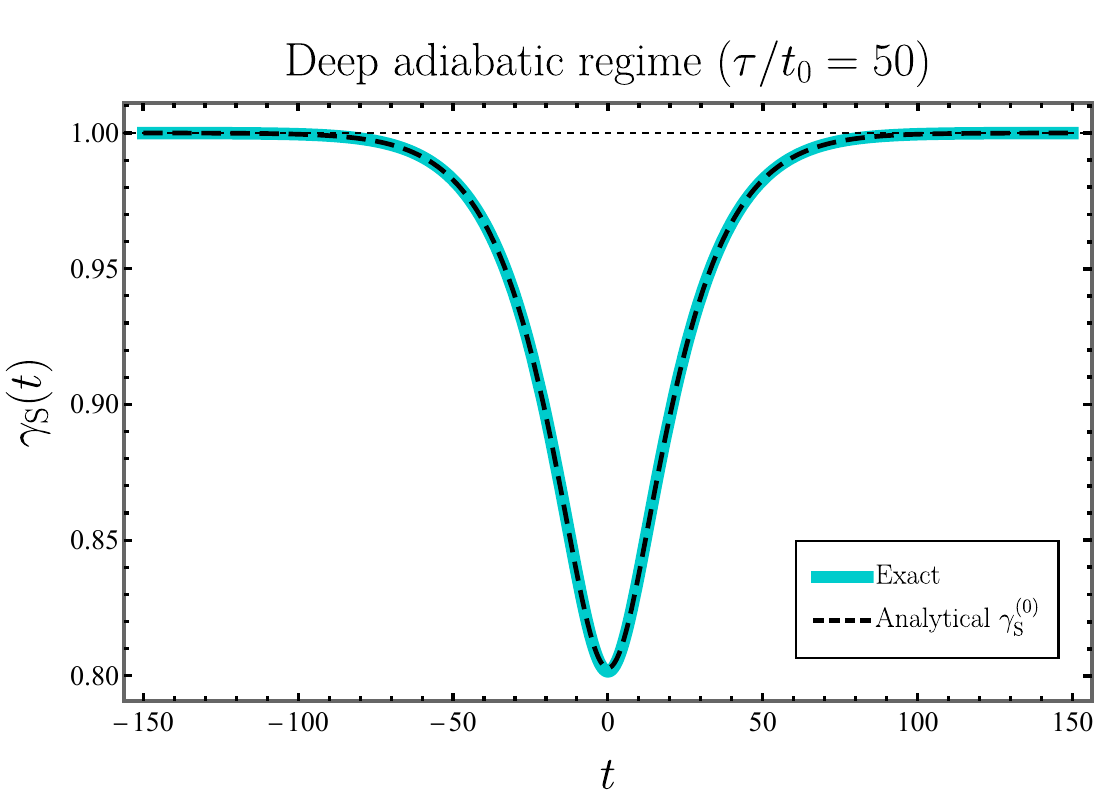}
\includegraphics[width=0.48\textwidth]{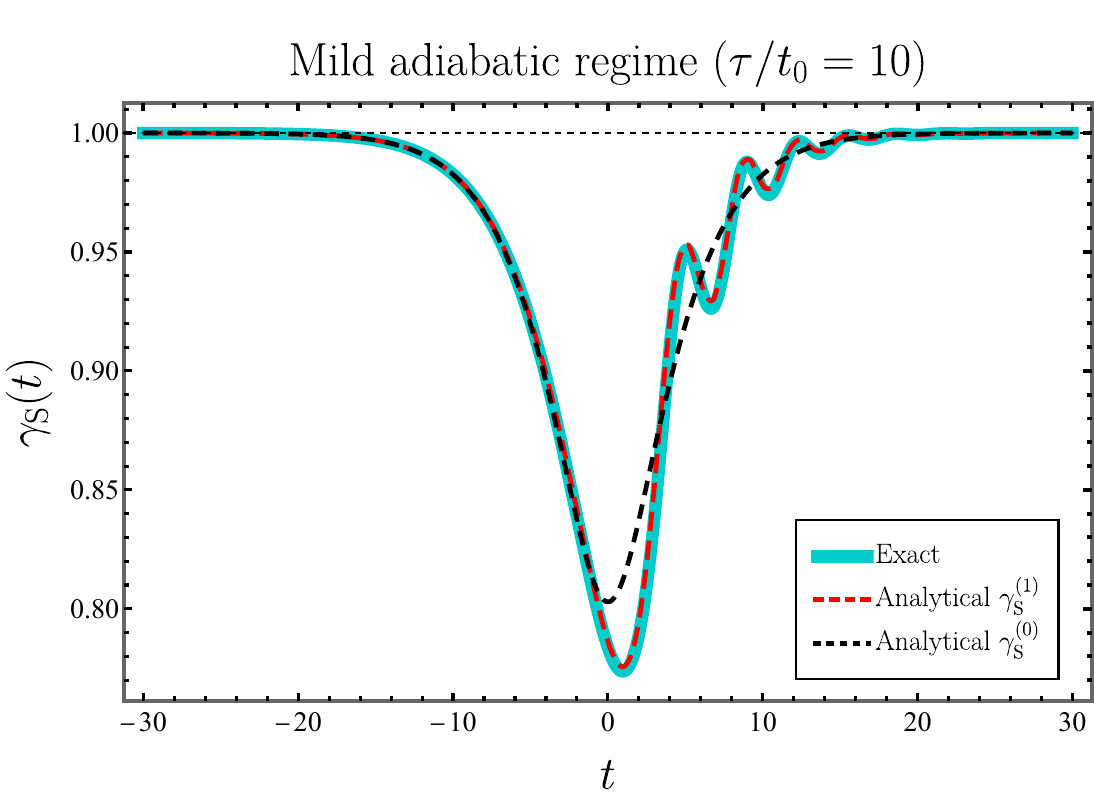}
\caption{Purity as a function of time for $\om_\uS =1$, $\om_\uE =2$, $\xio=0.9\xic $, $\tO=1$, $\tau=50$ in the left panel and $\tau=10$ in the right panel. The exact result corresponds to a numerical integration of the transport equations~\eqref{transport.eq}, the adiabatic leading order to \cref{gS.adiabatic} and the next-to-leading order to \cref{delta.gS.ep1}. 
}
\label{fig.gS.adiabatic.case}
\end{figure}

This expression is compared to the numerical integration of the transport equations in \cref{fig.gS.adiabatic.case}, where one can check that for small $t_0/\tau$ it provides an excellent approximation indeed. The formula~\eqref{gS.adiabatic} explains why recoherence takes place: at late times, $\theta\to 0$ as the interaction is switched off, see \cref{theta}, hence $\gamma_\uS^{(0)}\to 1$. This is because, at leading order in the adiabatic expansion, the setup merely tracks the adiabatic vacuum, so it goes back to the initial vacuum at late times, where the system and the environment are not entangled. This can be seen explicitly in \cref{eq:sigma:adiab:1}, which reduces to $\bm{\sigma}(t)\to\bm{\sigma}(\tin)$ at late times. 

As a consequence, recoherence is ubiquitous if the interaction is smoothly switched off, and decoherence (namely $\gamma_\uS(t\to\infty)<1$) is in essence a non-adiabatic effect~\cite{Das:2014hqa, Agon:2017oia, Gundhi:2024fqv}. 

\subsection{Non-adiabatic corrections}

For these reasons, in order to characterise the purity loss at late times, one needs to include higher-order contributions in the adiabatic expansion. Along similar lines as those of \cref{sec:Perturbation:theory}, this can be done by going to the interaction picture 
\begin{equation}
\hat{\widetilde{\bm{Z}}}(t)=\bm{G}_{\tad}^{-1}(t,\tin)\hat{\bm{Z}}(t),
\label{nad.basis}
\end{equation}
in which the equation of motion~\eqref{eq:Z:eom} reads
\bea
\frac{\dd \hat{\widetilde{\bm{Z}}}}{\dd t}=\widetilde{\bm{K}}_\tnad \hat{\widetilde{\bm{Z}}}_\tnad.
\label{Ztilde.eom}
\eea
Here,
\begin{align}
\nonumber
\widetilde{\bm{K}}_\tnad(t)=&\bm{G}^{-1}_{\tad}(t,\tin)\bm{K}_\tnad(t)\bm{G}_{\tad}(t,\tin)\\
=&\begin{pmatrix}
\beta_1(t) \bm{\bar{r}}[2W_1(t)] & -\dot{\theta}(t)\bm{r}[-W_1(t)]\bm{s}^{-1}[\om_1(t)]\bm{s}[\om_2(t)]\bm{r}[W_2(t)]\\
\dot{\theta}(t)\bm{r}[-W_2(t)]\bm{s}[\om_1(t)]\bm{s}^{-1}[\om_2(t)]\bm{r}[W_1(t)] & \beta_2(t) \bm{\bar{r}}[2W_2(t)] 
\end{pmatrix}, \label{Ktilde.nad}
\end{align}
and we have defined the improper rotation matrix
\begin{align}
\bm{\bar{r}}(\varphi):=\begin{pmatrix}
\cos\varphi & \sin\varphi\\
\sin\varphi & -\cos\varphi
\end{pmatrix} .
\end{align}
The solution to \cref{Ztilde.eom} is given by
\begin{equation}
\hat{\widetilde{\bm{Z}}}(t)=\widetilde{\bm{G}}_\tnad(t,\tin)\hat{\widetilde{\bm{Z}}}(\tin) 
\quad\text{where}\quad
\widetilde{\bm{G}}_\tnad(t,\tin)=\mathrm{T}\exp\left[\int_{\tin}^t \dd t^\pr\,\widetilde{\bm{K}}_\tnad(t^\pr)\right]. 
\label{G.tilde}
\end{equation}
Taylor expanding the time-ordered exponential function corresponds to performing a Dyson series, the first order of which reads
\begin{equation}
\widetilde{\bm{G}}_\tnad^{(1)}(t,\tin)=\bm{I}+
\bm{\delta} \widetilde{\bm{G}}^{(1)}(t,\tin)
\quad\text{where}\quad 
\bm{\delta} \widetilde{\bm{G}}^{(1)}(t,\tin)=
\int_{\tin}^t\dd t^\pr\,\widetilde{\bm{K}}_{\tnad}(t^\pr). \label{Gtilde1}
\end{equation}

In the original basis $\hat{\bm{R}} = \bm{S}_\theta \hat{\bm{Y}} = \bm{S}_\theta \bm{S}_{\om_1,\om_2}\hat{\bm{Z}}=\bm{S}_\theta \bm{S}_{\om_1,\om_2} \bm{G}_\tad \hat{\widetilde{\bm{Z}}}$, \cref{G.tilde} leads to
\bea
\hat{\bm{R}}(t) =& \bm{S}_\theta(t) \bm{S}_{\om_1,\om_2}(t) \bm{G}_\tad(t,\tin) \widetilde{\bm{G}}_\tnad(t,\tin)\bm{S}^{-1}_{\om_\uS,\om_\uE}\hat{\bm{R}}(\tin)\, .
\eea 
For the covariance matrix $\bm{\sigma}=\bm{\sigma}^{(0)}+\bm{\delta\sigma}^{(1)}+\cdots$ where $\bm{\sigma}^{(0)}$ was computed in \cref{eq:sigma:adiab:1}, this leads to
\bea
\bm{\delta\sigma}^{(1)} = \bm{S}_\theta(t) \bm{S}_{\om_1,\om_2}(t) \bm{G}_\tad(t,\tin)\left[\delta\widetilde{\bm{G}}^{(1)}(t,\tin) +\delta\widetilde{\bm{G}}^{(1)\mathrm{T}}(t,\tin)\right]\bm{G}_\tad^{\mathrm{T}}(t,\tin)\bm{S}_{\om_1,\om_2}(t) \bm{S}_\theta^{\mathrm{T}}(t)\, .
\eea 
It thus remains to evaluate this matrix. As intermediate steps in the calculation, let us note that
\bea
\bm{S}_\theta\bm{S}_{\om_1,\om_2}=\begin{pmatrix}
\cos(\theta)\bm{s}(\om_1) & \sin(\theta)\bm{s}(\om_2)\\
-\sin(\theta)\bm{s}(\om_1) & \cos(\theta)\bm{s}(\om_2)
\end{pmatrix}
\eea
and that
\bea
\bm{G}_\tad(t,\tin)\left[\delta\widetilde{\bm{G}}^{(1)}(t,\tin) +\delta\widetilde{\bm{G}}^{(1)\mathrm{T}}(t,\tin)\right]\bm{G}_\tad^{\mathrm{T}}(t,\tin)
=\begin{pmatrix}
\bm{V}_{11}(t) & \bm{V}_{12}(t)\\
\bm{V}_{12}(t) & \bm{V}_{22}(t)
\end{pmatrix},
\eea
where
\begin{align}
\bm{V}_{11}(t)&:=\int_{\tin}^t\dd t^\pr\, \frac{\dot{\om}_1(t^\pr)}{\om_1(t^\pr)}\bm{\bar{r}}\left\lbrace 2\left[W_1(t^\pr)-W_1(t)\right]\right\rbrace ,\\
\bm{V}_{22}(t)&:=\int_{\tin}^t\dd t^\pr\,\frac{\dot{\om}_2(t^\pr)}{\om_2(t^\pr)}\bm{\bar{r}}\left\lbrace 2\left[W_2(t^\pr)-W_2(t)\right]\right\rbrace ,\\
\bm{V}_{12}(t)&:=\int_{\tin}^t\dd t^\pr\,\dot{\theta}(t^\pr)\left[\sqrt{\frac{\om_2(t^\pr)}{\om_1(t^\pr)}}-\sqrt{\frac{\om_1(t^\pr)}{\om_2(t^\pr)}}\right]\bm{\bar{r}}[W_1(t^\pr)-W_1(t)+W_2(t^\pr)-W_2(t)] .
\end{align}

The correction to the system's covariance matrix corresponds to the upper-left block of $\bm{\delta\sigma}^{(1)}$, which is given by
\bea
\bm{\delta \sigma}_\uS ^{(1)}(t)=&\cos^2\left[\theta(t)\right]\bm{s}[\om_1(t)]\bm{V}_{11}(t)\bm{s}[\om_1(t)]+\sin^2\left[\theta(t)\right]\bm{s}[\om_2(t)]\bm{V}_{22}(t)\bm{s}[\om_2(t)]\\
&+\frac{\sin\left[2\theta(t)\right]}{2}\left\lbrace \bm{s}[\om_1(t)]\bm{V}_{12}(t)\bm{s}[\om_2(t)]+\bm{s}[\om_2(t)]\bm{V}_{12}(t)\bm{s}[\om_1(t)]\right\rbrace. \label{delta.sigmaS.ep1}
\eea
Finally, using the generic formula $\delta[(\det \bm{M})] = \det(\bm{M})\mathrm{Tr}(\bm{M}^{-1}\delta \bm{M})$ for any matrix $\bm{M}$, \cref{purity.def} leads to
\bea
\gamma_\uS^{(1)}= &
\gamma_\uS^{(0)}\left\lbrace1-\frac{1}{2}\mathrm{Tr}\left[\left(\bm{\sigma}_\uS^{(0)}\right)^{-1}\bm{\delta\sigma}_\uS^{(1)}\right]\right\rbrace
=\gamma_\uS ^{(0)}-\frac{1}{2}\left(\gamma_\uS ^{(0)}\right)^3\left[\sigma_{\uS,11}^{(0)}\delta\sigma_{\uS,22}^{(1)}+\sigma_{\uS,22}^{(0)}\delta\sigma_{\uS,11}^{(1)}\right] . \label{gS.nad.ep1.aux}
\eea 
Only the diagonal elements of $\bm{\delta\sigma}_\uS^{(1)}$ are involved in that expression, and from \cref{delta.sigmaS.ep1} they are given by
\bea
\delta\sigma_{\uS,11}^{(1)}=&\frac{\cos^2(\theta)}{\om_1}\,\mI_{\om_1}+\frac{\sin^2(\theta)}{\om_2}\,\mI_{\om_2}+\frac{\sin(2\theta)}{\sqrt{\om_1\om_2}}\,\mI_\theta , \\
\delta\sigma_{\uS,22}^{(1)}=&-\om_1\cos^2(\theta)\,\mI_{\om_1}-\om_2\sin^2(\theta)\,\mI_{\om_2}-\sqrt{\om_1\om_2}\sin(2\theta)\,\mI_\theta ,\label{delta.sigmaS11.sigmaS22.ep1}
\eea
where
\begin{align}
\mI_{\om_i}&:=\int_{{\tin}}^t\dd t^{\pr}\,\frac{\dot{\om}_i(t^\pr)}{\om_i(t^\pr)}\cos\left\lbrace 2\left[W_i(t)-W_i(t^\pr)\right]\right\rbrace,  \label{mI.omega}\\
\mI_{\theta}&:=\int_{{\tin}}^t\dd t^{\pr}\,\dot{\theta}(t^{\pr})\left[\sqrt{\frac{\om_2(t^{\pr})}{\om_1(t^{\pr})}}-\sqrt{\frac{\om_1(t^{\pr})}{\om_2(t^{\pr})}}\right]\cos[W_1(t)-W_1(t^{\pr})+W_2(t)-W_2(t^{\pr})] \label{mI.theta}.
\end{align}
Inserting \cref{eq:sigma:adiab:1,delta.sigmaS11.sigmaS22.ep1} into \cref{gS.nad.ep1.aux}, one finds
\bea
\delta\gamma_{\uS}^{(1)}=\frac{\sin(2\theta)}{2}\,\left(\gamma_\uS ^{(0)}\right)^3\left[\frac{\sin(2\theta)}{4}\left(\frac{\om_2}{\om_1}-\frac{\om_1}{\om_2}\right)\left(\mI_{\om_2}-\mI_{\om_1}\right)-\cos(2\theta)\left(\sqrt{\frac{\om_2}{\om_1}}-\sqrt{\frac{\om_1}{\om_2}}\right)\mI_{\theta}\right], \label{delta.gS.ep1}
\eea
which is displayed in the right panel of \cref{fig.gS.adiabatic.case} where the adiabatic condition is only mildly satisfied ($\tau/\tO=10$). 

One can check that the agreement is excellent and that the deviations from the leading-order adiabatic purity are correctly captured. These deviations include oscillations in the recohering phase, which make the purity profile time-asymmetric, contrary to the coupling function profile. This implies that at this order, the dynamics not only tracks the local adiabatic vacuum, but particle creation arising from the time dependence of the rotated frequencies $\om_1$ and $\om_2$ is also accounted for. The spontaneous creation of entangled particles within the sectors 1 and 2 is described by the term proportional to $\mI_{\om_2}-\mI_{\om_1}$ in \cref{delta.gS.ep1}, while the term involving $\mI_\theta$ describes the effect of the rotation between the two sectors when reaching the $(\uS,\uE)$ basis.

\begin{figure}[t]
\centering 
\includegraphics[width=0.72\textwidth]{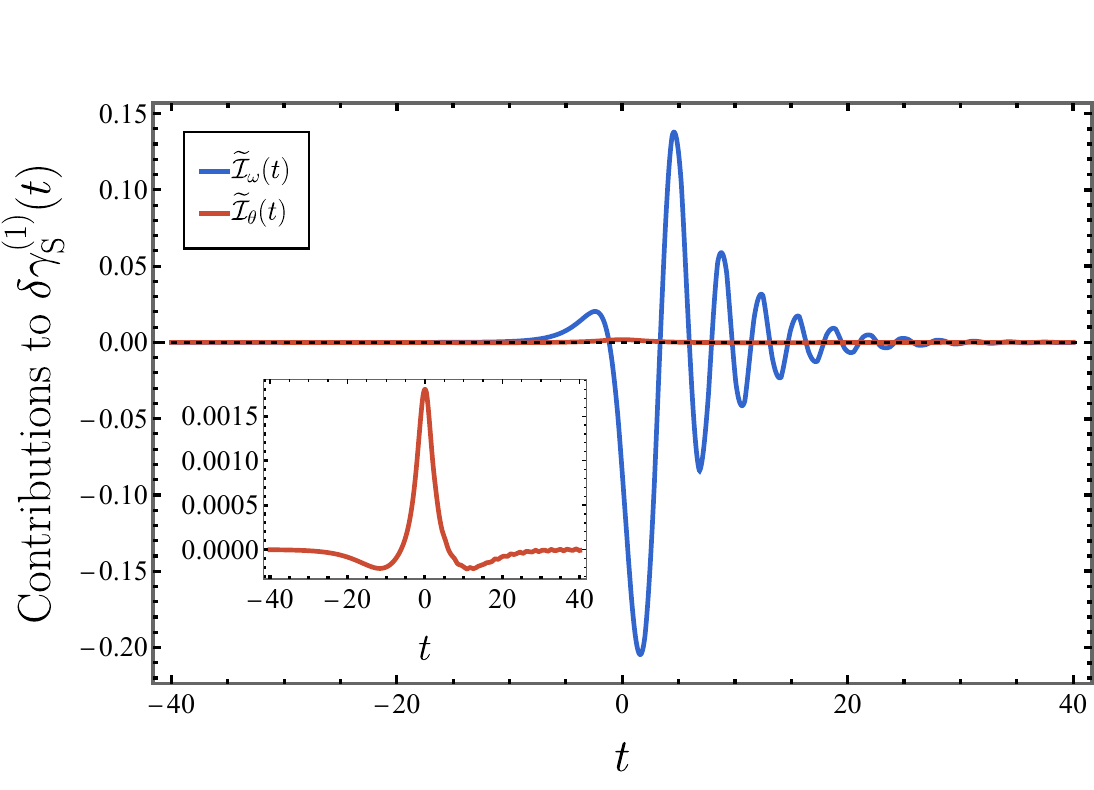} 
\caption{Comparison between the two contributions~\eqref{eq:Itildew:Itildetheta:def} arising in the first non-adiabatic correction to the purity: $\widetilde{\mI}_\om$ describes the effect of spontaneous particle creation and $\widetilde{\mI}_{\theta}$ accounts for the rotation between the adiabatic and the original bases. Particle creation provides the dominant contribution. We have used $\om_\uS=1$, $\om_\uE=2$, $\xio=0.9\xic$, $\tO=1$ and $\tau=10$, as in the right panel of \cref{fig.gS.adiabatic.case}.}
\label{fig.J.comparison} 
\end{figure}

At leading order, \cref{gS.adiabatic} only accounts for the effect of the rotation; hence, to check whether or not particle creation becomes relevant at next-to-leading order, let us compare the two contributions
\bea
\widetilde{\mI}_\om:=\frac{\sin(2\theta)}{4}\left(\frac{\om_2}{\om_1}-\frac{\om_1}{\om_2}\right)(\mI_{\om_2}-\mI_{\om_1})
\quad\text{and}\quad
\widetilde{\mI}_{\theta}:=-\cos(2\theta)\left(\sqrt{\frac{\om_2}{\om_1}}-\sqrt{\frac{\om_1}{\om_2}}\right)\mI_{\theta}
\label{eq:Itildew:Itildetheta:def}
\eea
in \cref{delta.gS.ep1}. They are displayed in \cref{fig.J.comparison}, for the case shown in the right panel of \cref{fig.gS.adiabatic.case}. One can see that particle creation is the dominant effect, and we have checked that this remains true when varying the parameters of the model. 

To summarise, at leading adiabatic order, only the basis rotation effect is accounted for, while at next-to-leading order, spontaneous particle creation provides the dominant contribution. 

\subsection{Late-time purity}
\label{sec:late:time:purity}

A striking consequence of \cref{delta.gS.ep1} is that, when the interaction is switched off, $\delta\gamma_{\uS}^{(1)}\to 0$ since it is controlled by $\sin(2\theta)$. Therefore, complete recoherence also takes place at next-to-leading order, which fails again to capture deviations from a pure state at late times.
This inability of the adiabatic expansion to capture late-time decoherence is in fact present at any finite order, as we now argue. 

Formally, the $n^{\mathrm{th}}$ order in the Dyson series of \cref{G.tilde} reads
\bea
\bm{\delta} \widetilde{\bm{G}}^{(n)}(t, {\tin})=\int_{\tin}^t\dd t_1\,\widetilde{\bm{K}}_{\tnad}(t_1)\int_{\tin}^{t_1}\dd t_2\,\widetilde{\bm{K}}_{\tnad}(t_2)\cdot\cdot\cdot \int_{\tin}^{t_{n-1}}\dd t_n\,\widetilde{\bm{K}}_{\tnad}(t_n). \label{delta.G.tilde.n}
\eea
As we are interested in the late-time purity, we wish to study the behaviour of the evolution matrix in the limit $ \bm{\delta} \widetilde{\bm{G}}^{(n)}(t\to\infty, {\tin})$. Notice that the matrix elements of the inner-most nested integral $\widetilde{\bm{K}}_\tnad(t)$ are of the form
\begin{equation}
\mathcal{I}=\int_{\tin}^{t_{n-1}}\dd t_n\, h_{\mathrm{f}}(t_n)h_{\mathrm{s}}(t_n), \quad \label{final.integral} 
\end{equation}
where $h_{\mathrm{s}}(t)$ is a slowly varying function, which is made by some combination of $\om_1(t)$, $\om_2(t)$, $\theta(t)$, $\dot{\om}_1(t)$, $\dot{\om}_2(t)$, $\dot{\theta}(t)$, and $h_{\mathrm{f}}(t)$ is a rapidly oscillating function involving $\cos[W_i(t)]$ and $\sin[W_i(t)]$. This structure is conserved upon multiplication by $\widetilde{\bm{K}}_\tnad(t_m)$ and integration,\footnote{This can be easily shown by, for example, a saddle point approximation.} hence the $n^{\mathrm{th}}$ propagator is of the form 
\begin{align}
\bm{\delta}\widetilde{\bm{G}}^{(n)}(t\to\infty,\tin)=\sum_{i}\int_{\tin}^{+\infty}\dd t_1\, h_{\mathrm{s},i}^{(n)}(t_1)h_{\mathrm{f},i}^{(n)}(t_1), \label{f.integral}
\end{align}
where $h_{\mathrm{s},i}^{(n)}$ varies over time scales of order $\tau$ and $h_{\mathrm{s},i}^{(n)}$ is an oscillating functions with typical frequencies $\om_1$ and $\om_2$. In the adiabatic limit, $\tau\gg \tO$, the period of the oscillating function is much shorter than the time scale over which the envelope $h_{\mathrm{s},i}^{(n)}$ varies. In \cref{slow.vs.fast.integral}, we show why an expansion in the ratio of these two time scales gives a vanishing result for the asymptotic value of the purity at all orders. This can be seen as a generalised Paley-Wiener theorem~\cite{Paley-Wiener}, which relates the analyticity of a function to decay properties of its Fourier transform.

\begin{figure}[t]
\centering   
\includegraphics[width=0.47\textwidth]{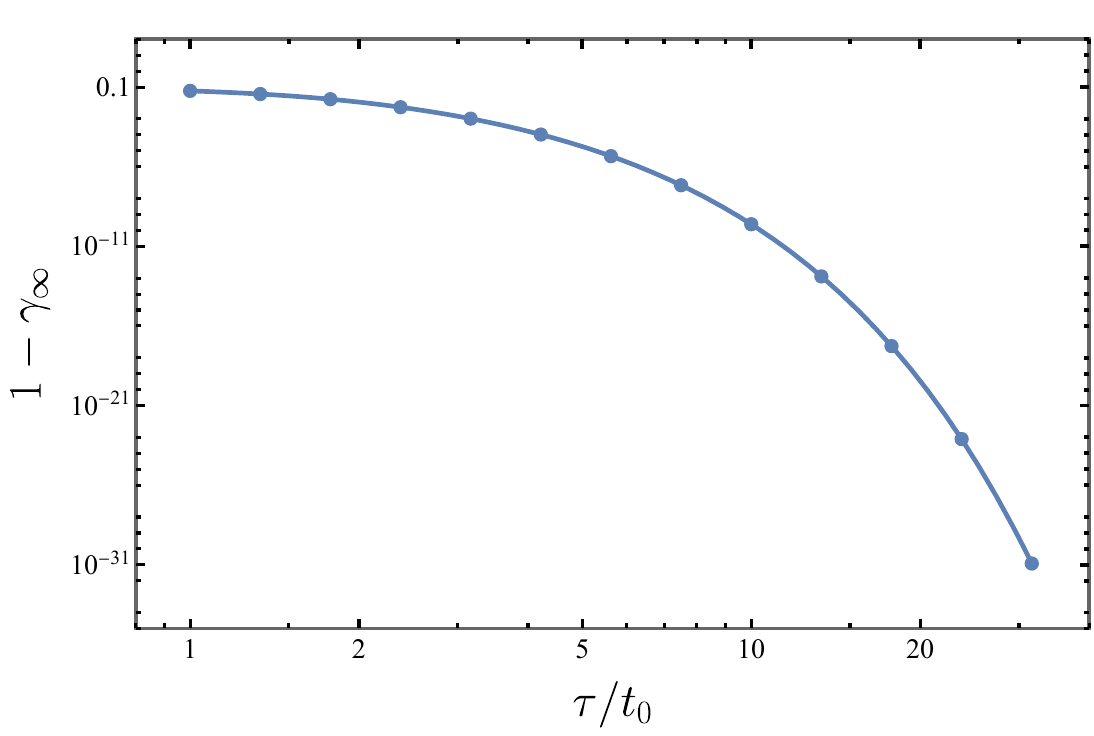} 
\includegraphics[width=0.48\textwidth]
{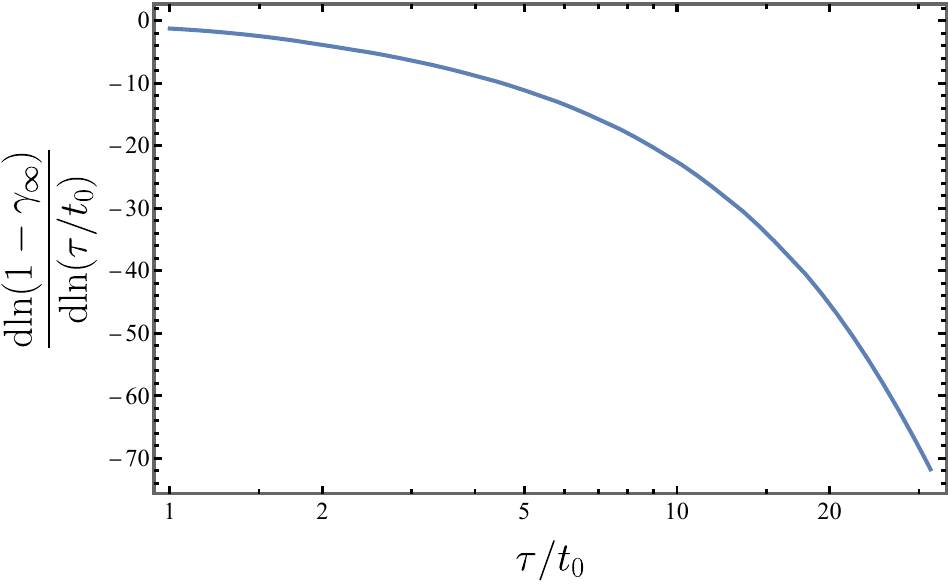}
\caption{Late-time purity $\gamma_\infty\equiv\gamma_\uS(t\to\infty)$ as a function of $\tau/\tO$, for $\om_\uS=1$, $\om_\uE=2$, $\tO=1$ and $\xio=0.9\xic$.
In practice, we stop the numerical integration when $\xi/\xic$ drops below $10^{-10}$, but we have checked that increasing the integration range does not change the result.
The functional dependence of $1-\gamma_\infty$ in terms of $\tO/\tau$ is non perturbative, which is confirmed in the right-panel: if $1-\gamma_\infty$ scaled as $(\tau/\tO)^{-p}$, $ {\vert \dd\ln(1-\gamma_\infty)}/{\dd\ln(\tau/\tO)\vert} $ would asymptote a finite value $p$ when $\tau/\tO\to \infty$, which is clearly not the case. 
}
\label{fig.gS.perturbativity.check} 
\end{figure}

To verify the validity of this conclusion, in \cref{fig.gS.perturbativity.check} we show the value of the late-time purity $\gamma_\infty\equiv\gamma_\uS(t\to\infty)$ computed in the asymptotic future from a numerical solution of the transport equations, for different values of $\tau$. One can check that $1-\gamma_\infty$ does not scale as a finite power of $\tau^{-1}$ when $\tau\to\infty$ but is much more suppressed, which confirms that the late-time purity is non-perturbative in the adiabatic parameter. In essence, decoherence is, therefore, a non-adiabatic phenomenon.\\

\begin{figure}[t]
\centering   
\includegraphics[width=0.525\textwidth]{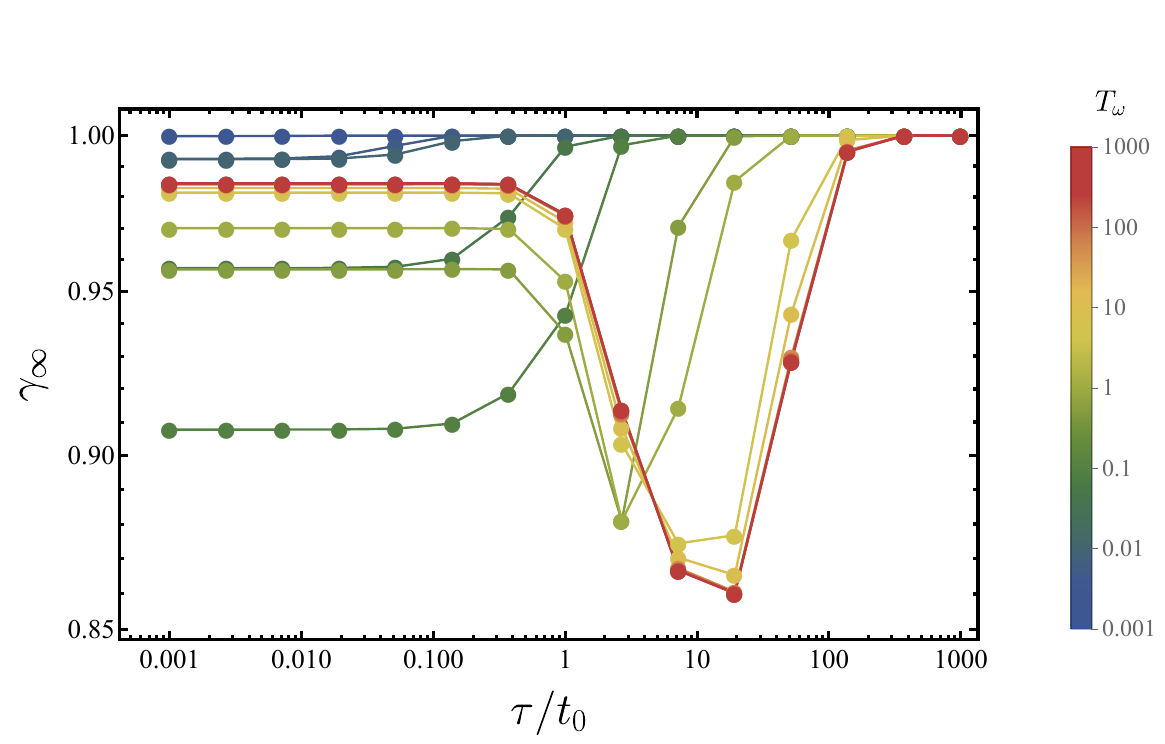} 
\includegraphics[width=0.46\textwidth]
{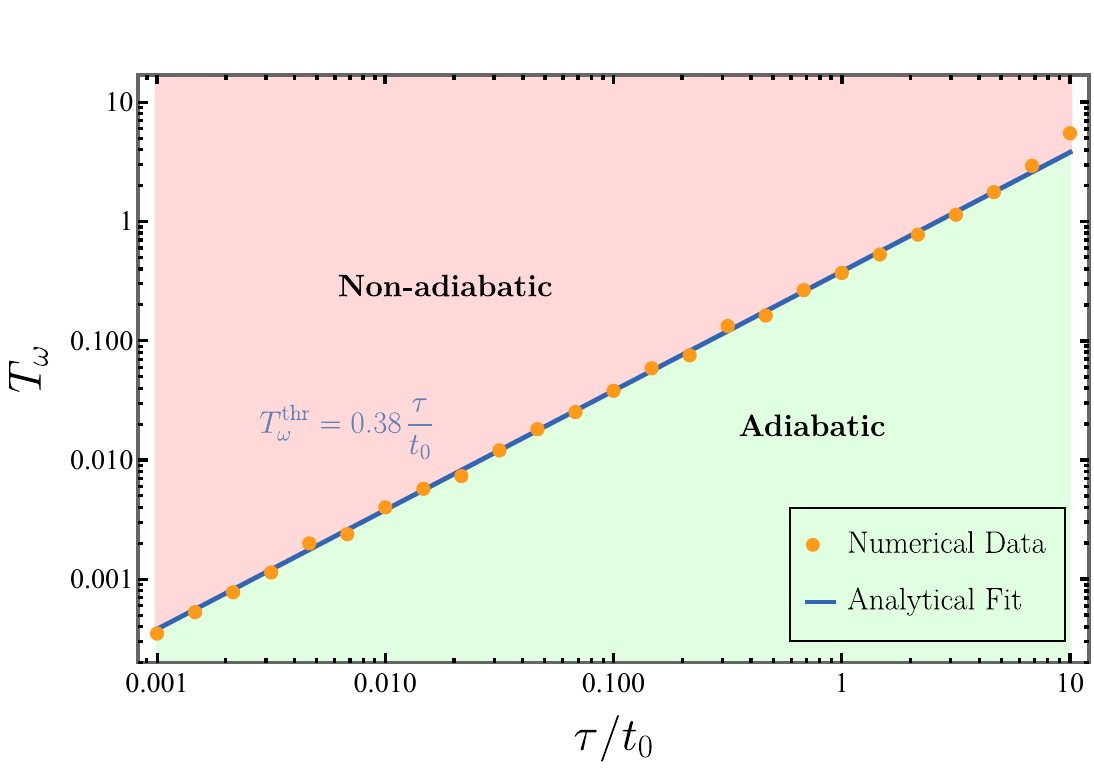} 
\caption{Left panel: Late-time purity $\gamma_\infty\equiv\gamma_\uS(t\to\infty)$ as a function of $\tau/\tO$ for different values of $T_{\omega}\equiv \omega_\uS/(\omega_\uE-\omega_\uS)$. Right panel: Threshold value $T_{\omega}^{\mathrm{thr}}$ below which $1-\gamma_\infty$ is smaller than $0.01(1-\gamma_{\mathrm{min}})$, where $\gamma_{\mathrm{min}}$ is the minimum value of $\gamma_\uS(t)$. The value $0.01$ is arbitrary and changing it (for instance to $0.001$) does not affect the linear scaling~\eqref{eq:Twthr:tau}, it only modifies the scaling coefficient.
The green region (``adiabatic'') is where recoherence takes place, while a finite amount of decoherence is obtained in the red region (``non-adiabatic''). We used $t_0=0.1$ in the left panel and $t_0=1$ in the right panel, while we fixed $\omega_{\uS}=1$ and $\xio=0.9\omega_{\uS}\omega_{\uE}$ in both panels.} 
\label{fig.gS.fin} 
\end{figure}

In order to further characterise the time scale $\tau$ above which the adiabatic regime is reached and $1-\gamma_\infty$ is suppressed, in \cref{fig.gS.fin} we show again $\gamma_\infty$ as a function of both $\tau/\tO$ and $T_\omega$, where $T_\omega=\om_\uS/(\om_\uE -\om_\uS )$. The analysis is restricted to the subcritical domain, where $\xio<\xic$. In the left panel, one can see that there are three regimes of interest. When $\tau/\tO\gg 1$ (in practice, $\tau/\tO\gtrsim 10^{2}-10^3$), recoherence takes place, regardless of  $T_\om$ and of the other parameters. This is consistent with the discussion above, where we showed that recoherence is ubiquitous in the adiabatic regime. When $1\lesssim \tau/\tO\lesssim 100$, recoherence only takes place if $T_\om$ is smaller than some threshold value $T_\om^{\mathrm{thr}}$. This implies that the critical time scale $\tau$ that delineates the adiabatic regime depends on other parameters, and on $T_\om$ in particular. Lastly, when $\tau/\tO\ll 1$, $\gamma_\infty$ reaches an asymptotic value that does not depend on $\tau/\tO$. Unless $T_\om$ is sufficiently small, that value deviates from one, \ie recoherence does not take place. This is because, in this regime, the purity oscillates during the interacting phase, and simply ``freezes'' to the (seemingly random) value it has when the interaction is sharply turned off. This is why $\gamma_\infty$ varies non-monotonically with $T_{\omega}$ at low $\tau/\tO$. Nonetheless, even in this regime of sharp turn off, there is still a value $T_\omega^{\mathrm{thr}}$ below which recoherence takes place, which is a non-trivial conclusion.

To better describe the regime in which recoherence takes place, in the right panel of \cref{fig.gS.fin} we show the plane $(\tau/\tO,T_\omega)$ where we display in green the parameters for which recoherence occurs and in red those for which this is not the case. One can see that the threshold value $T_\omega^{\mathrm{thr}}$ below which recoherence takes place scales linearly with $\tau/\tO$, \ie recoherence takes place if
\bea 
\label{eq:Twthr:tau}
\frac{\tau}{\tO}> \mathcal{O}(1) \frac{\om_\uS}{\om_\uE -\om_\uS}\, .
\eea 
This explains the behaviour previously found in \cref{fig.gS.var.omE}.
The scaling coefficient $\mathcal{O}(1)$ depends on the value of $\xio$, and in general increases with it. This can be related to the fact that in the mixing angle $\theta$, controlling the amount of decoherence, a smaller $\xio$ can be compensated by a smaller difference $\omega_\uE-\omega_\uS$, corresponding to a larger $T_{\omega}$.

\section{Non-Markovianity}
\label{sec:Markov}

The last property of our toy model we wish to investigate is the Markovian nature of its dynamical evolution. Indeed, if recoherence is interpreted as information backflow from the environment into the system, it should qualify as a ``non-Markovian'' mechanism in the sense of a memory effect. Our goal is to study the relationship between recoherence and non-Markovianity more closely, and to determine the conditions under which a Markovian dynamics emerges.

\subsection{Non-Markovian criterion for Gaussian maps}

First, let us introduce a Markovian criterion for the evolution of Gaussian states. In general, necessary and sufficient conditions for Markovian evolution are difficult to establish, but in the case of Gaussian maps, the problem can be solved as follows~\cite{non-Markov_PRL_origin, non-Markov_CV_dynamical_maps, non-Markov_CV_fidelity, 2015PhRvL.115g0401T, non-Markov_CV_interferometric, non-Markov_CV_steering, non-Markov_review}-- here, we follow \cite{2015PhRvL.115g0401T}. Let $\{\mathcal{E}(t_1,t_2), t_2\geq t_1 \geq t_0\}$ be a family of trace-preserving linear maps for the evolution of the system's state $\hat{\rho}_\uS$
\begin{equation}
\hat{\rho}_\uS(t_2)=\mathcal{E}(t_1,t_2)\hat{\rho}_\uS(t_1) . \label{dynamics}   
\end{equation}
These maps are said to be Markovian if they satisfy the semi-group property
\begin{equation}
\mathcal{E}(t_0,t_2)=\mathcal{E}(t_1,t_2)\mathcal{E}(t_0,t_1) \quad \text{for all } t_2\geq t_1 \geq t_0,      \label{semigroup}
\end{equation}
and if the map $\mathcal{E}(t_1,t_2)$ is completely positive (\ie~it maps positive density matrices to positive density matrices only) for all $t_2\geq t_1$.

As explained in \cref{sec:Model}, Gaussian maps can be expressed in terms of transport equations for the covariance matrix, see \cref{transport.eq}, the upper-left block of which gives rise to
\bea
\frac{\dd \bm{\sigma}_\uS (t)}{\dd t}=\bm{\Om}\bm{\mH}_\uS (t)\bm{\sigma}_\uS (t)-\bm{\sigma}_\uS(t) \bm{\mH}_\uS (t)\bm{\Om}+\bm{B}(t)
\label{transport.equation.S}
\eea
for the system, where
\bea
\label{eq:B:def}
\bm{B}(t) = -\xi(t)
\begin{pmatrix}
0 & \sigma_{\uS\uE,11}(t)\\
\sigma_{\uS\uE,11}(t) & 2 \sigma_{\uS\uE,21}(t)
\end{pmatrix}\, .
\eea 
The transport equation for the system has solutions of the form
\begin{equation}
\bm{\sigma}_\uS(t)=\bm{X}(\tO,t)\bm{\sigma}_\uS(t_0)\bm{X}^\T(\tO,t)+\bm{Y}(\tO,t)\, . \label{sigma.evolution}   
\end{equation}
Indeed, by inserting \cref{sigma.evolution} into \cref{transport.equation.S} one can readily check that the transport equations are satisfied provided the matrices $\bm{X}(\tO,t)$ and $\bm{Y}(\tO,t)$ satisfy
\bea
\label{eq:eom:X:Y}
\frac{\dd}{\dd t}{\bm{X}}(\tO,t) =& \bm{\Omega}\bm{\mH}_\uS(t)\bm{X}(\tO,t)\, ,\\
\frac{\dd}{\dd t}{\bm{Y}} (\tO,t)=&\bm{\Omega}\bm{\mH}_\uS(t)\bm{Y}(\tO,t)-\bm{Y}(\tO,t)\bm{\mH}_\uS(t)\bm{\Omega} + \bm{B}(t)\, ,
\eea
with initial conditions $\bm{X}(\tO,\tO)=\bm{1}$ and $\bm{Y}(\tO,\tO)=\bm{0}$. Note also that, upon composing \cref{sigma.evolution} at times $(\tO,t_1)$ with itself at times $(t_1,t_2)$, one can readily show that 
\bea
\label{X:Y:composition}
\bm{X}(t_0,t_2)=& \bm{X}(t_1,t_2) \bm{X}(t_0,t_1)\, , \\
\bm{Y}(t_0,t_2)=&\bm{X}(t_1,t_2)\bm{Y}(t_0,t_1)\bm{X}^\T(t_1,t_2)+\bm{Y}(t_1,t_2)\, .
\eea
This provides Gaussian maps with a semi-group structure
\bea
(\bm{X}(t_0,t_1),\bm{Y}(t_0,t_1))\circ(\bm{X}(t_1,t_2),\bm{Y}(t_1,t_2))=(\bm{X}(t_0,t_2), \bm{Y}(t_0,t_2))\, ,
\label{semigroup2}
\eea
so that condition is always fulfilled. Moreover, a positive density matrix is such that $\bm{\sigma}_\uS+i\bm{\Omega}\geq 0$, and for this condition to be conserved over time one requires~\cite{Lindblad_cloning}
\bea
\bm{Y}+i\bm{\Omega}-i\bm{X}\bm{\Omega} \bm{X}^\T\geq 0\, . \label{CP_criterion}    
\eea

An infinitesimal version of this positivity condition can be derived as follows. Between times $t$ and $t+\delta t$, \cref{eq:eom:X:Y} gives rise to $\bm{X}(t,t+\delta t)=\bm{1}+\bm{\Omega} \bm{\mH}_\uS(t) \delta t$ and $\bm{Y}(t,t+\delta t)=\bm{B}(t)\delta t$ at leading order in $\delta t$, hence \cref{CP_criterion} reduces to 
\bea
\bm{B} \geq 0\, . 
\eea 
From \cref{eq:B:def}, it is clear that $\det\bm{B}<0$, hence it is a non-positive matrix (with one positive and one negative eigenvalue). Therefore, the dynamics of the system is {\textit{always}} non-Markovian. This is in agreement with the findings in \cite{Kading:2025cwg}.

\subsection{Non-Markovian measure for Gaussian maps}

Having established that the dynamics is always non-Markovian, our next goal is to quantify the size of non-Markovian effects in the different phases of the evolution of the system. Our strategy is to measure the distance between $\hat{\rho}_\uS$ and a reference state $\hat{\tilde{\rho}}_\uS$ that evolves under a Markovian map. To that end, let us introduce the Bures distance $B(\hat{\rho}_1,\hat{\rho}_2)$ between two quantum states $\hat{\rho}_1$ and $\hat{\rho}_2$
\begin{equation}
B(\hat{\rho}_1,\hat{\rho}_2)=\sqrt{2[1-\sqrt{F(\hat{\rho}_1,\hat{\rho}_2)}]}
\, ,\quad\text{where}\quad
F(\hat{\rho}_1,\hat{\rho}_2)=\left(\mathrm{Tr}\sqrt{\sqrt{\rho_1}\rho_2\sqrt{\rho_1}}\right)^2
\end{equation}
is called fidelity. If $\hat{\rho}_1$ and $\hat{\rho}_2$ are two Gaussian states with vanishing first moments and covariance matrices given by $\bm{\sigma}_{1}$ and $\bm{\sigma}_{2}$ respectively, the fidelity is given by~\cite{1998JPhA...31.3659S,Nha:2005PRA}
\begin{equation}
F(\hat{\rho}_1,\hat{\rho}_2) = \frac{2}{\sqrt{\det(\bm{\sigma}_{1}+\bm{\sigma}_{2})+\left(\det\bm{\sigma}_{1}-1\right)\left(\det\bm{\sigma}_{2}-1\right)}-\sqrt{\left(\det\bm{\sigma}_{1}-1\right)\left(\det\bm{\sigma}_{2}-1\right)}}.
\end{equation}
The Bures distance has a number of interesting properties~\cite{2018PhRvA..98a2120A,2019NJPh...21e3036C}. It is conserved through unitary evolution, so its time evolution is a direct tracer of non-unitarity. More generally, it is a contraction under the action of completely positive trace-preserving maps. It also satisfies the triangular inequality, and when evaluated between two pure states $\vert\psi_1\rangle$ and $\vert\psi_2\rangle$ it boils down to the trace distance $T(\hat{\rho}_1,\hat{\rho}_2)=\mathrm{Tr}[\sqrt{(\hat{\rho}_1-\hat{\rho}_2)^\dagger(\hat{\rho}_1-\hat{\rho}_2)}]/2=\sqrt{1-\vert \langle\psi_1\vert\psi_2\rangle \vert^2}$.

As a reference Markovian map, we consider \cref{transport.equation.S} where the negative eigenvalue of $\bm{B}$ is dropped. In practice, denoting by 
\bea 
\lambda_\pm= \xi\left[-\sigma_{\uS\uE,21}\pm\sqrt{\sigma_{\uS\uE,21}^2+\sigma_{\uS\uE,11}^2}\right]
\eea
the two eigenvalues of $\bm{B}$, we thus perform the replacement
\bea
\label{eq:B:Btilde}
\bm{B}=
\begin{pmatrix}
0 & - \sqrt{-\lambda_- \lambda_+}\\
- \sqrt{-\lambda_- \lambda_+} & \lambda_+ + \lambda_-
\end{pmatrix}
\longrightarrow
\tilde{\bm{B}}=
\begin{pmatrix}
0 & 0 \\
0 & \lambda_+
\end{pmatrix}\, .
\eea 
When $\bm{B}$ is replaced by $\tilde{\bm{B}}$ in \cref{transport.equation.S}, the evolved covariance matrix is denoted $\tilde{\bm{\sigma}}_\uS$. We wish to measure the distance between $\bm{\sigma}_\uS$ and its Markovian counterpart $\tilde{\bm{\sigma}}_\uS$. More precisely, in order to diagnose the amount of non-Markovianity at a given instant in time rather than from cumulative effects, we consider the ``Bures velocity'' $v_B$ that we define as follows. At time $t$, we consider the state of the system $\bm{\sigma}_\uS(t)$. If we evolve it with the map~\eqref{transport.equation.S} until $t+\delta t$, we obtain $\bm{\sigma}_\uS(t+\delta t)$. Likewise, if we evolve it with the Markovian reference map, we obtain $\tilde{\bm{\sigma}}_\uS(t+\delta t)$. The distance between $\bm{\sigma}_\uS(t+\delta t)$ and $\tilde{\bm{\sigma}}_\uS(t+\delta t)$ measures the size of non-Markovian effects between $t$ and $t+\delta t$, which leads us to defining 
\bea
\label{eq:Bures:velocity}
v_B(t) &=\lim_{\delta t\to 0} \frac{B\left[\bm{\sigma}_\uS(t+\delta t),\tilde{\bm{\sigma}}_\uS(t+\delta t)\right]}{\delta t}\\ 
&=\frac{\sqrt{2}}{4}\sqrt{\frac{\det(\bm{\sigma}_\uS)}{\det^2(\bm{\sigma}_\uS)-1}\left(\mathrm{Tr}^2\left[\bm{\sigma}_\uS^{-1} \left(\bm{B}-\tilde{\bm{B}}\right)\right]+\left[\det(\bm{\sigma}_\uS)-1\right]\mathrm{Tr}\left\lbrace\left[\bm{\sigma}_\uS^{-1} \left(\bm{B}-\tilde{\bm{B}}\right)\right]^2\right\rbrace\right)} 
\eea 
where we have used \cref{transport.equation.S} to derive the second expression.\footnote{We also made use of the generic formula
\bea
\det(\bm{M}+\epsilon\bm{\delta M})=&\det(\bm{M})\left(1+\epsilon\Tr(\bm{M}^{-1}\bm{\delta M})+\frac{\epsilon^2}{2}\left\lbrace\Tr^2(\bm{M}^{-1}\bm{\delta M})-\Tr\left[(\bm{M}^{-1}\bm{\delta M})^2\right]\right\rbrace\right)
+\mO(\epsilon^3)\, .
\eea} For any $2\times 2$ matrix $\bm{M}$, from Cayley-Hamilton's theorem one has $\Tr^2(\bm{M})-\Tr(\bm{M}^2)=2\det(\bm{M})$, hence in the present case where the system is comprised of a single degree of freedom \cref{eq:Bures:velocity} reduces to 
\bea
\label{eq:Bures:velocity:2}
v_B(t)
&=\frac{\sqrt{2}}{4}\sqrt{\frac{\det^2(\bm{\sigma}_\uS)}{\det^2(\bm{\sigma}_\uS)-1}\mathrm{Tr}^2\left[\bm{\sigma}_\uS^{-1} \left(\bm{B}-\tilde{\bm{B}}\right)\right] - \frac{2 \det(\bm{\sigma}_\uS)}{\det(\bm{\sigma}_\uS)+1} \det \left[\bm{\sigma}_\uS^{-1} \left(\bm{B}-\tilde{\bm{B}}\right)\right]} .
\eea 

\begin{figure}[t]
\centering 
\includegraphics[width=0.48\textwidth]{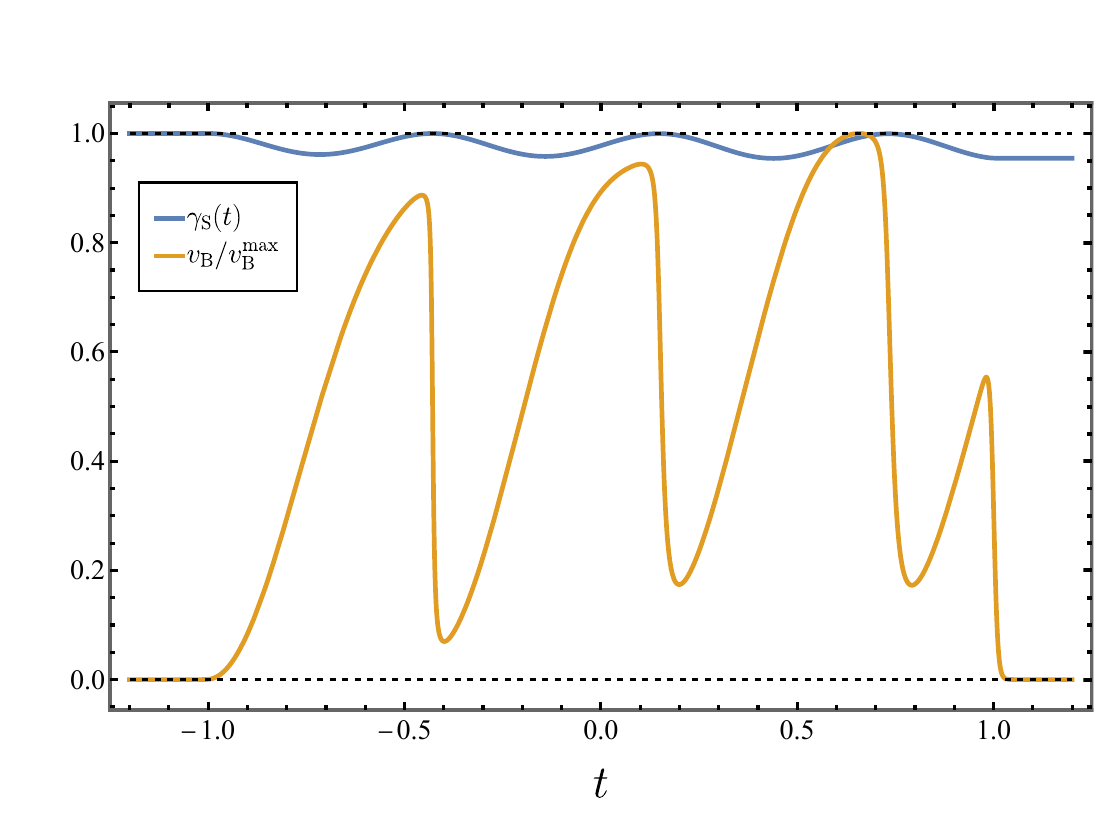} 
\includegraphics[width=0.48\textwidth]{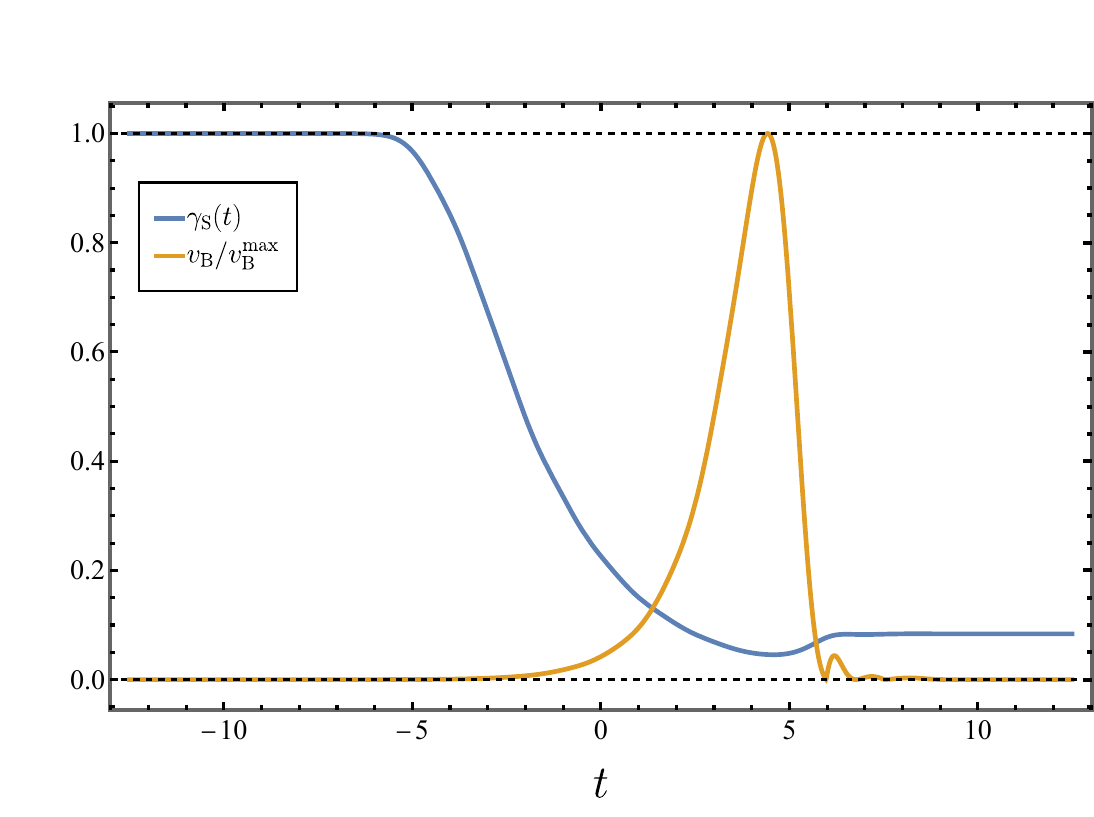} 
\includegraphics[width=0.48\textwidth]{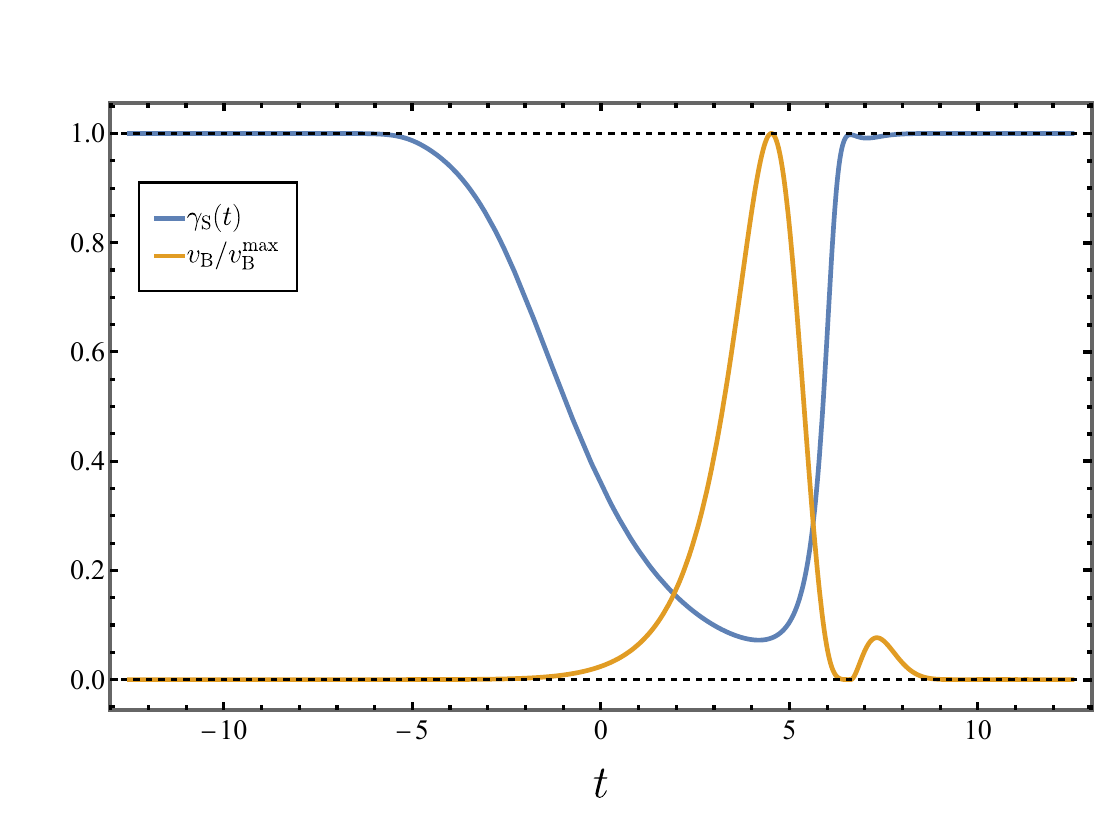} 
\caption{Purity, and Bures velocity~\eqref{eq:Bures:velocity:2} normalised to its maximal value, as a function of time in the subcritical regime ($\om_\uS=1$, $\om_\uE=10$, $\xio=0.5\xic$, $\tau=0.01$, $\tO=1$; first panel), in the supercritical decohering regime ($\om_\uS=1$, $\om_\uE=2$, $\xio=1.1\xic$, $\tau=1$, $\tO=5$, second panel) and in the supercritical recohering regime ($\om_\uS=1$, $\om_\uE=10$, $\xio=1.1\xic$, $\tau=1$, $\tO=5$, third panel).
}
\label{fig.purity.and.rates} 
\end{figure}

The result is displayed in  \cref{fig.purity.and.rates} for a few cases of interest. The first panel describes the subcritical regime, where the purity oscillates. When $\gamma_\uS$ is below one, the Bures velocity continuously increases, and abruptly decays when $\gamma_\uS$ reaches one again. It is thus larger in the recohering phase than in the decohering phase, which is consistent with the expectation that recoherence is a strongly non-Markovian effect. In the second panel, decoherence takes place in the supercritical regime, and we find that the Bures velocity grows as long as decoherence takes place, and only goes back to zero at late times when the evolution of the system becomes unitary (hence Markovian) again. Finally, the third panel also belongs to the supercritical regime, but this time, recoherence takes place at late times. Here as well, one can see that the Bures velocity increases in the decohering phase, and decreases in the recohering phase, crossing a maximum when the purity crosses a minimum. This leads us to conclude that, although recoherence is a non-Markovian mechanism by nature, the decohering phases of the system's evolution also proceed in the non-Markovian regime according to our measure, and there is no obvious relationship between de/re-coherence and Bures velocity.

\subsection{Markovian approximation}

The above considerations show that, although natural, the replacement~\eqref{eq:B:Btilde} does not provide a reliable Markovian approximation of the system's dynamics, even in the decohering phase. In this section, we show how the \emph{optimal} Markovian approximation can be found, still based on the Bures-velocity measure. Our goal is to identify the positive symmetric matrices $\tilde{\bm{B}}$ that minimise the right-hand side of \cref{eq:Bures:velocity:2}. 

A first remark is that the only matrix $\tilde{\bm{B}}$ that cancels out the Bures velocity is the matrix $\bm{B}$ itself. Indeed, denoting $\bm{M}=\bm{\sigma}_\uS^{-1} (\bm{B}-\tilde{\bm{B}})$, and recalling that $\det(\bm{\sigma}_\uS) = {\gamma}_\uS^{-2}$, see \cref{purity.def}, the vanishing of the right-hand side of \cref{eq:Bures:velocity} implies that $\Tr^2(\bm{M}) = 2(1-\gamma_\uS^2)\det(\bm{M})$. Denoting $\mu_{\pm}$ the two eigenvalues of $\bm{M}$, this leads to $(\mu_++\mu_-)^2=2(1-\gamma_\uS^2)\mu_- \mu_+$, hence $\mu_-^2+\mu_+^2=-\frac{\gamma_\uS^2}{1-\gamma_\uS^2} (\mu_++\mu_-)^2 $. With $\gamma_\uS <1$,\footnote{The case $\gamma_\uS=1$ is singular since the Bures velocity~\eqref{eq:Bures:velocity:2} diverges in this limit. If $\gamma_\uS$ remains equal to one the best Markovian approximation is obviously the unitary map $\tilde{\bm{B}}=0$, otherwise $\gamma_\uS$ deviates from one as soon as non-unitary effects take place.} the left-hand side is manifestly non-negative and the right-hand side non-positive, hence both sides vanish when $\mu_+=\mu_-=0$. This implies that $\bm{M}=0$, hence $\tilde{\bm{B}}=\bm{B}$, as stated above. 

One can then show that  $\tilde{\bm{B}}=\bm{B}$ is the only local minimum of the Bures velocity, seen as a function of $\tilde{\bm{B}}$. Indeed, from \cref{eq:Bures:velocity:2} one can show that $\partial v_B/\partial M_{11} \propto  M_{11}+ \gamma_\uS^2 M_{22}$, $\partial v_B/\partial M_{12} \propto M_{12}$ and $\partial v_B/\partial M_{22} \propto \gamma_\uS^2 M_{11}+ M_{22}$. With $\gamma_\uS <1$, these three partial derivatives vanish at $M_{11}=M_{12}=M_{22}=0$ only, hence $\bm{M}=0$ and $\tilde{\bm{B}}=\bm{B}$ again.

Let us now minimise the Bures velocity over the set of \emph{positive} $\tilde{\bm{B}}$ matrices. If $\tilde{\lambda}_{\pm}$ denote the two eigenvalues of $\tilde{\bm{B}}$, we have just shown that the only local minimum is at $\tilde{\bm{B}}=\bm{B}$, for which $\tilde{\lambda}_{-}=\lambda_-<0$ and $\tilde{\lambda}_{+}=\lambda_+>0$. This means that, in the quarter plane $\tilde{\lambda}_{\pm} > 0$, there is no local minimum. As a consequence, when minimising the Bures velocity over $\tilde{\lambda}_{\pm} \geq 0$, the minimum must lie within the boundary of the quarter plane, \ie at $\tilde{\lambda}_-=0$. If $\phi$ denotes the angle of the rotation that diagonalises $\tilde{\bm{B}}$, that matrix is thus of the form
\bea
\tilde{\bm{B}} = \tilde{\lambda}_+ 
\begin{pmatrix}
\cos^2\phi & \cos\phi\sin\phi\\
 \cos\phi\sin\phi & \sin^2\phi
\end{pmatrix}\, .
\eea
Plugging this ansatz into \cref{eq:Bures:velocity:2}, one finds that 
\bea
8 v_B^2 = A_0 + A_1 \tilde{\lambda}_+ + A_2 \tilde{\lambda}_+^2
\eea
with
\begin{align}
\label{eq:A0}
A_0= &\frac{1}{1-\gamma_\uS^4}\Tr^2\left(\bm{\sigma}_\uS^{-1} \bm{B}\right)-\frac{2 \gamma_\uS^2}{1+\gamma_\uS^2}\det(\bm{B})\, ,\\
A_1=& \frac{2 \gamma_\uS^2}{1+\gamma_\uS^2}\left(B_{22}\cos^2\phi - 2B_{12}\cos\phi\sin\phi+B_{11}\sin^2\phi\right) 
\nonumber \\  &
\label{eq:A1}
- \frac{2\gamma_\uS^2}{1-\gamma_\uS^4} \Tr\left(\bm{\sigma}_\uS^{-1} \bm{B}\right)\left(\sigma_{\uS,22}\cos^2\phi - 2\sigma_{\uS,12}\cos\phi\sin\phi+\sigma_{\uS,11}\sin^2\phi\right)\, ,\\
\label{eq:A2}
A_2= & \frac{\gamma_\uS^4}{1-\gamma_\uS^4}\left(\sigma_{\uS,22}\cos^2\phi - 2\sigma_{\uS,12}\cos\phi\sin\phi+\sigma_{\uS,11}\sin^2\phi\right)^2\, .
\end{align}
Since $\det B<0$ and $\gamma_\uS<1$, one has $A_0>0$ and $A_2>0$. This implies that, if $A_1>0$ then $v_B$ is minimal at $\tilde{\lambda}_+=0$ where $8v_B^2=A_0$, while if $A_1<0$ then $v_B$ is minimal at $\tilde{\lambda}_+=-A_1/(2 A_2)$ where $8 v_B^2=A_0-A_1^2/(4A_2)$. The optimal situation is thus when $A_1$ is negative and $A_1^2/(4A_2)$ is maximal. 

Although it is straightforward to maximise this quantity over $\phi$ numerically,
it is instructive to consider the case where $\phi=\pi/2$, which corresponds to having the positive eigendirections of $\bm{B}$ and $\tilde{\bm{B}}$ aligned, as was done in \cref{eq:B:Btilde}. Since $B_{11}=0$, it implies that the first term in  \cref{eq:A1} vanishes. Moreover, it is also worth noting that \cref{transport.equation.S} leads to 
\bea
\label{eq:gamma:dot}
\dot{\gamma}_\uS = - \frac{\gamma_\uS}{2}\mathrm{Tr}\left(\bm{\sigma}_\uS^{-1} \bm{B}\right)\, ,
\eea 
hence $A_1= 4 \gamma_\uS \dot{\gamma}_\uS \sigma_{\uS,11}/(1-\gamma_\uS^4)$. The sign of $A_1$ is therefore determined by the sign of $\dot{\gamma}_\uS$: if recoherence takes place, $\dot{\gamma}_\uS>0$, then $A_1>0$, hence $v_B$ is minimised at $\tilde{\lambda}_+=0$, \ie $\tilde{\bm{B}}=0$ and the best Markovian approximation turns out to be merely unitary evolution. In contrast, if decoherence takes place, $\dot{\gamma}_\uS<0$, $A_1<0$, hence $v_B$ is minimised at $\tilde{\lambda}_+=-A_1/(2 A_2)$ where 
\bea
v_B=\frac{\gamma_\uS}{2\sqrt{1+\gamma_\uS^2}}\sqrt{\vert \det(\bm{B})\vert}\, .
\eea

To summarise, we have found that the dynamics of the system can never be described by a Markovian map, regardless of whether recoherence or decoherence takes place. We have exhibited a family of Markovian maps for which unitary evolution is the best Markovian approximation of recohering systems. This is consistent with the intuitive understanding of recoherence as a memory effect. More generally, we have shown how the best Markovian approximation can be found by minimising the Bures velocity over the set of positive $\tilde{\bm{B}}$ matrices. Let us stress that $\bm{\sigma}_\uS$ appears explicitly in the optimal values of $\phi$ and $\tilde{\lambda}_+$, which does not define a bona-fide ``map''. How to optimise Markovian approximations to \emph{maps} is a more general question, but it is worth noting that this optimisation problem can be formulated by means of the Bures velocity.

\section{Summary and conclusion}
\label{sec:Conclusions}

The purpose of this work was to study the relationships between recoherence, adiabaticity, and Markovianity in open quantum systems. This analysis is motivated by physical situations where recoherence occurs, \ie cases where the system purifies at late times after a transient phase of decoherence.  Such situations are notably found in cosmology, where the field describing density fluctuations couples to other degrees of freedom (either other fields or unobserved scales) during cosmic inflation. As the expansion of the universe proceeds, scales of physical interest become larger than the Hubble radius (\ie the scale associated with the expansion rate of the background), which leads to explosive particle production and the increase in entanglement between all interacting degrees of freedom. For this reason, decoherence has long been thought to be a ubiquitous outcome of inflation and to play an important role in the quantum-to-classical transition of cosmological perturbations. The recent discovery of setups where recoherence takes place~\cite{Colas:2022kfu, Colas:2024ysu} thus appears surprising, and the precise conditions under which it occurs remain to be clarified. 

In cosmology, the system and the environment interact in a dynamical background that comes with its own time scale (namely, the Hubble time), and the interaction does not formally switch off at late times since the background dynamics always proceeds in the super-Hubble regime during the late stage of inflation. This is why, in order to leave us the possibility to manipulate the various time scales independently, in this work we considered a simple toy model made of two linearly-coupled harmonic oscillators. The two oscillators, with frequencies $\om_\uS$ and $\om_\uE$, are initially uncoupled, and the coupling function $\xi(t)$ vanishes both in the asymptotic past and the asymptotic future. Our goal was to compute the purity of either oscillator at late times. Despite the seemingly simplicity of the setup, it presents a strikingly diverse phenomenology. The other advantage of the model is that it can be cast in terms of transport equations that can be solved numerically, and it is prone to several analytical approximations, which we have developed.

Let us summarise our findings. We first showed that there exists a critical value for the interaction strength, $\xic=\om_\uS \om_\uE$. If $\xi(t)$ remains below $\xic$, the purity merely oscillates during the interacting phase, at a frequency $\om_\uE$ and with an amplitude controlled by $\om_\uS/\om_\uE (\xi/\xi_\uc)^2$, while it decays exponentially when $\xi(t)$ exceeds $\xic$.

We then noted that there also exists an upper bound on $\xi$ below which the interaction behaves perturbatively, \ie the dynamics is well approximated by a systematic expansion in the coupling strength $\xi$. This bound is given by \cref{eq:gp:w:psi}, and it not only depends on $\xio/\xic$ but also $\omega_\uS/\omega_\uE$. Therefore, there is a subcritical, perturbative regime, as well as a supercritical, non-perturbative regime, but there also exists a supercritical, perturbative regime (as well as a mildly subcritical, non-perturbative regime). The four possibilities are sketched in the parameter space of \cref{fig.phase.diagram}, and we sampled 8 points in that parameter space in order to investigate all possible behaviours. For simplicity, we assumed that the interaction instantaneously switches on and off. This makes a piecewise, fully analytical solution available, which we then expanded in different limits. Our results are summarised in \cref{tab.cases}. Although seemingly surprising, the formal validity of the perturbative approach in the supercritical regime was found to be restricted to the early stage of the dynamics, where the behaviour of the purity is dominated by rapid oscillations that perturbation theory is able to reproduce well. At late times, however, the slow decay of the purity is not properly accounted for, hence perturbation theory breaks down due to secular effects. We thus concluded that \textbf{\textit{decoherence is always a non-perturbative phenomenon}}, in the sense that it only takes place in the non-perturbative regime.

If the interaction is instantaneously switched off, the purity simply freezes to its value prior to the switch off, and recoherence never takes place. This is why we then considered the opposite limit where the time scale $\tau$ over which the interaction is turned on and off, is large. We developed a new, systematic adiabatic expansion, which is based on working in the field basis where the two oscillators would decouple if the coupling strength $\xi(t)$ were constant, and implement perturbation theory there. In that (so-called ``adiabatic'') basis, the interaction is mediated by $\dot{\xi}$ rather than $\xi$, hence it is indeed suppressed for slowly-varying coupling functions. At leading order, the setup tracks the vacuum state in the adiabatic basis, and our scheme reduces to the WKB expansion. It shows its full power at higher order where particle production occurs even in the adiabatic basis, and where non-adiabatic corrections are captured by the simple tools of standard perturbation theory. Regarding recoherence, we found that the system always fully recoheres (\ie its purity asymptotes one at late times) at leading order in the adiabatic expansion, and in fact, at any finite order. We confirmed numerically that the late-time purity is indeed non-perturbative in $\tau^{-1}$, and concluded that \textbf{\textit{decoherence is always a non-adiabatic phenomenon}}. 

We then characterised more precisely the conditions under which recoherence takes place in the subcritical regime. When $\tau\gg \tO$, where $\tO$ is half the duration of the interacting phase, recoherence always occurs, since this regime is manifestly adiabatic. Otherwise, we found that there is still a lower bound for $\tau$ above which recoherence takes place, see \cref{eq:Twthr:tau}. In summary, the adiabatic condition reads
\bea
\tau\gg \mathrm{min} \left(\tO, \frac{\om_\uS}{\om_\uE-\om_\uS}\tO \right) .
\eea 
This determines when recoherence occurs, which was one of the main goals of this paper.

Recoherence can be interpreted as a backflow of information from the environment into the system, and as such, it is, by essence, a non-Markovian phenomenon. We thus went on to focus more closely on the relationship between recoherence and Markovianity. A trace-preserving quantum map is said to be Markovian if it satisfies the semi-group property and if it is completely positive (\ie if it maps positive density matrices to positive density matrices only). Gaussian maps always satisfy the semi-group property, and the positivity condition reduces to requiring that the $\bm{B}$ matrix appearing in the transport equation is positive. We found that, in the family of toy models under consideration, the $\bm{B}$ matrix always possesses one positive and one negative eigenvalue. Therefore, although recoherence is in essence non-Markovian, \textbf{\textit{the dynamics of the system is never Markovian, even when it decoheres}}.

We then proposed to measure the amount of non-Markovianity a given Gaussian map features by tracking the speed at which the state of the system deviates from one that evolves under a reference Markovian map. That reference map was obtained from the original map by switching off the negative eigenvalue of $\bm{B}$ and setting it to zero. Since the distance between the states evolved with both maps was measured with the Bures distance, constructed from quantum fidelity, we dubbed this non-Markovian measure the ``Bures velocity''. In the subcritical regime, the purity oscillates, and the Bures velocity is larger in the recohering phase than in the decohering phase, in agreement with the fact that recoherence is a strongly non-Markovian effect. In the supercritical regime, however, the Bures velocity increases as long as decoherence takes place, and decreases at late times if recoherence occurs. Therefore, there is no clear relationship between de/re-coherence and Bures velocity. Finally, we showed how to construct the best Markovian approximation to a given map. We found that, even in the decohering phase, there is no Markovian map (\ie a positive $\bm{B}$ matrix) such that the Bures velocity vanishes. We have exhibited one family of Markovian maps within which, in the recohering regime, the best Markovian approximation consists in taking $\bm{B}=\bm{0}$, thus neglecting all non-unitary effects. The best Markovian approximation is nothing but the unitary approximation in that case. To summarise, \textbf{\textit{the dynamics of the system is never Markovian, even when decoherence takes place. We have introduced the Bures velocity and showed how it can be used to find the best Markovian approximation to a given map}}.

 Let us conclude by mentioning that it remains to apply these findings to physical situations where recoherence is likely to occur, in particular in the cosmological context. Our results should allow us to assess when to expect recoherence to take place, which regimes can be described by an adiabatic expansion, and how to develop a Markovian approximation in the decohering regime. Finally, some of the technical tools developed in this work would be worth investigating further. In particular, the adiabatic expansion, which may be viewed as a WKB-like scheme for interacting systems, has the potential to unveil subtle non-adiabatic effects in a broader range of situations than those discussed here. Note that for explicitness, we implemented the expansion in the subcritical regime, but it can also be used to describe the supercritical regime, if properly matched at the critical-point crossings. This could be useful to better probe the cases where recoherence takes place in the supercritical regime. Moreover, if the system and the environment interact non linearly, it may not always be possible to introduce an adiabatic basis explicitly, unless the non-linear part of the interaction is treated perturbatively. Generalising our approach to non-linear setups is therefore another interesting prospect. Finally, we proposed a new measure of non-Markovianity dubbed the Bures velocity, which we used to optimise the design of Markovian approximations. Such an optimisation problem often arises in the study of open quantum systems and we thus anticipate that the Bures velocity could have other applications.

\section*{Acknowledgements}
 We warmly thank Thomas Colas for insightful discussions and very interesting comments on this manuscript. We thank the DIM ORIGINES program from {\textit{R\'egion \^Ile de France}} for financial support.

\appendix

\section{The ISOSO limit} 
\label{app:ISOSO}
In this Appendix, we provide detailed formulas for the instantaneous switch on and switch off (ISOSO) limit studied in \cref{sec:ISOSO}. Inserting \cref{eq:ISOSO:x:A,eq:ISOSO:x:B} into the matching condition~\eqref{eq:ISOSO:matching}, the $\hat{b}_i$ operators can be expressed in terms of the $\hat{a}_i$ operators as follows,
\bea
\hat{b}_1=&\alpha_{1\uS}\hat{a}_{\uS}+\alpha_{1\uE}\hat{a}_{\uE} +\beta_{1\uS}\hat{a}_{\uS}  ^\dagger+\beta_{1\uE}\hat{a}_{\uE}^\dagger\, , \quad \hat{b}_1^\dagger=\alpha_{1\uS}^*\hat{a}_{\uS}^\dagger+\alpha_{1\uE}^*\hat{a}_{\uE}^\dagger+\beta_{1\uS}^*\hat{a}_{\uS}  +\beta_{1\uE}^*\hat{a}_{\uE}\, ,\\
\hat{b}_2=&\alpha_{2\uS}\hat{a}_{\uS}+\alpha_{2\uE}\hat{a}_{\uE} +\beta_{2\uS}\hat{a}_{\uS} ^\dagger+\beta_{2\uE}\hat{a}_{\uE}^\dagger\, , \quad \hat{b}_2^\dagger=\alpha_{2\uS}^*\hat{a}_{\uS}^\dagger+\alpha_{2\uE}^*\hat{a}_{\uE}^\dagger+\beta_{2\uS}^*\hat{a}_{\uS}+\beta_{2\uE}^*\hat{a}_{\uE}\, ,
\eea
where
\bea
\alpha_{1\uS}=\frac{\cos\theta}{2}\left(\sqrt{\frac{|\om_1|}{\om_\uS }}+i^{-\delta}\sqrt{\frac{\om_\uS }{|\om_1|}}\right)e^{i\om_\uS(\tO+\tin)}&, \quad \alpha_{1\uE}=-\frac{\sin\theta}{2}\left(\sqrt{\frac{|\om_1|}{\om_\uE }}+i^{-\delta}\sqrt{\frac{\om_\uE }{|\om_1|}}\right)e^{i\om_\uE(\tO+\tin)},\\
\beta_{1\uS}=\frac{\cos\theta}{2}\left(\sqrt{\frac{|\om_1|}{\om_\uS }}-i^{-\delta}\sqrt{\frac{\om_\uS }{|\om_1|}}\right)e^{-i\om_\uS(\tO+\tin)}&, \quad \beta_{1\uE}=-\frac{\sin\theta}{2}\left(\sqrt{\frac{|\om_1|}{\om_\uE }}-i^{-\delta}\sqrt{\frac{\om_\uE }{|\om_1|}}\right)e^{-i\om_\uE(\tO+\tin)},\\
\alpha_{2\uS}=\frac{\sin\theta}{2}\left(\sqrt{\frac{\om_2}{\om_\uS }}+\sqrt{\frac{\om_\uS }{\om_2}}\right)e^{i\om_\uS(\tO+\tin)}&, \quad \alpha_{2\uE}=\frac{\cos\theta}{2}\left(\sqrt{\frac{\om_2}{\om_\uE }}+\sqrt{\frac{\om_\uE }{\om_2}}\right)e^{i\om_\uE(\tO+\tin)},\\
\beta_{2\uS}=\frac{\sin\theta}{2}\left(\sqrt{\frac{\om_2}{\om_\uS }}-\sqrt{\frac{\om_\uS }{\om_2}}\right)e^{-i\om_\uS(\tO+\tin)}&, \quad \beta_{2\uE}=\frac{\cos\theta}{2}\left(\sqrt{\frac{\om_2}{\om_\uE }}-\sqrt{\frac{\om_\uE }{\om_2}}\right)e^{-i\om_\uE(\tO+\tin)}\, .
\eea
Here, we have written $\om_1=i^\delta |\om_1|$, hence $\delta=0$ if $\xio\leq\xic$ (subcritical regime) and $\delta=1$ if $\xio>\xic$ (supercritical regime). This leads to the following two-point correlation function for the $\hat{b}_i$ operators, when evaluated on the vacuum state 
\bea
\langle \hat{b}_1\hat{b}_1 \rangle=&\frac{\cos^2(\theta)}{4}\left[\frac{|\om_1|}{\om_\uS }-(-1)^\delta \frac{\om_\uS }{|\om_1|}\right]+\frac{\sin^2(\theta)}{4}\left[\frac{|\om_1|}{\om_\uE }-(-1)^\delta \frac{\om_\uE }{|\om_1|}\right],
\eea
\bea
\langle \hat{b}_1\hat{b}_1^\dagger \rangle=&\frac{1}{8}\left[4\cos\left(\frac{\pi\delta}{2}\right)+\left(\om_\uS+\om_\uE\right)\left(\frac{|\om_1|}{\om_\uS\om_\uE}+\frac{1}{|\om_1|}\right)
\right. \\ & \left.
+\left(\om_\uE-\om_\uS\right)\left(\frac{|\om_1|}{\om_\uS\om_\uE}-\frac{1}{|\om_1|}\right)\cos(2\theta)\right],
\eea 
\bea 
\langle \hat{b}_1\hat{b}_2 
\rangle =&\frac{\sin(2\theta)}{8}\sqrt{|\om_1|\om_2}\left(\frac{1}{\om_\uS}-\frac{1}{\om_\uE}+i^{-\delta}\,\frac{\om_\uE-\om_\uS}{|\om_1|\om_2}\right),
\eea 
\bea 
\langle \hat{b}_1\hat{b}_2^\dagger \rangle =&\frac{\sin(2\theta)}{8}\sqrt{|\om_1|\om_2}\left(\frac{1}{\om_\uS}-\frac{1}{\om_\uE}-i^{-\delta}\,\frac{\om_\uE-\om_\uS}{|\om_1|\om_2}\right),
\eea
\bea
\langle \hat{b}_1^\dagger\hat{b}_1 \rangle =&\frac{1}{8}\left[-4\cos\left(\frac{\pi\delta}{2}\right)+\left(\om_\uS+\om_\uE\right)\left(\frac{|\om_1|}{\om_{\uS}\om_{\uE}}+\frac{1}{|\om_1|}\right)
\right. \\ & \left.
+\left(\om_\uE-\om_\uS\right)\left(\frac{|\om_1|}{\om_{\uS}\om_{\uE}}-\frac{1}{|\om_1|}\right)\cos(2\theta)\right],
\eea
\bea
\langle \hat{b}_1^\dagger\hat{b}_1^\dagger \rangle=&\frac{\cos^2(\theta)}{4}\left[\frac{|\om_1|}{\om_\uS }-(-1)^\delta \frac{\om_\uS }{|\om_1|}\right]+\frac{\sin^2(\theta)}{4}\left[\frac{|\om_1|}{\om_\uE }-(-1)^\delta \frac{\om_\uE }{|\om_1|}\right],
\eea
\bea
\langle \hat{b}_1^\dagger\hat{b}_2\rangle =&\frac{\sin(2\theta)}{8}\sqrt{|\om_1|\om_2}\left(\frac{1}{\om_\uS}-\frac{1}{\om_\uE}-i^\delta\,\frac{\om_\uE-\om_\uS}{|\om_1|\om_2}\right),
\eea
\bea
\langle \hat{b}_1^\dagger\hat{b}_2^\dagger \rangle =&\frac{\sin(2\theta)}{8}\sqrt{|\om_1|\om_2}\left(\frac{1}{\om_\uS}-\frac{1}{\om_\uE}+i^\delta\,\frac{\om_\uE-\om_\uS}{|\om_1|\om_2}\right),
\eea
\bea
\langle \hat{b}_2\hat{b}_1 \rangle =&\frac{\sin(2\theta)}{8}\sqrt{|\om_1|\om_2}\left(\frac{1}{\om_\uS}-\frac{1}{\om_\uE}+i^{-\delta}\,\frac{\om_\uE-\om_\uS}{|\om_1|\om_2}
\right),
\eea
\bea
\langle \hat{b}_2\hat{b}_1^\dagger \rangle=&\frac{\sin(2\theta)}{8}\sqrt{|\om_1|\om_2}\left(\frac{1}{\om_\uS}-\frac{1}{\om_\uE}-i^{\delta}\,\frac{\om_\uE-\om_\uS}{|\om_1|\om_2}\right),
\eea
\bea
\langle \hat{b}_2\hat{b}_2\rangle=&\frac{\cos^2(\theta)}{4}\left(\frac{\om_2}{\om_\uE }-\frac{\om_\uE }{\om_2}\right)+\frac{\sin^2(\theta)}{4}\left(\frac{\om_2}{\om_\uS }-\frac{\om_\uS }{\om_2}\right),
\eea
\bea
\langle \hat{b}_2\hat{b}_2^\dagger \rangle=&\frac{\cos^2(\theta)}{4}\left(\frac{\om_2}{\om_\uE }+\frac{\om_\uE }{\om_2}+2\right)+\frac{\sin^2(\theta)}{4}\left(\frac{\om_2}{\om_\uS }+\frac{\om_\uS }{\om_2}+2\right),
\eea
\bea
\langle \hat{b}_2^\dagger\hat{b}_1\rangle =&\frac{\sin(2\theta)}{8}\sqrt{|\om_1|\om_2}\left(\frac{1}{\om_\uS}-\frac{1}{\om_\uE}-i^{-\delta}\,\frac{\om_\uE-\om_\uS}{|\om_1|\om_2}\right),
\eea
\bea
\langle \hat{b}_2^\dagger\hat{b}_1^\dagger \rangle=&\frac{\sin(2\theta)}{8}\sqrt{|\om_1|\om_2}\left(\frac{1}{\om_\uS}-\frac{1}{\om_\uE}+i^\delta\,\frac{\om_\uE-\om_\uS}{|\om_1|\om_2 }\right),
\eea
\bea
\langle \hat{b}_2^\dagger\hat{b}_2\rangle=&\frac{\cos^2(\theta)}{4}\left(\frac{\om_2}{\om_\uE }+\frac{\om_\uE }{\om_2}-2\right)+\frac{\sin^2(\theta)}{4}\left(\frac{\om_2}{\om_\uS }+\frac{\om_\uS }{\om_2}-2\right),\\
\langle \hat{b}_2^\dagger\hat{b}_2^\dagger \rangle=&\frac{\cos^2(\theta)}{4}\left(\frac{\om_2}{\om_\uE }-\frac{\om_\uE }{\om_2}\right)+\frac{\sin^2(\theta)}{4}\left(\frac{\om_2}{\om_\uS }-\frac{\om_\uS }{\om_2}\right)  .    
\eea
This allows one to compute the two-point correlation functions of the position and momentum operators in the phase $-\tO<t<\tO$, since from \cref{eq:ISOSO:x:B} they are related to the ladder-operator correlation functions according to
\bea
\langle\hat{x}_\uS (t)\hat{x}_\uS (t)\rangle=&\frac{\cos^2(\theta)}{2|\om_1|}\left(\langle \hat{b}_1\hat{b}_1^\dagger\rangle+ \langle \hat{b}_1^\dagger\hat{b}_1\rangle+\langle \hat{b}_1^2\rangle e^{-i2\om_1\Delta t}+\langle \hat{b}_1^{\dagger 2}\rangle e^{+i2\om_1\Delta t}\right)\\
&+\frac{\sin^2(\theta)}{2\om_2}\left(\langle \hat{b}_2\hat{b}_2^\dagger\rangle+\langle \hat{b}_2^\dagger\hat{b}_2\rangle+\langle \hat{b}_2^2\rangle e^{-i2\om_2\Delta t}+\langle \hat{b}_2^{\dagger2}\rangle e^{+i2\om_2\Delta t}\right)\\
&+\frac{\sin(2\theta)}{4\sqrt{|\om_1|\om_2}}\left[\left(\langle \hat{b}_1^\dagger\hat{b}_2\rangle+\langle \hat{b}_2\hat{b}_1^\dagger\rangle \right)e^{-i(\om_2-\om_1)\Delta t}+\left(\langle \hat{b}_1\hat{b}_2^\dagger\rangle+\langle \hat{b}_2^\dagger\hat{b}_1\rangle \right)e^{+i(\om_2-\om_1)\Delta t}\right]\\
&+\frac{\sin(2\theta)}{4\sqrt{|\om_1|\om_2}}\left[\left(\langle \hat{b}_1\hat{b}_2\rangle+\langle \hat{b}_2\hat{b}_1\rangle\right) e^{-i(\om_1+\om_2)\Delta t}+\left(\langle \hat{b}_1^\dagger\hat{b}_2^\dagger\rangle+\langle \hat{b}_2^\dagger\hat{b}_1^\dagger\rangle\right) e^{+i(\om_1+\om_2)\Delta t}\right],\label{xSB.xSB}
\eea
\bea
\langle\hat{p}_\uS (t)\hat{p}_\uS (t)\rangle=&\frac{\cos^2(\theta)}{2}(-1)^\delta|\om_1|\left(\langle \hat{b}_1\hat{b}_1^\dagger\rangle+ \langle \hat{b}_1^\dagger\hat{b}_1\rangle-\langle \hat{b}_1^2\rangle e^{-i2\om_1\Delta t}-\langle \hat{b}_1^{\dagger 2}\rangle e^{+i2\om_1\Delta t}\right)\\
&+\frac{\sin^2(\theta)}{2}\om_2\left(\langle \hat{b}_2\hat{b}_2^\dagger\rangle+\langle \hat{b}_2^\dagger\hat{b}_2\rangle-\langle \hat{b}_2^2\rangle e^{-i2\om_2\Delta t}-\langle \hat{b}_2^{\dagger2}\rangle e^{+i2\om_2\Delta t}\right)\\
&+\frac{\sin(2\theta)}{4}i^\delta \sqrt{|\om_1|\om_2}\left[\left(\langle \hat{b}_1^\dagger\hat{b}_2\rangle+\langle \hat{b}_2\hat{b}_1^\dagger\rangle \right)e^{-i(\om_2-\om_1)\Delta t}+\left(\langle \hat{b}_1\hat{b}_2^\dagger\rangle+\langle \hat{b}_2^\dagger\hat{b}_1\rangle \right)e^{+i(\om_2-\om_1)\Delta t}\right]\\
&-\frac{\sin(2\theta)}{4}i^\delta \sqrt{|\om_1|\om_2}\left[\left(\langle \hat{b}_1\hat{b}_2\rangle+\langle \hat{b}_2\hat{b}_1\rangle\right) e^{-i(\om_1+\om_2)\Delta t}+\left(\langle \hat{b}_1^\dagger\hat{b}_2^\dagger\rangle+\langle \hat{b}_2^\dagger\hat{b}_1^\dagger\rangle\right) e^{+i(\om_1+\om_2)\Delta t}\right], \label{pSB.pSB}
\eea
\bea
\langle\hat{x}_\uS (t)\hat{p}_\uS (t)\rangle=&i\frac{\cos^2(\theta)}{2}i^\delta\left(\langle \hat{b}_1\hat{b}_1^\dagger\rangle-\langle \hat{b}_1^\dagger\hat{b}_1\rangle-\langle \hat{b}_1^2\rangle e^{-i2\om_1\Delta t}+\langle \hat{b}_1^{\dagger 2}\rangle e^{+i2\om_1\Delta t}\right)\\
&+i\frac{\sin^2(\theta)}{2}\left(\langle \hat{b}_2\hat{b}_2^\dagger\rangle-\langle \hat{b}_2^\dagger\hat{b}_2\rangle-\langle \hat{b}_2^2\rangle e^{-i2\om_2\Delta t}+\langle \hat{b}_2^{\dagger2}\rangle e^{+i2\om_2\Delta t}\right)\\
&+i\frac{\sin(2\theta)}{4\sqrt{|\om_1|\om_2}}\left[\left(\om_1\langle \hat{b}_2\hat{b}_1^\dagger\rangle-\om_2\langle \hat{b}_1^\dagger\hat{b}_2\rangle\right)e^{-i(\om_2-\om_1)\Delta t}-\left(\om_1\langle \hat{b}_2^\dagger\hat{b}_1\rangle-\om_2\langle \hat{b}_1\hat{b}_2^\dagger\rangle \right)e^{+i(\om_2-\om_1)\Delta t}\right]\\
&-i\frac{\sin(2\theta)}{4\sqrt{|\om_1|\om_2}}\left[\left(\om_1\langle \hat{b}_2\hat{b}_1\rangle+\om_2\langle \hat{b}_1\hat{b}_2\rangle\right) e^{-i(\om_1+\om_2)\Delta t}-\left(\om_1\langle \hat{b}_2^\dagger\hat{b}_1^\dagger\rangle+\om_2\langle \hat{b}_1^\dagger\hat{b}_2^\dagger\rangle\right) e^{+i(\om_1+\om_2)\Delta t}\right], \label{xSB.pSB}
\eea
\bea
\langle\hat{p}_\uS (t)\hat{x}_\uS (t)\rangle=&-i\frac{\cos^2(\theta)}{2}i^\delta\left(\langle \hat{b}_1\hat{b}_1^\dagger\rangle-\langle \hat{b}_1^\dagger\hat{b}_1\rangle+\langle \hat{b}_1^2\rangle e^{-i2\om_1\Delta t}-\langle \hat{b}_1^{\dagger 2}\rangle e^{+i2\om_1\Delta t}\right)\\
&-i\frac{\sin^2(\theta)}{2}\left(\langle \hat{b}_2\hat{b}_2^\dagger\rangle-\langle \hat{b}_2^\dagger\hat{b}_2\rangle+\langle \hat{b}_2^2\rangle e^{-i2\om_2\Delta t}-\langle \hat{b}_2^{\dagger2}\rangle e^{+i2\om_2\Delta t}\right)\\
&-i\frac{\sin(2\theta)}{4\sqrt{|\om_1|\om_2}}\left[\left(\om_2\langle \hat{b}_2\hat{b}_1^\dagger\rangle-\om_1\langle \hat{b}_1^\dagger\hat{b}_2\rangle\right)e^{-i(\om_2-\om_1)\Delta t}-\left(\om_2\langle \hat{b}_2^\dagger\hat{b}_1\rangle-\om_1\langle \hat{b}_1\hat{b}_2^\dagger\rangle \right)e^{+i(\om_2-\om_1)\Delta t}\right]\\
&-i\frac{\sin(2\theta)}{4\sqrt{|\om_1|\om_2}}\left[\left(\om_2\langle \hat{b}_2\hat{b}_1\rangle+\om_1\langle \hat{b}_1\hat{b}_2\rangle\right) e^{-i(\om_1+\om_2)\Delta t}-\left(\om_2\langle \hat{b}_2^\dagger\hat{b}_1^\dagger\rangle+\om_1\langle \hat{b}_1^\dagger\hat{b}_2^\dagger\rangle\right) e^{+i(\om_1+\om_2)\Delta t}\right]  \label{pSB.xSB}
\eea
where $\Delta t=t+t_0$.

\section{Generalised Paley–Wiener theorem} \label{slow.vs.fast.integral}
Let us consider an integral of the form
\begin{align}
\label{eq:I:1}
\mathcal{I}=\int_{\varphi_-}^{\varphi_+} e^{i G(\varphi')} F(\varphi') \dd \varphi'\, ,
\end{align}
where $F$ and $G$ are two analytic functions. We assume that $G$ is monotonic and varies over $\varphi$ scales much shorter than the one of $F$, while $G'$ is as slow as $F$. Our goal is to perform an expansion of the integral $\mathcal{I}$ in the ratio of these two $\varphi$ scales. The situation is indeed analogous to the one discussed in \cref{sec:late:time:purity}, where $F$ contains combinations of $\om_1(t)$, $\om_2(t)$, $\theta(t)$, $\dot{\om}_1(t)$, $\dot{\om}_2(t)$, $\dot{\theta}(t)$, hence varies over time scales of order $\tau$, and $G$ contains $W_i(t)$ for $i=1,2$, hence it varies over time scales $\om_i^{-1}$ while $G'\propto \om_i$ varies over time scales $\tau$, see \cref{W}.

We first perform a change of integration variable $\varphi \to G(\varphi)$ such that \cref{eq:I:1} takes the form
\begin{align}
\int_{\varphi_-}^{\varphi_+} e^{i \varphi'} F(\varphi') \dd \varphi'\, ,
\end{align}
where $F$ has been redefined according to $F\to F/G'$ and is still a slow function, \ie~it varies over $\varphi$ scales much larger than one. Then, we split the integral according to
\begin{align}
\int_{\varphi_-}^{\varphi_+} e^{i \varphi'} F(\varphi') \dd \varphi' = \sum_\ell \int_{\varphi_\ell}^{\varphi_{\ell+1}} e^{i \varphi'} F(\varphi') \dd \varphi'
\end{align}
where $\varphi_\ell = 2 \ell \pi$. Within each interval, let us expand
\begin{align}
F(\varphi') = \sum_{n=0}^\infty \frac{F^{(n)}\left(\varphi_\ell\right)}{n!}\left(\varphi'-\varphi_\ell\right)^n\, ,
\end{align}
hence
\begin{align}
\int_{\varphi_\ell}^{\varphi_{\ell+1}} e^{i \varphi'} F(\varphi') \dd \varphi' = \sum_{n=1}^\infty\frac{F^{(n)}\left(\varphi_\ell\right)}{n!} \mathcal{I}_n
 \end{align}
 where 
 \begin{align}
 \mathcal{I}_n = \int_0^{2\pi} \varphi^n e^{i\varphi} \dd\varphi\, .
 \end{align}
 Integrating by parts $n$ times, one finds
 \begin{align}
  \mathcal{I}_n = \sum_{k=1}^n \left(2\pi\right)^k i^{n-k-1} \frac{n!}{k!}\, .
 \end{align}
 Therefore, one has
 \begin{align}
\int_{\varphi_\ell}^{\varphi_{\ell+1}} e^{i \varphi'} F(\varphi') \dd \varphi' = & 
  \sum_{n=1}^\infty \sum_{k=1}^n \frac{(2\pi)^k}{k!} i^{n-k-1}F^{(n)}(\varphi_\ell)\\
  = & \sum_{k=1}^\infty \sum_{n=k}^\infty \frac{(2\pi)^k}{k!} i^{n-k-1}F^{(n)}(\varphi_\ell)\, .
 \end{align}
 We would like to rewrite this as 
 \begin{align}
\int_{\varphi_\ell}^{\varphi_{\ell+1}} H(\varphi') \dd \varphi' = & \sum_{k=0}^\infty H^{(k)}(\varphi_\ell) \int_0^{2\pi} \frac{(\varphi')^k}{k!}\dd\varphi'\\
   = &  \sum_{k=0}^\infty H^{(k)}(\varphi_\ell) \frac{(2\pi)^{k+1}}{(k+1)!}\\
   = & \sum_{k=1}^\infty H^{(k-1)}(\varphi_\ell) \frac{(2\pi)^{k}}{k!}\, ,
 \end{align}
where $H$ is a function to determine. By identifying the two expressions, one can show that 
\begin{align}
H(\varphi) = - \sum_{n=1}^\infty i^n F^{(n)} (\varphi) 
\end{align}
fulfils that condition, \ie~by deriving the above expression $k-1$ times one finds indeed that $H^{(k-1)}(\varphi) = \sum_{n=k}^\infty i^{n-k-1}F^{(n)}(\varphi) $. Combining the above results, we have shown that
\begin{align}
\mathcal{I}=\int_{\varphi_-}^{\varphi_+} e^{i \varphi'} F(\varphi') \dd\varphi' =  \sum_{n=0}^\infty i^{n-1} \left[F^{(n)} (\varphi_+) - F^{(n)} (\varphi_-)\right].
\end{align}

Let us now consider the case where $\varphi_{\pm}\to\pm\infty$. If $F$ and all its derivatives asymptote zero quickly enough, the above sum vanishes, and one finds $\int_{-\infty}^{+\infty} e^{i \varphi'} F(\varphi') \dd \varphi' =0$. This is the case for the model considered in this work, since the interaction~\eqref{xi} is exponentially suppressed in the asymptotic past and future. Obviously, this does not mean $\mathcal{I}$ strictly vanishes, but simply that it is a non-analytic function of the inverse $\varphi$ scale of $F$.  For instance, using the residue theorem, it is easy to see that
\begin{align}
\int_{-\infty}^\infty \frac{e^{i t/\epsilon}}{1+t^2} \dd t =\pi e^{-\frac{1}{\epsilon}}\, ,
\end{align}
which is indeed non-analytic around $\epsilon=0$. It explains the exponential, non-perturbative behaviour encountered in \cref{fig.gS.perturbativity.check}. This generalises the Paley–Wiener theorem~\cite{Paley-Wiener} (originally derived for functions of compact support).

\bibliographystyle{JHEP}
\bibliography{refs} 

\end{document}